\documentclass[twocolumn, accepted=2025-03-26, letterpaper]{quantumarticle}
\pdfoutput=1
\usepackage[numbers,sort&compress]{natbib}
\usepackage{amsmath}
\usepackage{graphicx}
\usepackage{refcount}
\usepackage{fmtcount}
\usepackage{amssymb}
\usepackage{dsfont}
\usepackage{hyperref}
\usepackage[usenames,dvipsnames]{color}
\usepackage[normalem]{ulem}
\usepackage{tikz}
\usetikzlibrary{calc}
\usepackage{bm}
\usepackage{quantikz}
\usepackage{tikz-cd}

\hypersetup{pdfstartview=FitH,pdfpagemode=UseOutlines,colorlinks,
citecolor=blue,linkcolor=blue,urlcolor=blue}

\graphicspath{{.}{./Figures/}}

\newcommand{\dg}{\dagger}
\newcommand{\sd}{\downarrow}
\newcommand{\su}{\uparrow}

\newcommand{\norm}[1]{|\!| #1 |\!|}

\definecolor{blue(munsell)}{rgb}{0.0, 0.5, 0.69}

\definecolor{aaron}{rgb}{0.6, 0.6, 0.8}

\begin{document}


\title{Ground state energy and magnetization curve of a frustrated magnetic system from real-time evolution on a digital quantum processor}

\author{Aaron Szasz}
	\email[]{aszasz@google.com}
	\affiliation{Applied Mathematics and Computational Research Division, Lawrence Berkeley National Laboratory, Berkeley, California 94720, USA}
	\orcid{0000-0002-1127-2111}
\author{Ed Younis}
	\affiliation{Applied Mathematics and Computational Research Division, Lawrence Berkeley National Laboratory, Berkeley, California 94720, USA}
	\orcid{0000-0002-1306-1860}
\author{Wibe Albert de Jong}
	\affiliation{Applied Mathematics and Computational Research Division, Lawrence Berkeley National Laboratory, Berkeley, California 94720, USA}
	\orcid{0000-0002-7114-8315}


\begin{abstract}
Models of interacting many-body quantum systems that may realize new exotic phases of matter, notably quantum spin liquids, are challenging to study using even state-of-the-art classical methods such as tensor network simulations.  Quantum computing provides a promising route for overcoming these difficulties to find ground states, dynamics, and more.  In this paper, we argue that recently developed hybrid quantum-classical algorithms based on real-time evolution are promising methods for solving a particularly important model in the search for spin liquids, the antiferromagnetic Heisenberg model on the two-dimensional kagome lattice.  We show how to construct efficient quantum circuits to implement time evolution for the model and to evaluate key observables on the quantum computer, and we argue that the method has favorable scaling with increasing system size.  We then restrict to a 12-spin star plaquette from the kagome lattice and a related 8-spin system, and we give an empirical demonstration on these small systems that the hybrid algorithms can efficiently find the ground state energy and the magnetization curve.  For these demonstrations, we use four levels of approximation: exact state vectors, exact state vectors with statistical noise from sampling, noisy classical emulators, and (for the 8-spin system only) real quantum hardware, specifically the Quantinuum H1-1 processor; for the noisy simulations and hardware demonstration, we also employ error mitigation strategies based on the symmetries of the Hamiltonian.  Our results strongly suggest that these hybrid algorithms present a promising direction for studying quantum spin liquids and more generally for resolving important unsolved problems in condensed matter theory and beyond.
\end{abstract}

\maketitle


\section{Introduction:}

Quantum spin liquids, insulating phases of matter with no magnetic order down to zero temperature and with long-range entanglement~\cite{Balents2010,Savary_2017,Zhou2017}, have been the focus of intense computational effort over the past 50 years.  The first theoretical proposal of a spin liquid came from Anderson in 1973, when he suggested that the true ground state of the antiferromagnetic Heisenberg model on a triangular lattice could be just such a state~\cite{ANDERSON1973153}.  Although the true ground state of that model has since been shown to be magnetically ordered~\cite{Huse1988,White2007}, the current community consensus is that related models, including the triangular lattice Heisenberg model with both first- and second-neighbor interactions~\cite{Kaneko2014, Hu2015, Zhu2015,Iqbal2016,Hu2019}, the triangular lattice Hubbard model~\cite{Motrunich2005,Yang2010,Mishmash2013,Shirakawa2017,Szasz2020PRX,Tocchio2020,Szasz2021PRB,Tocchio2021,Chen2022}, and the kagome lattice Heisenberg model~\cite{Yan2011,Depenbrock2012,Jiang2012,Iqbal2013,Iqbal2014,Liao2017,He2017, Lauchli2019, Motruk2023}, do have spin liquid ground states; on the other hand, the precise nature of these spin liquid states remains in dispute.  Other important problems in condensed matter physics, notably the mechanism for high-temperature superconductivity~\cite{High_Tc_review}, have likewise not been conclusively resolved despite extensive study.

This remaining uncertainty is due to the limits of each of our classical computational methods for studying ground states: exact diagonalization is limited to small systems sizes, quantum Monte Carlo suffers from the sign problem for frustrated models, tensor networks are limited to low-to-medium entanglement and thus struggle with gapless ground states, and variational Monte Carlo methods are biased by the choice of ansatz.  In principle, a ground state search on a quantum computer could avoid all of these pitfalls: a quantum algorithm could scale polynomially in the number of lattice sites, would not suffer from the sign problem, would have no limitation on allowable entanglement, and could be unbiased.  

However, many proposed quantum algorithms such as quantum phase estimation~\cite{qpe_Kitaev,QPE_prl_1997,qpe_prl_1999} require very deep circuits or many qubits and are thus not suitable for near-term intermediate-scale quantum (NISQ) devices~\cite{Preskill2018quantumcomputingin}.  On the other hand, the most commonly discussed NISQ-friendly method for ground state search, the variational quantum eigensolver (VQE)~\cite{McClean_2016,vqe_prl_2019,cerezo2021variational}, may not realize a quantum advantage over classical algorithms.  In the VQE approach, a parameterized quantum circuit prepares a state on the quantum device, and the parameters are optimized classically to find the parameters giving the lowest energy for the chosen circuit ansatz.  Unfortunately, VQE suffers from the barren plateau problem~\cite{mcclean2018barren,Wang2021,vqe_prl_2021,larocca2024}.  Namely, for parameterized circuits that are sufficiently expressive to capture the true ground state of a complex quantum system, the classical parameter optimization has exponential cost due to exponentially small gradients of the variational energy with respect to the circuit ansatz parameters.  Proposed strategies for getting around the barren plateau problem have thus far proven unsuccessful in general.  For example, a circuit ansatz inspired by the Hamiltonian of interest might need many fewer parameters to approximately describe the ground state, and hence would be easier to optimize~\cite{Wiersema2020}; however, this approach only avoids barren plateaus for certain problems~\cite{Wiersema2024}, and also an efficiently optimizable ansatz may be efficient to emulate classically~\cite{cerezo2024}.  Another hope would be to avoid barren plateaus by initializing the optimization with a state sufficiently close to the ground state; however, this ``warm start'' strategy requires increasingly good initial states with increasing system size~\cite{mhiri2025}.  Thus it remains unclear whether VQE can achieve quantum advantage for many problems of interest.

Our goal in this paper is to investigate a different class of hybrid quantum-classical algorithms, one based on real-time evolution and Krylov subspaces~\cite{stair2020multireference,UVQPE_Klymko,ding2022even,ding2023simultaneous,UVQPE_Yizhi,ODMD,motta2023subspace}.  We demonstrate (a) that they have a potential advantage over existing classical algorithms and (b) that they could be practical on NISQ devices.  In the algorithms we consider, we use a quantum computer to compute matrix elements $\langle \psi_0|e^{-iHt}|\psi_0\rangle$ for different values of the time delay $t$, where $|\psi_0\rangle$ is a state that ideally has a significant overlap with the true ground state.  These overlap matrix elements are then classically post-processed to estimate the ground state energy.  We specifically consider two such algorithms, unitary variational quantum phase estimation (UVQPE)~\cite{UVQPE_Klymko, UVQPE_Yizhi} and observable dynamic mode decomposition (ODMD)~\cite{ODMD}.  In each algorithm, the classical post-processing step is given by a small linear algebra problem: a generalized eigenvalue problem for UVQPE and a linear least squares problem for ODMD.  Both algorithms can be made noise-resilient by a simple regularization of the linear algebra problem, making them suitable for noisy quantum devices before the advent of full fault-tolerance. 

\begin{figure}
    \centering
    \includegraphics[width=\columnwidth]{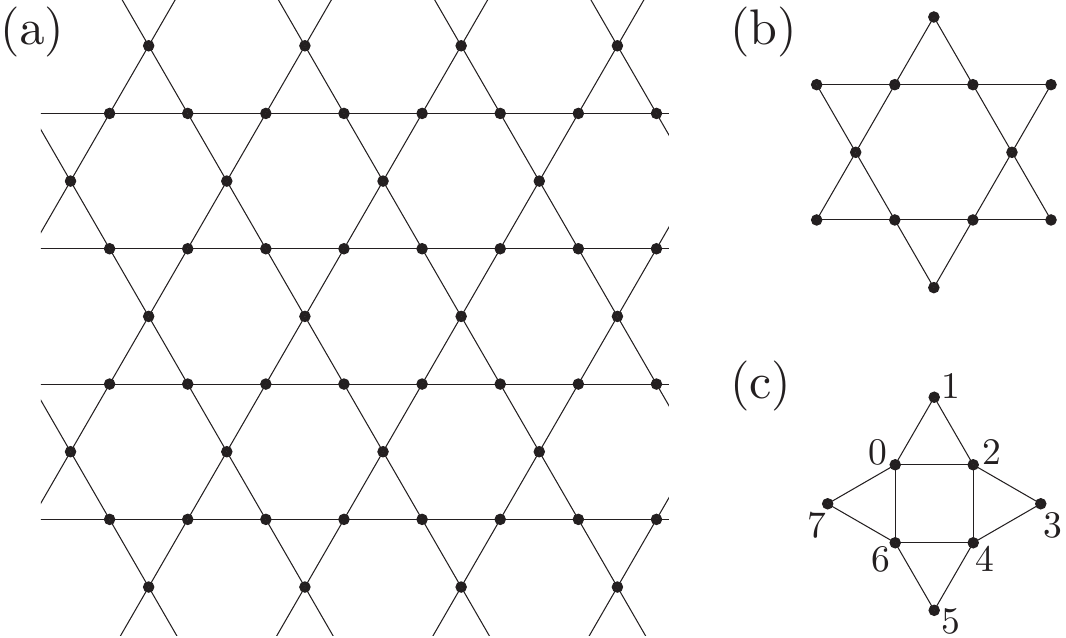}
    \caption{(a) The kagome lattice, on which antiferromagnetic interactions between particles are believed to lead to exotic phases of matter such as quantum spin liquids.  (b) A 12-spin star plaquette taken from the kagome lattice, which we use for a practical demonstration of hybrid algorithms.  (c) An analogous 8-spin plaquette, also studied in the present work.  Spins are numbered for later reference.}
    \label{fig:kl_and_plaquettes}
\end{figure}

The particular model we use for our study is the aforementioned Heisenberg model on the kagome lattice shown in Fig.~\ref{fig:kl_and_plaquettes}(a).  The current consensus in the community is that the ground state of the model is indeed a spin liquid, but there is debate as to the precise nature of the state.  The leading candidate for the ground state is a Dirac spin liquid~\cite{Iqbal2013,Iqbal2014,He2017}, which has no gap in the energy spectrum above the ground state, while other candidate ground states do have an energy gap~\cite{Yan2011,Depenbrock2012,Jiang2012}.  Thus the nature of the ground state could be largely settled just by finding conclusively whether the low-energy spectrum is gapped or gapless.  The model is also one of the hardest to study classically, as can be clearly seen in a recent survey paper~\cite{wu2023vscore}.  Because the question of gaplessness only involves examining low-lying energy eigenvalues, an apparently simple task that is nevertheless very difficult classically, the kagome lattice Heisenberg model is an ideal problem to study via quantum computer.  Indeed, VQE has also been demonstrated for studying this model~\cite{Bosse2022, Kattemolle2022, Wang2023} on clusters of up to 24 spins, finding accurate ground state energy with a classical emulator and moderately accurate energy with quantum hardware. However, due to the barren plateau problem discussed above, VQE will likely not scale well to system sizes large enough to answer the unresolved questions about the ground state of this classically challenging model.

We show how UVQPE and ODMD can be applied to find the ground state energy of the full 2D kagome lattice Heisenberg model with a lower computational cost compared with purely classical approaches.  However, to reduce noise to a manageable level and to gain access to classical benchmarks, for a practical demonstration using quantum hardware or a classical emulator we need a smaller system.  We thus opt for a proof-of-principle demonstration on a single 12-spin star plaquette of the kagome lattice and an analogous smaller star plaquette with 8 spins, shown in Fig.~\ref{fig:kl_and_plaquettes}(b) and (c), respectively.  For both plaquettes, we run UVQPE and ODMD using an exact classical emulator both with and without statistical noise from measurement sampling.  For the 8-spin plaquette, we also use a noisy classical emulator and the Quantinuum H1-1 processor.  In all cases, both algorithms converge rapidly to the true ground state energy.

In addition to ground state energy, another interesting property of candidate spin liquid models is the magnetization curve.  As an external magnetic field is applied and progressively strengthened, the zero-field spin liquid state can break down, giving rise to increasing magnetization and ultimately leading to a fully-polarized state.  However, frustrated materials often demonstrate a series of ``plateaus'' in the spectrum, where a particular state with fixed magnetization is stable across a range of applied field strengths.  These plateaus can also correspond to quantum spin liquids, or to a variety of interesting magnetic phases.  Making use of the spin-symmetry of the Heisenberg model, we also use UVQPE and ODMD to find the magnetization curve for the 8- and 12-spin star plaquettes with a computational cost only polynomially higher than that of finding just the ground state energy.

We emphasize that the search for quantum spin liquids is far from the only condensed matter application of UVQPE, ODMD, and related algorithms.  However, we select the kagome lattice Heisenberg model as our test case for several key reasons.  First, the problem is one of the most challenging for classical computational methods, so a quantum algorithm that can make progress here would likely be useful for many other important problems.  Second, the symmetries of the model, such as spin conservation, allow us to test algorithmic simplifications and error mitigation strategies that depend on those symmetries.  Finally, the ground state of the model is itself an important unsolved problem, hence even a new algorithm specific to this problem would be important.

The organization of the rest of the paper is as follows.  In Section \ref{sec:algs} we introduce the hybrid quantum-classical algorithms in some detail, including a discussion of the ``mirror circuit'' approach to computing state overlaps.  In Section \ref{sec:full_2D} we discuss the application of the methods to the Heisenberg model on the full 2D kagome lattice.  We focus on how to efficiently compute the expectation values $\langle \psi_0|e^{-iHt}|\psi_0\rangle$ on a quantum computer, including a model-specific discussion of an efficient symmetry-respecting implementation of the time evolution operator and a discussion of symmetry-based error mitigation.  We also argue that the hybrid algorithms scale well with increasing system size and are thus a promising approach to solving this challenging model.  In Section \ref{sec:star_plaquettes} we discuss the restriction of the model to the smaller star plaquettes, including how that restriction allows us to dramatically shorten the time-evolution circuits for use on quantum hardware.  In Section \ref{sec:results}, we show detailed results for the ground state energy from running UVQPE and ODMD on the star plaquettes, both with classical emulators and on hardware.  In Section \ref{sec:mag_curve}, we likewise present results for the magnetization curve.  Finally, we conclude in Section \ref{sec:discussion}.


\section{Hybrid algorithms based on real-time evolution\label{sec:algs}}

We use two hybrid quantum-classical algorithms to obtain ground state energies from real-time evolution on a quantum computer, UVQPE~\cite{UVQPE_Klymko} and ODMD~\cite{ODMD}.  For both algorithms, the quantum computer (or classical emulator) is used to find expectation values of the time evolution operator $U = e^{-iHt}$ in some initial state $|\psi_0\rangle$ for a range of times $t$.  The expectation values are then used to construct either a generalized eigenvalue problem or a linear least squares problem, which is solved classically to obtain an approximate ground state energy and a representation of the corresponding eigenstate.

We provide practical introductions to UVQPE and ODMD in \ref{sec:algs_UVQPE} and \ref{sec:algs_ODMD}, respectively.  We also provide, in \ref{sec:algs_mirror}, a thorough guide to using the mirror circuit approach~\cite{Cortes2022} to measure expectation values of the time evolution operator, of the form $\langle \psi_0|e^{-iHt}|\psi_0\rangle$.  In this section we focus on the general methods and postpone problem-specific details, such as how to find circuits to implement $U$ or to prepare $|\psi_0\rangle$, to Sec.~\ref{sec:full_2D} below.

\vspace{0.1cm}

\subsection{Unitary variational quantum phase estimation\label{sec:algs_UVQPE}}

The ground state of a Hamiltonian is far from generic.  For example, in one dimension, the ground state of a gapped Hamiltonian has area law entanglement~\cite{Hastings_2007, arad2013area}, compared with volume law entanglement for typical states.  Thus we might expect the ground state to have a high overlap with states that are relatively simple to prepare on a quantum computer.  With this insight in mind, one promising approach to estimating ground state energy is to (a) find a set of states that can be feasibly generated on a quantum computer and which may have large overlap with the ground state, and (b) find the lowest energy of a state in the subspace spanned by those states by solving a generalized eigenvalue problem.

Variational quantum phase estimation (VQPE)~\cite{UVQPE_Klymko} is a specific case of this approach, using a Krylov subspace in which the basis states are given by some initial state $|\psi_0\rangle$ that has significant overlap with the ground state and time-evolved versions of the state:
\begin{equation}
    |\psi_0\rangle,\,\,e^{-iH\Delta t}|\psi_0\rangle,\,\,e^{-2iH\Delta t}|\psi_0\rangle,\,\,e^{-3iH\Delta t}|\psi_0\rangle,\,\,\cdots\label{eq:basis_states}
\end{equation}
for some fixed time interval $\Delta t$.  The lowest energy within the subspace spanned by these states is the smallest eigenvalue of the generalized eigenvalue problem
\begin{widetext}
    \begin{equation}
    \left(\begin{array}{cccc} 
    \langle H\rangle & \langle e^{-iH\Delta t}H\rangle & \langle e^{-2iH\Delta t}H\rangle & \cdots \\
    \langle e^{iH\Delta t}H\rangle & \langle H\rangle & \langle e^{-iH\Delta t}H\rangle & \cdots \\
    \langle e^{2iH\Delta t}H\rangle & \langle e^{iH\Delta t}H\rangle & \langle H\rangle & \cdots\\
    \vdots & \vdots & \vdots & \ddots
    \end{array}\right) \mathbf{v} = 
    \left(\begin{array}{cccc} 
    1 & \langle e^{-iH\Delta t}\rangle & \langle e^{-2iH\Delta t}\rangle & \cdots \\
    \langle e^{iH\Delta t}\rangle & 1 & \langle e^{-iH\Delta t}\rangle & \cdots \\
    \langle e^{2iH\Delta t}\rangle & \langle e^{iH\Delta t}\rangle & 1 & \cdots\\
    \vdots & \vdots & \vdots & \ddots
    \end{array}\right) \lambda \mathbf{v}
    \end{equation}
where all expectation values are measured in the initial state $|\psi_0\rangle$ and we can truncate the basis \eqref{eq:basis_states} at any number of states that we choose.  Note that the matrices have a Toeplitz structure, with all entries on each diagonal being equal.  The energy estimate progressively improves as the number of basis states, i.e. the number of time evolution steps, increases.

A slight variant on this approach is to instead look for eigenvalues of the time evolution operator $U=e^{-i H T}$ for some fixed $T$.  If $T$ is small enough, so that the spectrum of $H\,T$ is bounded by an interval of size less than $2\pi$, then the eigenvalues of $H$ can be unambiguously determined from the eigenvalues of $U$.  In the special case that $T=\Delta t$, the same time step size used in generating the basis states, we get a significant improvement in the efficiency of the algorithm.  The generalized eigenvalue problem becomes
    \begin{equation}
    \left(\begin{array}{cccc} 
    \langle e^{-iH\Delta t}\rangle & \langle e^{-2iH\Delta t}\rangle & \langle e^{-3iH\Delta t}\rangle & \cdots \\
    1 & \langle e^{-iH\Delta t}\rangle & \langle e^{-2iH\Delta t}\rangle & \cdots \\
    \langle e^{iH\Delta t}\rangle & 1 & \langle e^{-iH\Delta t}\rangle & \cdots\\
    \vdots & \vdots & \vdots & \ddots
    \end{array}\right) \mathbf{v} = 
    \left(\begin{array}{cccc} 
    1 & \langle e^{-iH\Delta t}\rangle & \langle e^{-2iH\Delta t}\rangle & \cdots \\
    \langle e^{iH\Delta t}\rangle & 1 & \langle e^{-iH\Delta t}\rangle & \cdots \\
    \langle e^{2iH\Delta t}\rangle & \langle e^{iH\Delta t}\rangle & 1 & \cdots\\
    \vdots & \vdots & \vdots & \ddots
    \end{array}\right) \tilde{\lambda} \mathbf{v}.\label{eq:UVQPE}
    \end{equation}
\end{widetext}
We can then generate the matrices on both the left-hand side and the right-hand side just by measuring the expectation values
\begin{equation}
    \langle \psi_0|e^{-iH\Delta t}|\psi_0\rangle,\,\,\langle \psi_0|e^{-2iH\Delta t}|\psi_0\rangle,\,\,\cdots\label{eq:U_expecs}
\end{equation}
whereas in the original formulation we also needed to compute Hamiltonian matrix elements such as $\langle \psi_0|e^{-iH\Delta t}H|\psi_0\rangle$.  This reduces the number of measurements that need to be performed on the quantum computer by at least a factor of two, and most likely by more since $H$ is in general non-unitary and thus requires multiple circuits to compute each expectation value.  This modified version of the algorithm is called unitary variational quantum phase estimation, or UVQPE. 

Note that, as remarked above, the eigenvalues of $H$ can be unambiguously inferred from the eigenvalues of $U$ only if the spectrum of $H\times \Delta t$ is limited to an interval of size $2\pi$.  As a result, in order to use UVQPE we must pick a time step size $\Delta t$ that is inversely proportional to the spectral range of $H$. See Sec.~IIIB of Ref.~\cite{ODMD} for details.

Both variants of the method, VQPE and UVQPE, are susceptible to noise because the overlap matrix on the right-hand side tends to have a very large condition number.  This can be mitigated by performing a singular value decomposition of the overlap matrix and discarding all singular values below some threshold; in practice, we find good results when discarding singular values up to roughly 10x the noise level.  After applying the singular value thresholding, UVQPE becomes relatively noise-resilient, as shown in Ref.~\cite{ODMD}, while VQPE remains highly susceptible to noise.

In addition to extracting the ground state energy, we also gain access to the corresponding eigenvector.  The eigenvector for the lowest eigenvalue of the generalized eigenvalue problem describes a linear combination of the original basis states, known as a Ritz vector, that approximates the ground state of the original Hamiltonian.  We will make use of this capability below to understand the convergence of the algorithm in the case of the Heisenberg model on the star plaquettes.

\subsection{Observable dynamic mode decomposition\label{sec:algs_ODMD}}

While UVQPE with singular value thresholding of the overlap matrix is reasonably noise-resilient, the robustness of classically post-processing the expectation values in Eq.~\ref{eq:U_expecs} to get the ground state energy can potentially be improved by replacing the generalized eigenvalue problem by a linear least squares problem.  This approach is exemplified by a recently developed algorithm, observable dynamic mode decomposition (ODMD).~\cite{ODMD}

ODMD is based on the classical dynamic mode decomposition (DMD) method, where linear least squares is used to find a linearization of the time-evolution operator for classical state vectors.  In ODMD, we replace the classical state vectors by ``observable vectors'' of the form 
\begin{equation}
    \left(\begin{array}{cccc}\!\!\langle e^{-miH\Delta t}\rangle \!&\! \langle e^{-(m+1)iH\Delta t}\rangle \!&\! \cdots \!&\! \langle e^{-(m+d-1)iH\Delta t}\rangle\!\!\end{array}\right)^\text{T}
\end{equation} 
consisting of expectation values of $d$ successive powers of $e^{-iH\Delta t}$.  We then look for a matrix $A$ that approximately advances each observable vector to the next time step, 
\begin{equation}
    \left(\begin{array}{cccc}\!\!\langle e^{-(m+1)iH\Delta t}\rangle \!&\! \langle e^{-(m+2)iH\Delta t}\rangle \!&\!\! \cdots \!\!&\! \langle e^{-(m+d)iH\Delta t}\rangle\!\!\end{array}\right)^\text{T}.
\end{equation}
We can find the best linearization $A$ across many time steps by solving the linear least square problem
\begin{widetext}
    \begin{equation}
    \left(\begin{array}{cccc} 
    \langle e^{-iH\Delta t}\rangle & \langle e^{-2iH\Delta t}\rangle & \langle e^{-3iH\Delta t}\rangle & \cdots \\
    \langle e^{-2iH\Delta t}\rangle & \langle e^{-3iH\Delta t}\rangle & \langle e^{-4iH\Delta t}\rangle & \cdots \\
    \langle e^{-3iH\Delta t}\rangle & \langle e^{-4iH\Delta t}\rangle & \langle e^{-5iH\Delta t}\rangle & \cdots\\
    \vdots & \vdots & \vdots & \ddots
    \end{array}\right)  = A
    \left(\begin{array}{cccc} 
    1 & \langle e^{-iH\Delta t}\rangle & \langle e^{-2iH\Delta t}\rangle & \cdots \\
    \langle e^{-iH\Delta t}\rangle & \langle e^{-2iH\Delta t}\rangle & \langle e^{-3iH\Delta t}\rangle & \cdots \\
    \langle e^{-2iH\Delta t}\rangle & \langle e^{-3iH\Delta t}\rangle & \langle e^{-4iH\Delta t}\rangle & \cdots\\
    \vdots & \vdots & \vdots & \ddots
    \end{array}\right).\label{eq:ODMD}
    \end{equation}
\end{widetext}
Note that in ODMD the matrices have a Hankel structure, with the entries on each anti-diagonal being equal. 
We then classically diagonalize $A$, which is a small and well-conditioned matrix, and the most negative eigenvalue phase is an approximation to the ground state energy.  Like in VQPE or UVQPE, ODMD also gives access to the ground state as a linear combination of the basis states of Eq.~\eqref{eq:basis_states}, as shown in Ref.~\cite{ODMD} Sec.~3C.  

Here too we can make the solution more robust to noise by performing singular value thresholding on the matrix on the right-hand side.  When using this thresholding approach, ODMD is slightly more resilient to statistical error/shot noise than is UVQPE.~\cite{ODMD}

\subsection{\texorpdfstring{Mirror circuits for $e^{-iHt}$ expectation values}{Mirror circuits for time evolution expectation values}\label{sec:algs_mirror}}

Expectation values like $O\equiv\langle \psi_0|e^{-iHt}|\psi_0\rangle$ that are used as input for the classical post-processing portion of UVQPE and ODMD are commonly computed using the Hadamard test~\cite{Cleve:1997dh}.  However, this approach can be prohibitively expensive because it involves controlled time evolution that replaces single-qubit gates by two-qubit ones and two-qubit gates by three-qubit ones.~\footnote{The multi-qubit gates can be avoided for certain nice models~\cite{Efekan}, but will be required in general.}  Especially in the near-term, we want to reduce circuit depth as much as possible, so a better solution is needed, one that allows for lower gate depth.

Fortunately, there is precisely just such a method for computing matrix elements on a quantum computer~\cite{Havlicek2019,Cortes2022}.  The first key insight is that we can view the state $|\psi_0\rangle$ as $U_0|\mathbf{0}\rangle$ for some state preparation unitary $U_0$, where $|\mathbf{0}\rangle = |0\rangle^{\otimes N}$ is the all-0 state.  Note that finding a circuit implementing $U_0$ is problem-specific; we outline the construction for the example case of the Heisenberg model in Sec.~\ref{sec:full_2D_state_prep} and Sec.~\ref{sec:star_plaquettes_psi0} below.  Then the expectation value becomes
\begin{equation}
    O = \langle\psi_0|e^{-iHt}|\psi_0\rangle =  \langle\mathbf{0}\,\bigg|\,U_0^\dg e^{-iHt} U_0^{ }|\mathbf{0}\rangle,
\end{equation}
which we view as an inner product between the state on the left of the large vertical line, $|\mathbf{0}\rangle$, and the state on the right, 
\begin{equation}
    |0(t)\rangle \equiv U_0^\dg e^{-iHt} U_0^{ }|\mathbf{0}\rangle.\label{eq:0t}
\end{equation}
Then the matrix element $O = \langle \mathbf{0}|0(t)\rangle$ is just the coefficient of the all-0 basis state when the state $|0(t)\rangle$ is written in the standard basis.  Thus if we measure many copies of $|0(t)\rangle$ in the standard basis, we can approximate $|O|^2$ as the fraction of measured bit strings that are all zeros.  Following Ref.~\cite{Cortes2022}, we refer to this sampled probability as $F_1$.

So far we have only found $|O|$ and not the full expectation value $O$ including phase, so we have not yet demonstrated a viable replacement for the Hadamard test.  One solution is proposed in \cite{Cortes2022}, making use of a reference state $|\text{Ref}\rangle$ that is easy to prepare, is orthogonal to $|\psi_0\rangle$, and is an eigenstate of $H$ with known eigenvalue $E_R^{ }$.  In other words, $|\text{Ref}\rangle$ must be an eigenstate of $H$ whose energy can be derived classically.  For the types of challenging problems where a quantum algorithm is needed to find the ground state, this type of classically solvable eigenstate can typically be found only by making use of symmetries of the particular Hamiltonian of interest; we give a concrete example for the Heisenberg model in Sec.~\ref{sec:full_2D} below.  For now, we assume such a reference state exists.  Then we can find a circuit that prepares 
\begin{equation}
    U_{R}^{ }|\mathbf{0}\rangle = \left(|\psi_0\rangle + |\text{Ref}\rangle\right)/\sqrt{2}.\label{eq:F2_superposition}
\end{equation}  
Using this circuit, we prepare the state 
\begin{equation}
    |0_R(t)\rangle = U_{R}^\dg e^{-iHt} U_{R}^{ }|\mathbf{0}\rangle\label{eq:0Rt}
\end{equation} 
and measure in the standard basis.  The observed probability that the measurement returns the all-0 bit string, which we call $F_2$, approaches a theoretical value of 
\begin{align}
    F_2 & \rightarrow |(\langle \psi_0|+\langle\text{Ref}|)e^{-iHt}(|\psi_0\rangle+|\text{Ref}\rangle)|^2/4 \\
    & = (r^2_{ } + 1 + 2r\cos(\theta + E_R^{ } t))/4 \nonumber
\end{align}
where $r$ and $\theta$ are defined by $O\equiv r e^{i\theta}$.

Ref~\cite{Cortes2022} stops here, with access to $|O|$ and the cosine of the phase $(\theta + E_R t)$.  However, there is still an ambiguity of the phase, $\theta \leftrightarrow -(\theta+2E^{ }_R t)$.  To resolve this, we also want access to the sine of the phase, which we can find with 
\begin{align}
    & |(-i\langle \psi_0|+\langle\text{Ref}|)e^{-iHt}(|\psi_0\rangle+|\text{Ref}\rangle)|^2/4\\
    = & (r^2 + 1 +2r\sin(\theta + E^{ }_R t)/4.\nonumber
\end{align}  
This quantity is approximated by $F_3$, the probability of measuring the all-0 bit string in the state
\begin{equation}
    |0_{Ri}(t)\rangle = U_{Ri}^\dg e^{-iHt} U_{R}^{ }|\mathbf{0}\rangle\label{eq:0Rit}
\end{equation}
where $U_{Ri}$ prepares the superposition
\begin{equation}
    U_{Ri}^{ }|\mathbf{0}\rangle = \left(i|\psi_0\rangle + |\text{Ref}\rangle\right)/\sqrt{2}\label{eq:F3_superposition}
\end{equation}
with an extra phase of $\pi/2$ between $|\psi_0\rangle$ and the reference state.

To summarize, $F_1$, $F_2$, and $F_3$ are given by the measured probability of the all-0 state in the three states $|0(t)\rangle$, $|0_R(t)\rangle$, and $|0_{Ri}(t)\rangle$ as given in Eqs.~\eqref{eq:0t}, \eqref{eq:0Rt}, and \eqref{eq:0Rit}, respectively.  From these three measured quantities we can get the overlap according to
\begin{equation}
    O = r e^{i\theta} = \left[ 2F_2 + 2iF_3 - \left(F_1+1\right)\cdot\left(\frac{i+1}{2}\right)\right]e^{-i E^{ }_R t}_{ }.\label{eq:mirror_circuit_computation}
\end{equation}
It turns out, however, that we can slightly improve this result.  Here the magnitude of $O$, $r=|O|$, is computed from the combined results of three circuits, for $F_1$, $F_2$, and $F_3$.  In general the error from convolving the probability distributions for three independent quantities will be larger than for any of the individual distributions, suggesting that we could do better by computing the magnitude from $F_1$ alone, $r=\sqrt{F_1}$, while keeping the angle from \eqref{eq:mirror_circuit_computation}.  

Indeed, using $F_1$ alone for the magnitude is already beneficial even if we consider just shot noise from sampling an otherwise noiseless simulation, as we show in Appendix~\ref{appendix:mirror_shot_noise}.  A further benefit is that the hardware noise for the $F_1$ circuit will be slightly lower than for $F_2$ and $F_3$, both because (a) the circuit is shorter since it does not involve the superposition with a reference state and (b) we can apply symmetry-based error mitigation that is much simpler than in the presence of a reference state (see Sec.~\ref{sec:star_plaquettes_error_miti}).

Another question one could ask is, if we have a fixed measurement budget, i.e. a fixed total number of shots to be distributed between $F_1$, $F_2$, and $F_3$, how should we distribute the shots among the three circuits?  As we show in Appendix~\ref{appendix:mirror_shot_noise}, the optimal choice for a noiseless simulation is to use approximately the same number of shots for each of the three circuits, even when we only use $F_1$ to determine $|O|$.  The reduced hardware noise for $F_1$ suggests that in practice we might want to allocate slightly more shots to that circuit.  Hence in the experimental demonstration reported in Sec.~\ref{sec:results} below, we divide our shots among the three circuits as 40\% to $F_1$ and 30\% each to $F_2$ and $F_3$.


\section{Application to the kagome lattice Heisenberg model\label{sec:full_2D}}

In this section, we first precisely define the Heisenberg model and discuss useful symmetries of the model.  We then show how the expectation values $\langle \psi_0|e^{-iHt}|\psi_0\rangle$ can be efficiently computed for this model using the mirror circuit approach.  Specifically, we provide an efficient quantum circuit implementation of the time evolution operator $e^{-iHt}$.  We also show how to choose a good initial state $|\psi_0\rangle$ that should have reasonable overlap with the ground state while being easy to prepare on a quantum computer, and we furthermore show how to efficiently find a circuit to prepare the superposition of $|\psi_0\rangle$ with a reference state.  We then derive bounds on the spectrum of the Hamiltonian in order to determine the maximum allowed value of the time step size $\Delta t$ for use in UVQPE [Eq.~\eqref{eq:UVQPE}] and ODMD [Eq.~\eqref{eq:ODMD}].  Finally, we argue that UVQPE and ODMD should scale well with increasing system size and thus are promising methods for finding the true ground state of the kagome lattice Heisenberg model.

\subsection{The Heisenberg model\label{sec:full_2D_model}}

We consider the antiferromagnetic Heisenberg model,
\begin{equation}
    H = J\sum_{\langle ij\rangle} \mathbf{S}_i^{ }\cdot \mathbf{S}_j^{ } - h\sum_i S^z_i\label{eq:Heisenberg_def}
\end{equation}
where $\langle ij\rangle$ indicates nearest-neighbor pairs of sites on the kagome lattice, illustrated by the bonds between sites drawn in Fig.~\ref{fig:kl_and_plaquettes}(a).  $\mathbf{S}$ is the spin-1/2 operator $\boldsymbol{\sigma}/2\equiv(\sigma^x_{ },\sigma^y_{ },\sigma^z_{ })/2$, $J>0$ is the antiferromagnetic coupling strength, and $h$ is the strength of an external magnetic field.  Note that with this definition of $\mathbf{S}$, we are effectively setting $\hbar$ to 1, and we will use those units throughout the paper.

The Heisenberg Hamiltonian conserves total $S^z$, $[H,\sum_i S_i^z]=0$.  This symmetry will be essential to our implementation of UVQPE and ODMD.  Most importantly, each eigenstate of $H$ lies within a subspace with fixed total spin ($S$), with several important consequences.  First, if an initial state $|\psi_0\rangle$ has a well-defined total $S^z$, any time-evolved state $e^{-iHt}|\psi_0\rangle$ will also live in the same spin sector; consequently, since the ground state found by UVQPE and ODMD is in the span of the basis states \eqref{eq:basis_states}, it will also be in the same fixed spin sector.  We can thus use the hybrid algorithms to independently find the ground state energy within each spin sector.  Since these subspaces are smaller than the full Hilbert space, it is easier to pick a good initial state with a high overlap with the ground state, and the algorithms should converge more quickly.  
Furthermore, in the important special case where there is no external field, $h=0$, we need only consider a single spin sector, with total $S^z=0$, since Lieb's theorem~\cite{Lieb1989} guarantees that the ground state of an antiferromagnetic model will have total spin 0.  

Second, the division of $H$ eigenstates into orthogonal subspaces with fixed total spin enables the mirror circuit approach to measuring expectation values, state overlaps, and other observables on a quantum computer.  As explained in Sec.~\ref{sec:algs_mirror} above, the mirror circuit method requires a reference state $|\text{Ref}\rangle$ that is guaranteed to be orthogonal to the state $|\psi_0\rangle$ in which expectation values are calculated, and that is furthermore an eigenstate of the Hamiltonian with known energy so that its time evolution is given by a simple known exponential.  In the case of the Heisenberg model, we can use a fully polarized state, with all spins up or all spins down, as the reference.  Each of these states is the only one in its symmetry sector, and hence must be an eigenstate of $H$ and orthogonal to all states with different total $S^z$.  Furthermore, the energy is easy to calculate analytically as simply the expectation value of $H$ in a product state.  On the kagome lattice, the energy of the fully polarized state with all spins pointing up is $-(J/2)N_\text{bonds} - (h/2)N_\text{spins}$, where $N_\text{bonds}$ is the number of nearest-neighbor pairs on the lattice and $N_\text{spins}$ the total number of spins.

Finally, the conservation of total $S^z$ during time evolution allows for simple approaches to error mitigation when using quantum hardware or a noisy emulator~\cite{BonetMonroig2018,Cai2021quantumerror,Tran2021}.  First, if an initial state $|\psi_0\rangle$ has a known total $S^z$, the states after time evolution should remain in the same symmetry sector, and thus measured bit strings not consistent with that symmetry sector can be thrown out.  The use of state preparation circuits such as $U_0$ which takes $|\mathbf{0}\rangle$ to $|\psi_0\rangle$ complicates this logic, since $U_0$ in general does not conserve $S^z$.  Fortunately, some bit strings can still be shown to be invalid even after applying $U_0$, hence some errors can be detected and thrown out in post-processing.  A less obvious consequence is that the magnitude of errors can be reduced by stochastically inserting single-qubit rotations that affect states in the wrong symmetry sector but not those in the correct symmetry sector, a method that was introduced in Ref.~\cite{Tran2021}.  We provide further details on the error mitigation strategies after restricting to small spin plaquettes for our empirical demonstration; see Sec.~\ref{sec:star_plaquettes} below.

In addition to the spin symmetry, the kagome lattice Heisenberg model also has spatial symmetries.  The lattice, as shown in Fig.~\ref{fig:kl_and_plaquettes}(a), is evidently symmetric under both horizontal and vertical reflections and under rotations by $\pi/3$, $2\pi/3$, and $\pi$ (i.e. $C_6$, $C_3$, and $C_2$ rotations, respectively) depending on the center of the rotation.  In principle these symmetries could also be used for error mitigation.  Because exact time evolution of a spatially symmetric state would preserve the symmetry, the true probabilities of certain symmetry-related bit strings should be equal.  Differences between the measured probabilities of such symmetry-related bit strings can be assumed to be due either to errors or to taking a finite number of samples, and either way the symmetry can be restored in classical post-processing at minimal cost.  We do not use this error mitigation strategy in the practical demonstrations reported below because our mirror circuit method only requires the probability of the all-0 bit string, which transforms to itself under any spatial symmetry.  However, we expect spatial symmetry-based error mitigation to play an important role in quantum studies of the kagome lattice that take other approaches.

\subsection{\texorpdfstring{Efficient circuits approximating $e^{-iHt}$}{Efficient circuits approximating the time evolution operator}\label{sec:full_2D_Trotter}}

To exactly implement the time evolution operator $U = e^{-iHt}$ with a quantum circuit, we would first need to know the unitary $U$ classically, which in general already requires solving for the spectrum of $H$, what we wanted to find in the first place.  Even if we could somehow find $U$ without solving $H$, we would still have to solve the unitary synthesis problem to find a circuit that implements it, which also scales exponentially in system size.

In lieu of such an exact method, we take the standard approach of approximating the time evolution using a Suzuki-Trotter decomposition~\cite{Suzuki1991, Childs2021_Trotter}.  We divide the terms of the Hamiltonian into groups, $H=\sum_i H_i$, where the an time evolution of each $H_i$ can be found exactly classically.  In the lowest-order decomposition, the overall time evolution is approximated as
\begin{equation}
    e^{-iHT} = \left[ \prod_i e^{-iH_i T/m} \right]^m + \mathcal{O}\left(T^2/m\right).\label{eq:Trotter_v1}
\end{equation}
Note that error per time step of size $T/m$ scales as $(T/m)^2$, and the overall error $T^2/m$ is the total, assuming $(T/m)^2$ is small.  As the discretization in time becomes very fine, $m\rightarrow\infty$, the error term scaling as $T^2/m$ goes to 0 and the Trotterized time evolution becomes exact.  

\begin{figure}
    \centering
    \includegraphics[width=0.7\columnwidth]{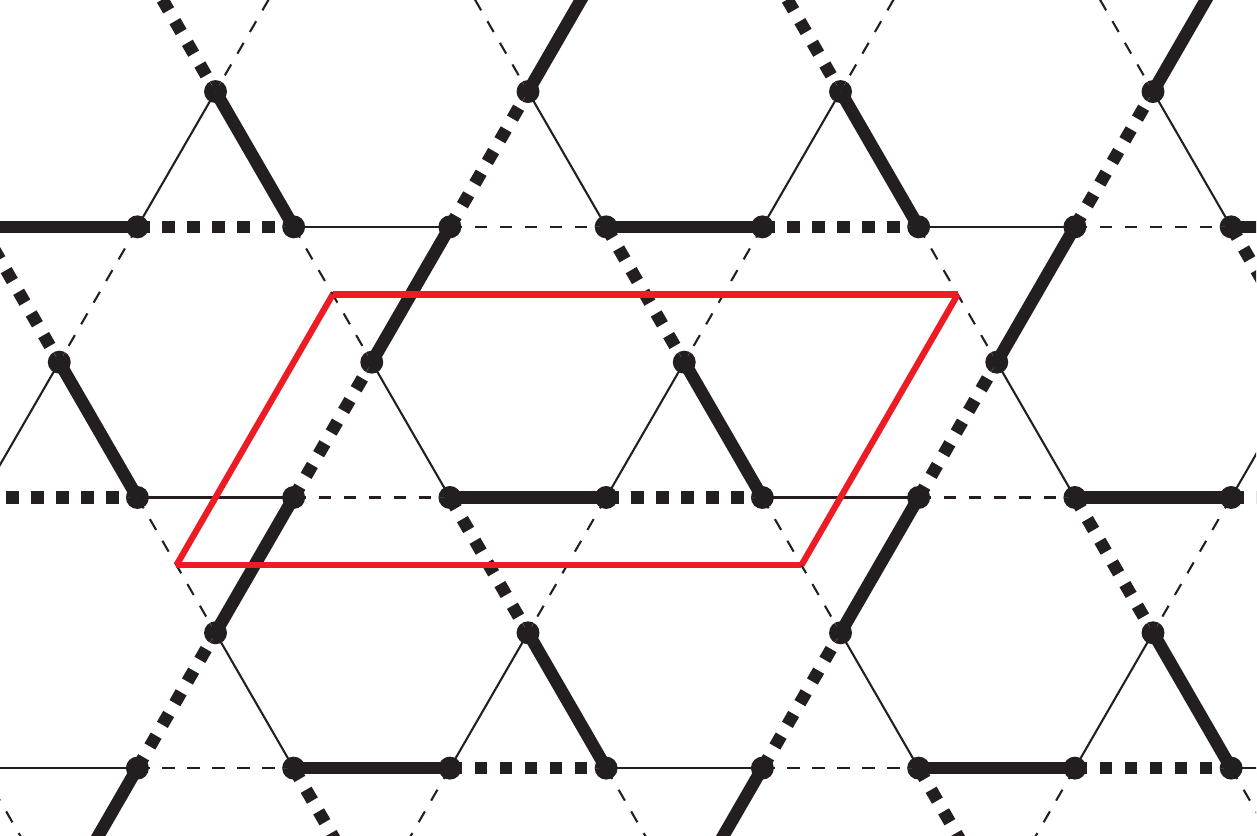}
    \caption{Bond-by-bond Trotter decomposition for the kagome lattice Heisenberg model.  The interactions are divided into four groups of non-overlapping bonds, indicated by thick solid lines, thick dashed lines, thin solid lines, and thin dashed lines.  Note that the resulting Floquet operator for the Trotterized time evolution has a doubled unit cell indicated by the red parallelogram, containing 6 sites while the original model's unit cell has just 3.  This decomposition also breaks mirror and rotation symmetries of the lattice.}
    \label{fig:bond-by-bond_Trotter}
\end{figure}

For the Heisenberg model, a common choice is what we call a bond-by-bond Suzuki-Trotter decomposition.  Each individual $\mathbf{S}\cdot\mathbf{S}$ term can be exponentiated classically, and in fact a simple quantum circuit with 3 CNOT operators can be found analytically for the operator $\exp(-i\theta\mathbf{S}\cdot\mathbf{S})$ for any value of $\theta$.    
Any two of these exponentiated operators commute with each other if they act on disjoint pairs of sites, and thus if each $H_i$ consists of $\mathbf{S}\cdot\mathbf{S}$ interactions on a collection of non-overlapping bonds, its exponential can easily be found classically and implemented as a quantum circuit.  On the kagome lattice, the bonds can be organized into four disjoint groups, $H_1$ through $H_4$, as shown in Fig.~1 of Ref.~\cite{Kattemolle2023}.  Another example is shown in Fig.~\ref{fig:bond-by-bond_Trotter}.  Note that because each vertex of the kagome lattice is connected to four edges, a bond-by-bond Trotter approach with three or fewer groups is not possible.  If $h$ is nonzero, a fifth step is needed, which includes all the on-site $S^z$ terms.

This naive implementation of Trotter evolution for the kagome lattice Heisenberg model is valid and the corresponding quantum circuits are easy to generate, but it has some drawbacks.  Notably, bond-by-bond Trotter evolution breaks all the spatial symmetries of the lattice, hence the symmetry or lack thereof in measured bit string counts can no longer be used to evaluate the reliability of a quantum computation or to perform symmetry-based error detection.  Even more importantly, the total number of two-qubit gates required for each Trotter step can actually be reduced.  For each time step, the bond-by-bond approach uses 3 CNOTs per bond, assuming the quantum hardware allows for gates to be applied between each pair of spins (i.e., qubits) connected by a bond.  We will now show an alternative approach that both preserves more of the spatial symmetries of the lattice and reduces the CNOT count to 8/3 CNOTs per bond, a more than 10\% reduction.

\begin{figure}
    \centering
    \includegraphics[width=0.95\columnwidth]{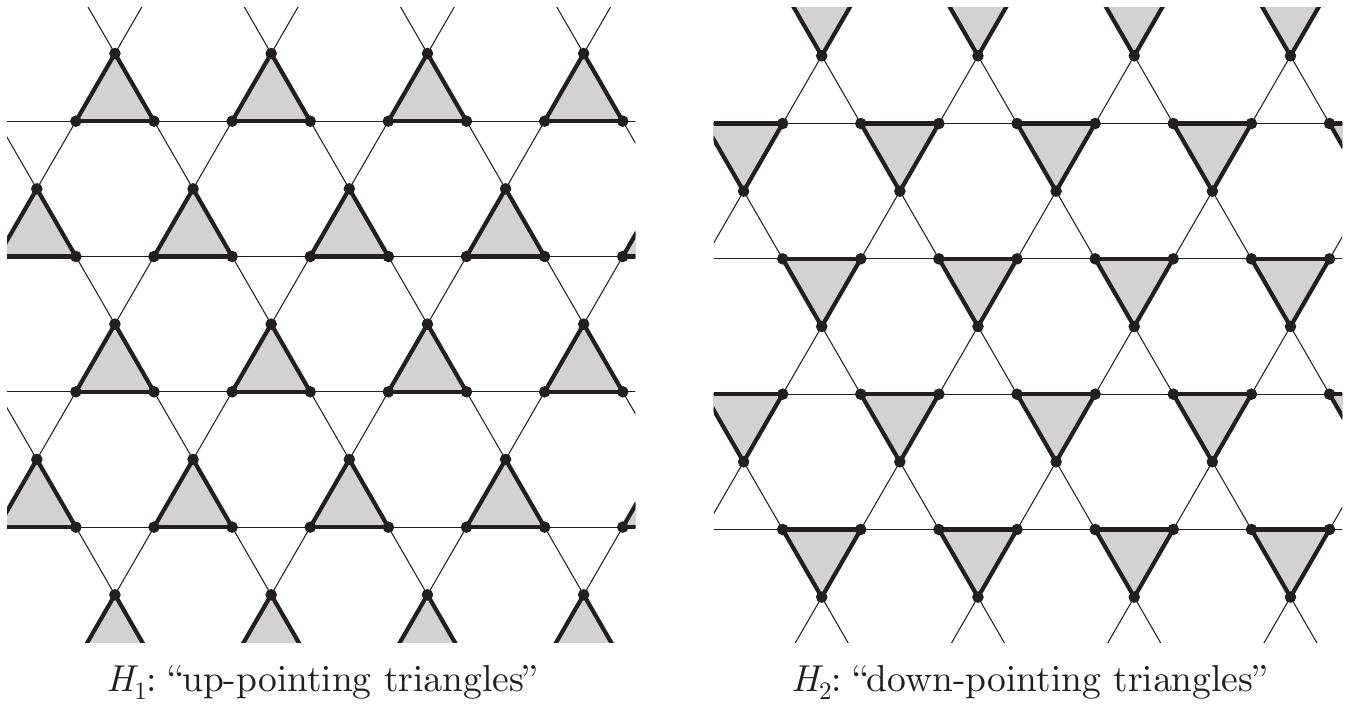}
    \caption{Triangle-by-triangle Trotter decomposition.  The spin-spin interactions are organized into groups of three corresponding to triangles in the kagome lattice.  The triangles are then grouped into up-pointing triangles (left panel) and down-pointing triangles (right).  Efficient quantum circuits can be found classically for the exponential of $H_1$ and the exponential of $H_2$.  This Trotter decomposition preserves some symmetries of the underlying Hamiltonian, including one mirror symmetry and a $C_3$ rotation.  As in the bond-by-bond decomposition, the unit cell is enlarged from 3 spins to 6.}
    \label{fig:tri-by-tri_Trotter}
\end{figure}

In the alternative ``triangle-by-triangle'' Suzuki-Trotter decomposition, instead of exponentiating individual $\mathbf{S}\cdot\mathbf{S}$ interactions, we group together the three bonds around each triangle in the lattice to form a three-qubit operator $H_\Delta$.  We can still compute $U_\Delta = e^{-i \theta H_\Delta}$ analytically to get a three-qubit unitary, and then we can numerically solve the unitary synthesis problem to find a circuit that implements $U_\Delta$.  We can then divide the full Hamiltonian into two sets of non-overlapping triangles: $H_1$ containing all the up-pointing triangles and $H_2$ containing all the down-pointing ones.  The two groups of terms are illustrated in Fig.~\ref{fig:tri-by-tri_Trotter}.

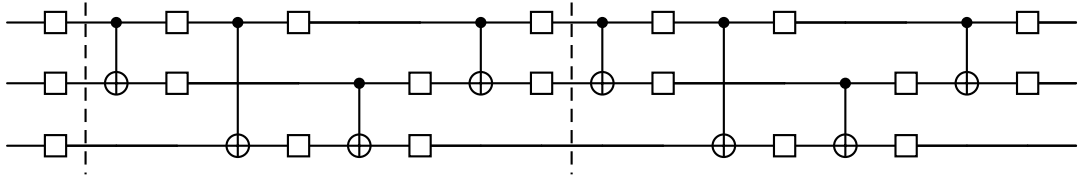
\begin{figure*}
    \centering
    \begin{tikzcd}[ampersand replacement=\&]
\qw \& \gate{} \slice[style=black]{}
\& \ctrl{1} \& \gate{} \& \ctrl{2} \& \gate{} \& \qw      \& \qw     \& \ctrl{1} \& \gate{} \slice[style=black]{}
\& \ctrl{1} \& \gate{} \& \ctrl{2} \& \gate{} \& \qw      \& \qw     \& \ctrl{1} \& \gate{} \& \qw \\
\qw \& \gate{} 
\& \targ{}  \& \gate{} \& \qw      \& \qw     \& \ctrl{1} \& \gate{} \& \targ{}  \& \gate{} 
\& \targ{}  \& \gate{} \& \qw      \& \qw     \& \ctrl{1} \& \gate{} \& \targ{}  \& \gate{} \& \qw \\
\qw \& \gate{} 
\& \qw      \& \qw     \& \targ{}  \& \gate{} \& \targ{}  \& \gate{} \& \qw      \& \qw 
\& \qw      \& \qw     \& \targ{}  \& \gate{} \& \targ{}  \& \gate{} \& \qw      \& \qw     \& \qw 
\end{tikzcd}
    \caption{Template for a quantum circuit implementing the exact time evolution for the Heisenberg Hamiltonian on one triangle of the kagome lattice.  The empty boxes indicate single-qubit rotations; some are fully arbitrary while others are simpler, for example of the form $H R_z(\phi)$ where $H$ is the Hadamard gate.  The dashed vertical lines are included to clarify the structure of the circuit---the CNOT layouts of the first and second halves are the same.  Note that the operator implemented by the circuit is symmetric with respect to permutations of the three qubits, so there are various equivalent templates. We use this particular qubit ordering in order to allow for a slightly increased degree of parallelism in applying gates when we consider just a single star plaquette, as discussed in Sec.~\ref{sec:star_plaquettes}.  }
    \label{fig:tri-by-tri_circuit}
\end{figure*}

Carrying out the numerical synthesis for $U_\Delta$ using the Berkeley Quantum Synthesis Toolkit (BQSKit)~\cite{BQSKit}, we find that for any $\theta$, it can be implemented using just 8 CNOTs, compared with 9 in total (3 per bond) when separately exponentiating the three spin-spin interactions. \footnote{While we do not have an analytical proof that an 8 CNOT circuit exactly implements the unitary for any $\theta$, we numerically find that 8 CNOTs are sufficient to reproduce the unitary at any level of precision.}  The structure of the 8 CNOT circuit is shown in Fig.~\ref{fig:tri-by-tri_circuit}.

Evidently, the triangle-by-triangle Trotter evolution is both more efficient than the naive bond-by-bond approach, using 8/9 the number of CNOTs, and respects more of the spatial symmetries of the underlying lattice and Hamiltonian.  For this reason, we will use the triangle-by-triangle approach in our experimental demonstration on the star plaquettes.  More generally, this approach to Trotterization, replacing bond-by-bond exponentiation with exponentiation of slightly larger groupings of qubits, should be beneficial for any lattice composed of small corner-sharing polygons, which are called bisimplex lattices~\cite{Henley2001}.

We briefly note that the triangle-by-triangle approach remains efficient even when the connectivity among the qubits is restricted.  For example, if the three qubits making up a triangle have only linear connectivity, the exact three-qubit time evolution can be still be implemented with just 12 CNOTs.  In comparison, the bond-by-bond decomposition requires 15 CNOTs to implement all three separate spin-spin interactions: 9 for the interactions themselves, and 6 for SWAP gates.  Thus the relative advantage of the triangle-by-triangle approach is maintained.

\subsection{State preparation: how to choose and efficiently prepare a good initial state\label{sec:full_2D_state_prep}}

Our next goal is first to choose a good initial state $|\psi_0\rangle$ that is likely to have high overlap with the ground state while being easy to prepare on quantum hardware; for concreteness, we focus on the case of $h=0$.  We will then show how to use the recently developed method of multi-state synthesis~\cite{Szasz2023multistate} to easily find circuits that implement the macroscopic superposition of $|\psi_0\rangle$ with the fully-polarized reference state.

\subsubsection{Choosing and preparing the initial state}

To find a good initial state that is likely to have high overlap with the ground state, we first consider once again the symmetries of the Hamiltonian.  As discussed in Sec.~\ref{sec:full_2D_model} above, $H$ conserves total $S^z$.  We will now specialize to the most-studied case, with $h=0$, in which case $H$ actually conserves not just $S^z$ but also the total spin $S$.  Then Lieb's theorem guarantees that the initial state has the minimal possible total spin, which assuming an even number of sites in total will be 0~\cite{Lieb1989}, so a good initial state should also be in this spin sector.  

While restricting to $S=0$ is a relatively strong constraint, the remaining Hilbert space still grows exponentially in the total number of spins; specifically, for a $2n$-spin Heisenberg model, the dimension of the $S=0$ eigenspace is $(2n)!/(n!(n+1)!)$~\cite{oeisA000108} $\sim 2^{2n}/n^{3/2}$.~\footnote{In comparison, the size of the $S^z=0$ eigenspace is $(2n)!/2n!$, larger by a factor of $n+1$.}  Furthermore, generic eigenstates of the total spin operator with small spin are highly entangled, and hence hard to prepare on a quantum computer~\cite{Carbone2022}.

Fortunately, there are some very simple, low-entangled states within the $S=0$ space.  Namely, a singlet dimer on two spins has total spin zero, and thus a tensor product of dimers does as well.  So any covering of singlet dimers will be in the correct spin sector.  A dimer covering is also easy to prepare on a quantum computer, since there is only entanglement between pairs of spins, and a singlet can be prepared with just one CNOT operator together with a few single-qubit rotations.  

In fact, the dimer covering idea is even more promising once we consider the specifics of the Hamiltonian.  Dimer coverings are known to be the ground states of some apparently frustrated models~\cite{SRIRAMSHASTRY19811069,Rokhsar1988,Moessner2011}. Furthermore, superpositions of dimer coverings, or resonating valence bond (RVB) states, are a type of spin liquid and have been proposed as the ground states of many frustrated models~\cite{ANDERSON1973153,Capriotti2001,baskaran2006resonating}.

To understand why dimer coverings are promising on the triangular lattice, the starting point is the same triangle-by-triangle decomposition of the Hamiltonian that we used for our efficient implementation of Trotterized time evolution: we write $H=\sum_\Delta H_\Delta$, where for each triangle in the lattice $H_\Delta$ includes the three $\mathbf{S}\cdot\mathbf{S}$ interactions on the edges of the triangle.  In the case we are considering, $h=0$, $H_\Delta$ has just two energy eigenvalues, namely $\pm 3J/2$, each with multiplicity four.  The ground state subspace corresponding to the eigenvalue $-3J/2$ is spanned by the states 
\begin{equation}
    \begin{tabular}{ccc}
\raisebox{-1.2em}{\begin{tikzpicture}
\draw[black] (0,0) -- (0.25,0.433);
\draw[black] (0,0) -- (0.5,0);
\draw[black] (0.5,0) -- (0.25,0.433);
\filldraw[black] (0,0) circle (1pt) ;
\filldraw[black] (0.5,0) circle (1pt) node[anchor=west]{$\!\su$};
\filldraw[black] (0.25,0.433) circle (1pt) ;
\draw[rotate around={60:(0.125,0.2165)},black] (0.125,0.2165) ellipse (0.4 and 0.1);
\end{tikzpicture}} 
, &
\raisebox{-1.2em}{\begin{tikzpicture}
\draw[black] (0,0) -- (0.25,0.433);
\draw[black] (0,0) -- (0.5,0);
\draw[black] (0.5,0) -- (0.25,0.433);
\filldraw[black] (0,0) circle (1pt) node[anchor=east]{$\su$};
\filldraw[black] (0.5,0) circle (1pt) ;
\filldraw[black] (0.25,0.433) circle (1pt) ;
\draw[rotate around={-60:(0.375,0.2165)},black] (0.375,0.2165) ellipse (0.4 and 0.1);
\end{tikzpicture}} 
\,\,\,,\,\,\, & 
\raisebox{-0.8em}{\begin{tikzpicture}
\draw[black] (0,0) -- (0.25,0.433);
\draw[black] (0,0) -- (0.5,0);
\draw[black] (0.5,0) -- (0.25,0.433);
\filldraw[black] (0,0) circle (1pt) ;
\filldraw[black] (0.5,0) circle (1pt) ;
\filldraw[black] (0.25,0.433) circle (1pt) node[anchor=east]{$\su$};
\draw[black] (0.25,0) ellipse (0.4 and 0.1);
\end{tikzpicture}}
, \\ \\
\raisebox{-1.2em}{\begin{tikzpicture}
\draw[black] (0,0) -- (0.25,0.433);
\draw[black] (0,0) -- (0.5,0);
\draw[black] (0.5,0) -- (0.25,0.433);
\filldraw[black] (0,0) circle (1pt) ;
\filldraw[black] (0.5,0) circle (1pt) node[anchor=west]{$\!\sd$};
\filldraw[black] (0.25,0.433) circle (1pt) ;
\draw[rotate around={60:(0.125,0.2165)},black] (0.125,0.2165) ellipse (0.4 and 0.1);
\end{tikzpicture}} 
, & 
\raisebox{-1.2em}{\begin{tikzpicture}
\draw[black] (0,0) -- (0.25,0.433);
\draw[black] (0,0) -- (0.5,0);
\draw[black] (0.5,0) -- (0.25,0.433);
\filldraw[black] (0,0) circle (1pt) node[anchor=east]{$\sd$};
\filldraw[black] (0.5,0) circle (1pt) ;
\filldraw[black] (0.25,0.433) circle (1pt) ;
\draw[rotate around={-60:(0.375,0.2165)},black] (0.375,0.2165) ellipse (0.4 and 0.1);
\end{tikzpicture}} 
\,\,\,,\,\,\, & 
\raisebox{-0.8em}{\begin{tikzpicture}
\draw[black] (0,0) -- (0.25,0.433);
\draw[black] (0,0) -- (0.5,0);
\draw[black] (0.5,0) -- (0.25,0.433);
\filldraw[black] (0,0) circle (1pt) ;
\filldraw[black] (0.5,0) circle (1pt) ;
\filldraw[black] (0.25,0.433) circle (1pt) node[anchor=east]{$\sd$};
\draw[black] (0.25,0) ellipse (0.4 and 0.1);
\end{tikzpicture}}
\end{tabular}\label{eq:tri_gs_states}
\end{equation}
where each ellipse denotes a pair of spins in the singlet state $|S\rangle = (|01\rangle-|10\rangle)/\sqrt{2}$.  Note that this basis for the ground subspace is overcomplete: in each row, one of the states can be written as a linear combination of the other two.

Then for any state $|\psi\rangle$ on the full kagome lattice, 
\begin{equation}
    \langle \psi|H|\psi \rangle = \sum_\Delta \langle \psi|H_\Delta|\psi\rangle \geq \sum_\Delta -3J/2,\label{eq:variational_principle_bound}
\end{equation}
giving a lower bound on the ground state energy.  If there was a singlet dimer covering of the kagome lattice such that each triangle had a dimer along one edge~\footnote{More formally, the reduced density matrix of the state on each triangle should lie within the span of the states listed in Eq.~\ref{eq:tri_gs_states}.}, the state would saturate the lower bound, and thus would necessarily be an exact ground state of the model.  Given the previously highlighted difficulty of solving the model, it should come as no surprise that no such dimer covering exists.  

\begin{figure}
    \centering
    \includegraphics[width = 0.98\columnwidth]{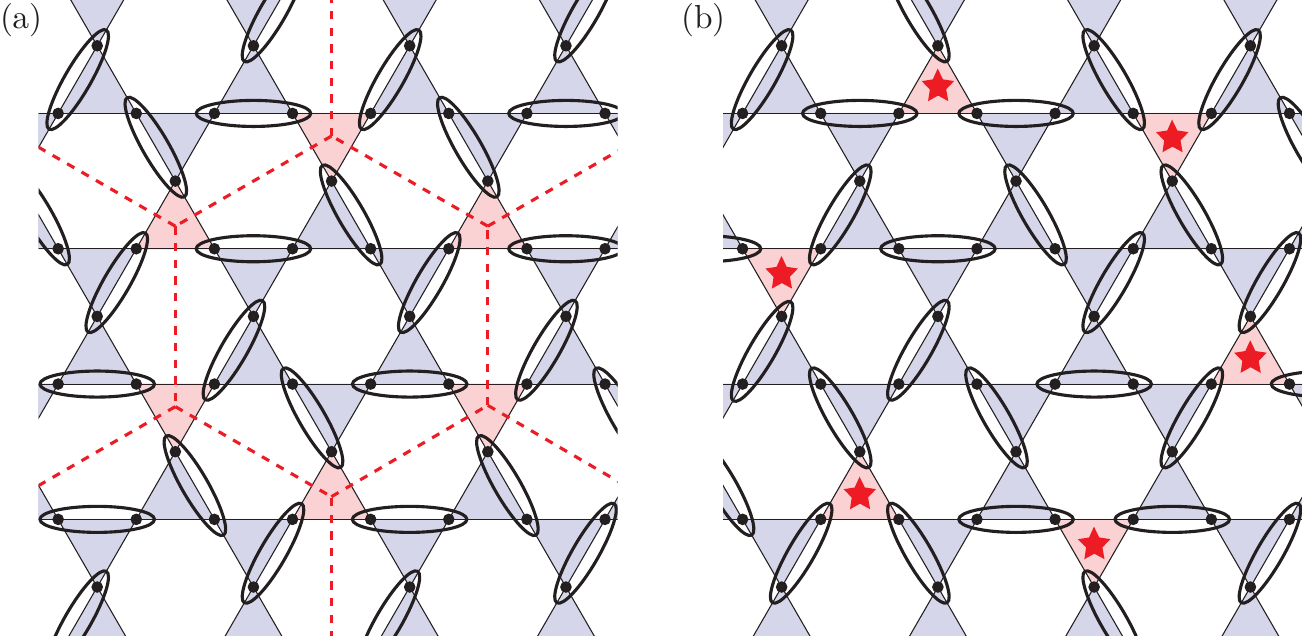}
    \caption{Possible dimer coverings of the kagome lattice.  (a) One simple covering has an enlarged 12-spin unit cell, as indicated by the dashed red lines.  The triangles shaded blue have a dimer along one edge and hence the lowest possible local energy.  ``Defect'' triangles with no dimer are located at the corners of the super-unit cell and are shaded in red.  There are two defect triangles for six triangles with a dimer.  (b) With a less regular pattern for the dimer covering, the density of defects (shaded in red and also marked with a red star) can be reduced, hence further reducing the energy of the state and presumably achieving a higher overlap with the true ground state.  There must be at least one defect triangle for every two hexagon loops in the kagome lattice, and the pattern shown here achieves that minimum, so is a strong candidate for the initial state.}
    \label{fig:kl_dimer_coverings}
\end{figure}

Nevertheless, we can find dimer coverings for which a large majority of the triangles do have a dimer on one edge, and these coverings are excellent candidates for the initial state $|\psi_0\rangle$.  Two such coverings are shown in Fig.~\ref{fig:kl_dimer_coverings}.  In each, ``defect'' triangles with no dimer are shaded in red.  A minimum density of defects is set by the fact that it is impossible to have two adjacent star plaquettes both with no defects
.  The dimer covering in the right panel of Fig.~\ref{fig:kl_dimer_coverings} is one of many that achieves this minimum.  

Finally, we comment on how such a singlet dimer covering would be prepared on a quantum computer.  Fortunately, this task is quite easy, since each dimer involves just two spins and thus all the dimers can be prepared from the state $|00\rangle$ in parallel, using a simple circuit with a single CNOT gate:
\begin{equation}
    \begin{tikzcd}[ampersand replacement=\&]
\qw \& \gate{X} \& \gate{H} \& \ctrl{1} \& \qw      \& \qw \\
\qw \& \qw      \& \qw      \& \targ{}  \& \gate{X} \& \qw 
    \end{tikzcd}\label{eq:U0_circuit}
\end{equation}
Since this circuit is applied on each pair of qubits, the total CNOT count for the state preparation circuit for a system size with $n$ total spins is $n/2$.

\subsubsection{Preparing the superposition with the reference state}

A remaining challenge is the preparation of the superposition of a dimer covering $|\psi_0\rangle$ with the reference state $|\text{Ref}\rangle$ for measuring the phase of observables in the mirror circuit approach. Recall from Sec.~\ref{sec:full_2D_model} that for the reference state we use the fully polarized state with all spins up, $|\text{Ref}\rangle = |\mathbf{0}\rangle$, since it is the unique state in its $S^z$ symmetry sector and hence orthogonal to $|\psi_0\rangle$ and an eigenstate of $H$ with known energy.  Focusing on the circuit for the quantity $F_2$, our goal is to prepare a state of the form \eqref{eq:F2_superposition}, specifically
\begin{equation}
    \frac{|\mathbf{0}\rangle + [(|01\rangle - |10\rangle)/\sqrt{2}]^{\otimes n/2}}{\sqrt 2}\label{eq:F2_superposition_dimer}
\end{equation}
where $n$ is the number of qubits, an equal weight superposition of the all-0 state with a tensor product of dimers.  

We perform the state preparation in two easy steps.  First, we prepare the Greenberger–Horne–Zeilinger (GHZ) state, $(|\mathbf{0}\rangle+|\mathbf{1}\rangle)/\sqrt{2}$, where $|\mathbf{1}\rangle=|1\rangle^{\otimes n}$ is the all-1 state.  With linear qubit connectivity, this state preparation can be achieved with $(n-1)$ sequential CNOT gates following one Hadamard gate, as shown here with four qubits.
\begin{equation}
    \begin{tikzcd}[ampersand replacement=\&, row sep={0.4cm,between origins}]
        \ket{0} \& \qw      \& \qw      \& \targ{}   \& \qw \\
        \ket{0} \& \gate{H} \& \ctrl{1} \& \ctrl{-1} \& \qw \\
        \ket{0} \& \qw      \& \targ{}  \& \ctrl{1}  \& \qw \\
        \ket{0} \& \qw      \& \qw      \& \targ{}   \& \qw
    \end{tikzcd}
    = \frac{|0000\rangle + |1111\rangle}{\sqrt 2}\label{eq:GHZ_circuit}
\end{equation}
With a two-dimensional connectivity grid, the CNOT gates can be parallelized into $\mathcal{O}(\sqrt{n})$ layers; with all-to-all connectivity, parallelization can be further improved to give $\mathcal{O}(\log(n))$ layers.  If we also make use of measurement and feed-forward, the GHZ state can be prepared in constant depth, though still requiring $\mathcal{O}(n)$ CNOT gates.~\cite{Briegel2001}  

In the second step, we apply a circuit that converts the GHZ state from Eq.~\ref{eq:GHZ_circuit} to the desired superposition \eqref{eq:F2_superposition_dimer}.  In general, finding a circuit that maps one state to another is exponentially hard in the number of qubits.  However, in this case, even though both states involve long-ranged entanglement, the transformation between the two can be accomplished entirely by local transformations on pairs of qubits.  Specifically, we need to find a two-qubit circuit that solves the ``multi-state mapping'' problem 
\begin{align}
|00\rangle & \mapsto |00\rangle \label{eq:multistate_mapping}\\
|11\rangle & \mapsto (|01\rangle - |10\rangle)/\sqrt{2}\nonumber
\end{align}  
with the images of $|01\rangle$ and $|10\rangle$ unconstrained.  When the resulting circuit is applied, in the GHZ state, to each pair of qubits that should be in a singlet dimer in $|\psi_0\rangle$, the result is precisely the state \eqref{eq:F2_superposition}.  The multi-state mapping problem \eqref{eq:multistate_mapping} can be easily solved using BQSKit~\cite{BQSKit, Szasz2023multistate}, giving the circuit
\begin{equation}
    \begin{tikzcd}[ampersand replacement=\&]
    \& \qw \& \ctrl{1} \& \gate{R_y\left(\frac{3\pi}{2}\right)} \& \ctrl{1} \& \gate{R_y\left(\frac{3\pi}{2}\right)} \& \qw \\
    \& \gate{R_y\left(\frac{7\pi}{4}\right)} \& \targ{} \& \gate{R_y\left(\frac{7\pi}{4}\right)} \& \targ{} \& \gate{R_y\left(\frac{\pi}{2}\right)} \& \qw
    \end{tikzcd}\label{eq:superposition_prep_circuit}
\end{equation}
which implements the transformation 
\begin{equation}
    \left(\begin{array}{cccc}
    1 &&&\\
    & 1/\sqrt{2} & & 1/\sqrt{2}\\
    & 1/\sqrt{2} && -1/\sqrt{2}\\
    &&1&
    \end{array}\right).
\end{equation}
Note that because we did not constrain the images of $|01\rangle$ and $|10\rangle$ under the transformation, there are other valid circuits for the desired state preparation that will map these two states differently.

Finally, to produce the slightly different superposition state \eqref{eq:F3_superposition}, we can just add a single qubit rotation between the GHZ preparation and the application of the local transformation circuits \eqref{eq:superposition_prep_circuit}.  Specifically, acting with the gate $R_Z(-\pi/2)$ on any one qubit will transform the GHZ state as
\begin{equation}
    \frac{|\mathbf{0}\rangle + |\mathbf{1}\rangle}{\sqrt{2}}\mapsto e^{-i\pi/4}\frac{|\mathbf{0}\rangle + i|\mathbf{1}\rangle}{\sqrt{2}},
\end{equation}
and then acting with \eqref{eq:superposition_prep_circuit} will give the desired superposition up to an irrelevant global phase.

In total, preparing the reference superposition states requires $(n-1)$ CNOT operators to generate the GHZ state and $n$ CNOTs to perform the transformation of $|\mathbf{1}\rangle$ to $|\psi_0\rangle$, for a relatively low total CNOT count of $2n-1$.  

\subsection{Bounds on Hamiltonian spectrum\label{sec:full_2D_H_bounds}}

As noted in Sec.~\ref{sec:algs_UVQPE} above, in order to infer the eigenvalues of $H$ from the eigenvalues of the time evolution operator $U_{\Delta t}=e^{-iH{\Delta t}}$, we need the spectrum of $H\Delta t$ to lie within an interval of size $2\pi$.  Thus to determine the maximum allowed size of $\Delta t$, we must derive bounds on the spectrum of the Hamiltonian.

With $h=0$, we have already demonstrated that the ground state energy is bounded from below by $(-3J/2)N_\Delta$ where $N_\Delta$ is the number of triangles in the lattice.  To bound the maximum energy, we can consider a single bond, with Hamiltonian $J\mathbf{S}\cdot\mathbf{S}$.  This operator has eigenvalues $-3J/2$ and $J/2$ ($\times 3$), so the maximum energy for the whole system is bounded from above by $J/2$ times the number of bonds, which is $3N_\Delta$.  This maximum energy is achieved by the fully polarized states with all spins up or all spins down. Thus we conclude that for $h=0$, $\norm{H}_2=(3J/2)N_\Delta$.

If we turn on an external field (without loss of generality we take $h>0$), the highest energy state will be the fully polarized state with spins pointing against the applied field, with total energy $(3J/2)N_\Delta + (h/2) n$ with $n$ the total number of spins.  A lower bound on the energy is $-(3J/2)N_\Delta - (h/2) n$.  To prove this, we use the triangle inequality:  
\begin{subequations}
\begin{align}
|E_0| &= \left|\text{min}_{|\psi\rangle} \langle \psi|H|\psi\rangle\right|\\
& \leq \text{max}_{|\psi\rangle} \left|\langle \psi|H|\psi\rangle\right|\\
& \leq J\,\text{max}\left|\left\langle \sum\mathbf{S}\cdot \mathbf{S}\right\rangle\right| + h\,\text{max}\left|\left\langle \sum S^z\right\rangle\right|\\
& \leq (3J/2)N_\Delta + (h/2) n.
\end{align}     
\end{subequations}
In the final line, we used the bound on the $h=0$ Hamiltonian from the previous paragraph and the fact that the maximal $|\langle S^z\rangle|$ on any given site is $1/2$, for the $|0\rangle$ and $|1\rangle$ states.  

Finally, we conclude that $\norm{H}_2=(3J/2)N_\Delta + (h/2)n$, so we require $\Delta t < 2\pi/(3JN_\Delta + h n)$.

\subsection{Favorable scaling with system size \label{sec:full_2D_scaling}}

Finally, we argue that ground state search for the kagome lattice Heisenberg model using UVQPE or ODMD has reasonably good scaling in system size.  Good scaling is essential because very large systems are needed in order to definitively address the question of whether the low-lying energy spectrum is gapped or gapless.  For a model which is gapless on the infinite two-dimensional plane, on a finite system with linear size $L$ there will still be a gap of size scaling with $1/L$.  Thus simulations on a finite size system will \emph{always} observe a gap, and without considering the scaling of the gap size with $L$ it will not be possible to distinguish a gap due to finite size from a gap that will persist as $L\rightarrow\infty$.  Indeed, the largest exact classical simulation of this model considers clusters of up to $48$ spins~\cite{Lauchli2019}, or $L\approx 7$, which is still too small to determine whether the observed gap is fundamental or a finite size effect.  For a definitive answer, we will likely need to simulate much larger systems with hundreds or thousands of spins.~\footnote{Of course, this reasoning assumes that we have access only to energies and not to the corresponding eigenstates.  Other signatures, including momentum of low-lying excitations, can help identify particular spin liquid phases even on somewhat smaller systems.~\cite{Savary_2017, He2017, Hu2019}  Nevertheless, at least hundreds of spins will still be required.}

We begin by estimating the total cost of measuring an expectation value $\langle\psi_0|e^{-iHT}|\psi_0\rangle$ using the mirror circuit approach.  This involves three circuits, for measuring $F_1$, $F_2$, and $F_3$, and each of those requires two copies of its respective state preparation circuit and an implementation of the time evolution $e^{-iHT}$.  As established above, for an $n$-spin system, the state preparation for $F_1$ requires $n/2$ CNOT gates, while for $F_2$ and $F_3$ it requires $\approx 2n$ CNOTs.  A single Trotter step using the triangle-by-triangle approach requires 8 CNOTs per triangle, and there are on average 3/2 spins per triangle on the infinite 2D lattice, so in total the Trotter step requires $\approx(16/3)n$ CNOTs.  Thus with $m$ Trotter steps in total to implement $e^{-iHT}$, the total CNOT counts are approximately
\begin{subequations}
\begin{align}
F_1:\,\,\,\, & [1+(16/3)m]\times n\\
F_2,\,\,F_3:\,\,\,\, & [4+(16/3)m]\times n
\end{align}
\end{subequations}
The corresponding Trotter error scales as $T^2/m$, but also scales with the commutator of $H_1$ and $H_2$, the two pieces of the Hamiltonian containg all the up-pointing triangles and all the down-pointing triangles, respectively.  But the commutator of $H_1$ and $H_2$ can be written as a sum over all the individual commutators of one up-pointing triangle with one down-pointing triangle.  Of these $\mathcal{O}(n^2)$ commutators, most are 0 because the triangles do not overlap; the number of nonzero commutators is (3/2) times the number of triangles, which is propostional to $n$, and all are equal.  This gives a total Trotter error scaling as $n \times T^2/m$.  Then to achieve a fixed small Trotter error, the number of Trotter steps $m$ should scale as $n T^2$.~\footnote{The dependence on $T^2$ rather than $T$ is somewhat unintuitive, since we typically fix a Trotter step size, so the number of Trotter steps scales linearly with the total evolution time.  However, in that case the error accumulates over time, and here we require that the \emph{total} Trotter error be approximately independent of the evolution time.\label{footnote:T2}}  In short, the total CNOT count for the circuits for measuring $\langle\psi_0|e^{-iHT}|\psi_0\rangle$ with fixed error should scale with $n^2 T^2$.  

How small does the fixed error need to be?  As shown for ODMD in App. A of Ref.~\cite{ODMD}, the singular value threshold used to filter out noise, which needs to be larger than the magnitude of the noise to be effective, must be smaller than the overlap between the chosen initial state $|\psi_0\rangle$ and the true ground state.  In other words, if $|\psi_0\rangle = \sqrt{p_0}|\text{GS}\rangle +\sqrt{1-p_0}|\phi\rangle$, then the Trotter error must be smaller than $p_0$.  This gives a required CNOT count of at least $n^2 T^2 p_0^{-1}$.  We will return shortly to the question of estimating $p_0$ for reasonable choices of the initial state. 

The next question is how many shots are required to get good estimates for $F_1$, $F_2$, and $F_3$.  As shown in Ref.~\cite{ODMD}, both UVQPE and ODMD are robust to a large degree of shot noise, which was modeled in that paper as Gaussian noise drawn from a distribution with fixed width independent of the size of the expectation value.  
However, we must still be able to resolve the scale of typical expectation values relative well. If we assume a worst case scenario that the overlap of $|\psi_0\rangle$ with the true ground state is small, $p_0\ll 1$, then typical magnitudes of expectation values are approximately $p_0$ because $e^{-iHT}$ will impart effectively random phases on the components of $|\phi\rangle$ that will cancel out on average.  Then the magnitude of $F_1$ is approximately $p_0^2$.  Since shot noise for $M$ samples scales as approximately $1/\sqrt{M}$, we need a number of shots that scales as $p_0^{-4}$.    

The dependence on $p_0$ of both the required number of Trotter steps and the number of shots suggests that the algorithm will still fundamentally have an exponential scaling in system size.  In particular, if the initial state $|\psi_0\rangle$ is made by tiling some translation-invariant unit cell as in Fig.~\ref{fig:kl_dimer_coverings}(a), there is a fixed overlap per unit cell of $|\psi_0\rangle$ with the true ground state, hence the total overlap is exponentially small in the number of unit cells.  If the overlap per unit cell is $p$ and each unit cell has $N_u$ spins, then the exponential cost $p_0^{-1}$ will be proportional to $p^{-n/N_u} = \left(p^{-1/N_u}\right)^n$.  Fortunately, if the unit cell is large and if $p$ is relatively close to 1, then $p^{-1/N_u}\ll 2$, so the hybrid algorithms will still have a significant exponential speedup compared with exact classical methods.  Additionally, we find when we empirically consider small systems that the requirements on initial state overlap and number of shots are not too stringent.  As we show in Sec.~\ref{sec:results} below, for an 8-spin system with an initial overlap of 0.286, only around 1000 shots are required for UVQPE and ODMD to both rapidly converge to the correct ground state energy.

The last important question is how many time steps are needed for convergence. For UVQPE, the required number of time steps to achieve a specified level of error is given by Eq.~(27) of Ref.~\cite{UVQPE_Yizhi}.  We assume a very small size for the energy gap above the ground state, since otherwise the question of gaplessness of the model could be resolved without going to large system sizes.  In that case, solving for the number of time steps $j$ in the aforementioned equation gives 
\begin{equation}
j \geq \log((E_{N-1}-E_0)\sin^2(\Xi)/\cos^2(\Xi)\varepsilon)/2\log(\tilde{\epsilon}_{1,2}).
\end{equation}
Here $E_{N-1}$ is the maximal eigenvalue of the Hamiltonian, $E_0$ the minimal eigenvalue, $\cos^2(\Xi)=p_0$, $\varepsilon$ is the desired level of error in the ground state energy, and $\tilde{\epsilon}_{1,2}=1+3(\Delta E\times \Delta t)/2\pi$. $\Delta E$ is the energy gap above the ground state.

To find the scaling with system size, $n$, we examine each quantity in the expression.  The full width of the energy spectrum, $E_{N-1}-E_0$, scales with $n$ because energy is an extensive quantity.  $p_0$ is exponentially small in $n$, though likely with a base close to 1, as explained above; we can therefore ignore $\sin^2(\Xi)$.  Assuming we want a fixed level of error as we increase system size, the error $\varepsilon$ is independent of $n$ and can thus also be ignored in the thermodynamic limit.  Furthermore, $\log(\tilde{\epsilon}_{1,2})\sim\Delta E\times\Delta t$.  This gives
\begin{equation}
j \sim n\log(n)/(\Delta E \times \Delta t).
\end{equation}
Finally, we consider the system-size dependence of $\Delta E$ and $\Delta t$.  $\Delta E$ is the energy gap above the ground state, which as a worst-case scenario in the case of the gapless spectrum would go as the inverse of the linear size of the system, or $n^{-1/2}$.  $\Delta t$ is the time step size from Eq.~\eqref{eq:basis_states}, which as discussed in Sec.~\ref{sec:algs_UVQPE} needs to scale as the inverse of the spectral range of the Hamiltonian to avoid aliasing when extracting the eigenvalues of $H$ from eigenvalues of $e^{-iH\Delta t}$; thus $\Delta t$ is proportional to $1/n$.  The result is that the required number of time steps scales as
\begin{equation}
j \sim n^{5/2}\log(n).
\end{equation}
Note that if we required only that energy error per spin was constant when scaling up, rather than total error, so that $\varepsilon\sim~n$, this would result in a slightly more favorable scaling of $n^{5/2}$ without the additional $\log(n)$.


Now recall that the cost, measured in number of CNOT gates, of finding one matrix element $\langle\psi_0|e^{-iHT}|\psi_0\rangle$ to sufficient accuracy using Trotter evolution scaled as $p_0^{-5} n^2 T^2$.  To find the total cost across all time steps, we can sum $T^2$ up to the number of steps, which we estimate as $n^{5/2}$:
\begin{equation}
    C = \sum_{s=1}^{n^{5/2}} (\Delta t \times s)^2 \sim \Delta t^2 \left(n^{5/2}\right)^3 \sim n^{11/2}.
\end{equation}
Thus in total the cost of running UVQPE to find the ground state energy, in terms of CNOT gates, will scale as roughly $n^{15/2}p_0^{-5}$.  The longest individual circuits will have on the order of $p_0^{-1} n^2 T^2 \sim p_0^{-1} n^2 (\Delta t)^2 \left(n^{5/2}\right)^2 \sim p_0^{-1} n^5$ CNOT gates.

We can do the same type of analysis for ODMD.  Starting from Eqs.~(23) and (24) of Ref.~\cite{ODMD} and again assuming that gap size $\Delta E = E_1 - E_0$ is small, we find that the required number of time steps scales as
\begin{equation}
    d \sim (\Delta E \times \Delta t)^{-1} \sim n^{3/2}.
\end{equation}
This gives a total CNOT cost of $n^{9/2}p_0^{-5}$ and longest individual circuits with $\mathcal{O}(p_0^{-1} n^3)$ CNOT gates.

We conclude the discussion of scaling with some important notes.  (1) As mentioned in footnote~$\mathit{6}$, we have assumed that we need the same level of error in all expectation values.  Possibly UVQPE and/or ODMD would still work even if the error increases in subsequent time steps, in which case the circuit depth would need to scale only with $T$ rather than $T^2$.  This would significantly reduce the practical cost of the algorithms.  (2) The potentially greatest contribution to the cost of the algorithms comes from the overlap between the initial state and the true ground state, $p_0$.  We argued that for a lattice model where both states are translation-invariant, $p_0$ is necessarily exponential in system size.  We emphasize that other systems, such as molecules, where neither the ground state nor the initial state are translation invariant, may not be subject to the same exponential scaling. (3) Our analysis here was based on a first-order Trotter decomposition.  A second-order Trotter decomposition~\cite{Suzuki1991} would result in a Trotter error scaling as $n T^3/m^2$, which would result in a lower number of Trotter steps needed and hence a smaller number of CNOT gates.  Other higher-order Trotter expansions might also reduce the resource requirements.


\section{Restriction to star plaquettes\label{sec:star_plaquettes}}

We now turn to a practical demonstration of the hybrid algorithms on small spin systems closely related to the full 2D kagome lattice model.  We consider a single 12-spin star plaquette of the kagome lattice, shown in Fig.~\ref{fig:kl_and_plaquettes}(b), as well as an analogous 8-spin system shown in Fig.~\ref{fig:kl_and_plaquettes}(c), which can be viewed as one plaquette from the square kagome, or shuriken, lattice~\cite{Siddharthan2001_squarekagome}.  In each case, the Hamiltonian is the same as in Eq.~\eqref{eq:Heisenberg_def} with spin-spin interactions of strength $J$ between nearest neighbors and an applied field of strength $h$, just restricted to the smaller system.

In this section, we highlight important features of these smaller systems, including classically solvable exact ground states and a resulting technique for approximating the time evolution with shorter circuits.  We also discuss some nuts-and-bolts details for an empirical demonstration with classical emulators and on quantum hardware, namely initial states and details of our symmetry-based error mitigation strategies.  The actual results for ground state energy can be found in Sec.~\ref{sec:results} and for the magnetization curve in Sec.~\ref{sec:mag_curve}.

\subsection{\texorpdfstring{Exact ground states when $h=0$}{Exact ground states when h=0}\label{sec:star_plaquettes_exact_GS}}

Recall from Sec.~\ref{sec:full_2D_state_prep}, where we discussed the choice of initial state for the full two-dimensional lattice, that we showed two key facts: (1) When $h=0$, the Heisenberg Hamiltonian on a single 3-spin triangle has a 4-fold degenerate ground subspace spanned by the states shown in \eqref{eq:tri_gs_states}, which have two spins in a singlet dimer and the last spin free.  (2) As shown in Eq.~\eqref{eq:variational_principle_bound}, a lower bound for the ground state energy is given by $-3J/2 \times N_\Delta$, where $N_\Delta$ is the number of triangles in the system.  

We then concluded that a covering of the kagome lattice by nearest-neighbor singlet dimers would achieve the minimal energy $-3J/2$ on each triangle with a dimer along one edge, and hence would saturate the bound on the total energy and thus would necessarily be an exact ground state of the full system.  While no such dimer covering exists for the full 2D lattice, it is possible on both star plaquettes.  The resulting exact ground states of the star plaquette Heisenberg models (with $h=0$) are the ``pinwheel'' states
\begin{equation}
    \begin{tabular}{ccc}
\begin{tikzpicture}
\coordinate (A) at (-0.25,0.433);
\coordinate (B) at (0,0.866);
\coordinate (C) at (0.25,0.433);
\coordinate (D) at (0.75,0.433);
\coordinate (E) at (0.5,0);
\coordinate (F) at (0.75,-0.433);
\coordinate (G) at (0.25,-0.433);
\coordinate (H) at (0,-0.866);
\coordinate (I) at (-0.25,-0.433);
\coordinate (J) at (-0.75,-0.433);
\coordinate (K) at (-0.5,0);
\coordinate (L) at (-0.75,0.433);
    \filldraw[black] (A) circle (1pt) ;
    \filldraw[black] (B) circle (1pt);
    \filldraw[black] (C) circle (1pt) ;
    \filldraw[black] (D) circle (1pt);
    \filldraw[black] (E) circle (1pt) ;
    \filldraw[black] (F) circle (1pt);
    \filldraw[black] (G) circle (1pt) ;
    \filldraw[black] (H) circle (1pt);
    \filldraw[black] (I) circle (1pt) ;
    \filldraw[black] (J) circle (1pt);
    \filldraw[black] (K) circle (1pt) ;
    \filldraw[black] (L) circle (1pt);
    
    \draw[black] (A) -- (C);
    \draw[black] (A) -- (B);
    \draw[black] (B) -- (C);
    \draw[rotate around={60:($(A)!0.5!(B)$)},black] ($(A)!0.5!(B)$) ellipse (0.4 and 0.1);

    \draw[black] (C) -- (E);
    \draw[black] (C) -- (D);
    \draw[black] (D) -- (E);
    \draw[black] ($(C)!0.5!(D)$) ellipse (0.4 and 0.1);

    \draw[black] (E) -- (G);
    \draw[black] (E) -- (F);
    \draw[black] (F) -- (G);
    \draw[rotate around={-60:($(E)!0.5!(F)$)},black] ($(E)!0.5!(F)$) ellipse (0.4 and 0.1);

    \draw[black] (G) -- (I);
    \draw[black] (G) -- (H);
    \draw[black] (H) -- (I);
    \draw[rotate around={60:($(G)!0.5!(H)$)},black] ($(G)!0.5!(H)$) ellipse (0.4 and 0.1);

    \draw[black] (I) -- (K);
    \draw[black] (I) -- (J);
    \draw[black] (J) -- (K);
    \draw[black] ($(I)!0.5!(J)$) ellipse (0.4 and 0.1);

    \draw[black] (K) -- (A);
    \draw[black] (K) -- (L);
    \draw[black] (L) -- (A);
    \draw[rotate around={-60:($(K)!0.5!(L)$)},black] ($(K)!0.5!(L)$) ellipse (0.4 and 0.1);
    
\end{tikzpicture}
& &
\begin{tikzpicture}
    \filldraw[black] (0,0) circle (1pt) ;
    \filldraw[black] (0.25,0.433) circle (1pt);
    \filldraw[black] (0.5,0) circle (1pt) ;
    \filldraw[black] (0.933,-0.25) circle (1pt);
    \filldraw[black] (0.5,-0.5) circle (1pt) ;
    \filldraw[black] (0.25,-0.933) circle (1pt);
    \filldraw[black] (0,-0.5) circle (1pt) ;
    \filldraw[black] (-0.433,-0.25) circle (1pt);
    
    \draw[black] (0,0) -- (0.5,0);
    \draw[black] (0,0) -- (0.25,0.433);
    \draw[black] (0.5,0) -- (0.25,0.433);
    \draw[rotate around={60:(0.125,0.2165)},black] (0.125,0.2165) ellipse (0.4 and 0.1);
    
    \draw[black] (0.5,0) -- (0.5,-0.5);
    \draw[black] (0.5,0) -- (0.933,-0.25);
    \draw[black] (0.5,-0.5) -- (0.933,-0.25);
    \draw[rotate around={-30:(0.7165,-0.125)},black] (0.7165,-0.125) ellipse (0.4 and 0.1);
    
    \draw[black] (0.5,-0.5) -- (0,-0.5);
    \draw[black] (0.5,-0.5) -- (0.25,-0.933);
    \draw[black] (0,-0.5) -- (0.25,-0.933);
    \draw[rotate around={60:(0.375,-0.7165)},black] (0.375,-0.7165) ellipse (0.4 and 0.1);
    
    \draw[black] (0,-0.5) -- (0,0);
    \draw[black] (0,-0.5) -- (-0.433,-0.25);
    \draw[black] (0,0) -- (-0.433,-0.25);
    \draw[rotate around={-30:(-0.2165,-0.375)},black] (-0.2165,-0.375) ellipse (0.4 and 0.1);
\end{tikzpicture}
\\
12 spins & \,\,\,\,\,\,\,\,\,\,\,\, & 8 spins
\end{tabular}
    \label{eq:pinwheel}
\end{equation}
and their mirror images.  Note that the two mirror image pinwheel states for each plaquette are linearly independent but not orthogonal.~\footnote{Although we only explicitly discuss the 8- and 12-spin star plaquettes, in principle we could make larger loops of $n$ corner-sharing triangles.  Analogous pinwheel states are exact ground states for any $n$.}  This exact solvability of the $h=0$ Heisenberg model on the star plaquettes is discussed further in Refs.~\cite{MONTI1991197,Miyahara2011}.

The exact solvability means that, although the kagome lattice has a very high degree of geometric frustration, the Heisenberg model on the star plaquettes is technically ``frustration-free;'' a Hamiltonian $H$ is called frustration-free if it can be written as a sum of local terms $H=\sum H_j$, such that the ground state of $H$ is simultaneously a lowest-energy eigenstate of each $H_j$.~\cite{Bravyi2010}  The most obvious cases, such as Kitaev's toric code model, clearly allow simultaneous diagonalization of all local terms because all the terms in the Hamiltonian commute.  The present example is less obvious because the terms in $H$ do not commute and furthermore the terms in the Hamiltonian as written in Eq.~\eqref{eq:Heisenberg_def} do \emph{not} all have a simultaneous ground state---rather, the frustration-free nature of the model only becomes apparent once the individual spin-spin interactions are grouped into triangles.

\subsection{Trotterized time evolution\label{sec:f-f_Trotter}\label{sec:star_plaquettes_Trotter}}

Just as for the full 2D kagome lattice, there are multiple choices for how to implement the time evolution operator $e^{-iHt}$ for the star plaquettes.  One option, since the systems are small enough to diagonalize exactly, would be to find the $2^n\times 2^n$ unitary for $e^{-iHt}$, then to perform unitary synthesis to find a circuit that implements the time evolution exactly up to some level of numerical precision.  However, this approach has both practical and fundamental issues.  On the practial side, ``bottom-up'' numerical synthesis is an exponentially expensive classical computation~\cite{Davis2020qsearch} while ``top-down'' matrix decomposition approaches produce extremely long circuits that are not practical on near-term devices~\cite{Mottonen2004CSD, Shende2005QSD}.  On the fundamental side, we would like to use a method such that our empirical approach at least in principle is scalable to the full 2D system.

We thus again consider possible Suzuki-Trotter decompositions.  As in 2D, one option is a bond-by-bond decomposition, with bonds divided into four groups as shown here for the 8-spin plaquette: even-numbered bonds along the outside of the plaqutte, odd-numbered outer bonds, even-numbered inner bonds, and odd-numbered inner bonds.
\begin{equation}
    \begin{tabular}{ccccccc}
\begin{tikzpicture}
        \filldraw[black] (0,0) circle (1pt) ;
        \filldraw[black] (0.25,0.433) circle (1pt);
        \filldraw[black] (0.5,0) circle (1pt) ;
        \filldraw[black] (0.933,-0.25) circle (1pt);
        \filldraw[black] (0.5,-0.5) circle (1pt) ;
        \filldraw[black] (0.25,-0.933) circle (1pt);
        \filldraw[black] (0,-0.5) circle (1pt) ;
        \filldraw[black] (-0.433,-0.25) circle (1pt);
        
        \draw[black, draw opacity = 0.3] (0,0) -- (0.5,0);
        \draw[black] (0,0) -- (0.25,0.433);
        \draw[black, draw opacity = 0.3] (0.5,0) -- (0.25,0.433);
        
        \draw[black, draw opacity = 0.3] (0.5,0) -- (0.5,-0.5);
        \draw[black] (0.5,0) -- (0.933,-0.25);
        \draw[black, draw opacity = 0.3] (0.5,-0.5) -- (0.933,-0.25);
        
        \draw[black, draw opacity = 0.3] (0.5,-0.5) -- (0,-0.5);
        \draw[black] (0.5,-0.5) -- (0.25,-0.933);
        \draw[black, draw opacity = 0.3] (0,-0.5) -- (0.25,-0.933);
        
        \draw[black, draw opacity = 0.3] (0,-0.5) -- (0,0);
        \draw[black] (0,-0.5) -- (-0.433,-0.25);
        \draw[black, draw opacity = 0.3] (0,0) -- (-0.433,-0.25);
    \end{tikzpicture} 
    & 
    \begin{tikzpicture}
        \filldraw[black] (0,0) circle (1pt) ;
        \filldraw[black] (0.25,0.433) circle (1pt);
        \filldraw[black] (0.5,0) circle (1pt) ;
        \filldraw[black] (0.933,-0.25) circle (1pt);
        \filldraw[black] (0.5,-0.5) circle (1pt) ;
        \filldraw[black] (0.25,-0.933) circle (1pt);
        \filldraw[black] (0,-0.5) circle (1pt) ;
        \filldraw[black] (-0.433,-0.25) circle (1pt);
        
        \draw[black, draw opacity = 0.3] (0,0) -- (0.5,0);
        \draw[black, draw opacity = 0.3] (0,0) -- (0.25,0.433);
        \draw[black] (0.5,0) -- (0.25,0.433);
        
        \draw[black, draw opacity = 0.3] (0.5,0) -- (0.5,-0.5);
        \draw[black, draw opacity = 0.3] (0.5,0) -- (0.933,-0.25);
        \draw[black] (0.5,-0.5) -- (0.933,-0.25);
        
        \draw[black, draw opacity = 0.3] (0.5,-0.5) -- (0,-0.5);
        \draw[black, draw opacity = 0.3] (0.5,-0.5) -- (0.25,-0.933);
        \draw[black] (0,-0.5) -- (0.25,-0.933);
        
        \draw[black, draw opacity = 0.3] (0,-0.5) -- (0,0);
        \draw[black, draw opacity = 0.3] (0,-0.5) -- (-0.433,-0.25);
        \draw[black] (0,0) -- (-0.433,-0.25);
    \end{tikzpicture} & 
    \begin{tikzpicture}
        \filldraw[black] (0,0) circle (1pt) ;
        \filldraw[black, draw opacity = 0.3, fill opacity = 0.3] (0.25,0.433) circle (1pt);
        \filldraw[black] (0.5,0) circle (1pt) ;
        \filldraw[black, draw opacity = 0.3, fill opacity = 0.3] (0.933,-0.25) circle (1pt);
        \filldraw[black] (0.5,-0.5) circle (1pt) ;
        \filldraw[black, draw opacity = 0.3, fill opacity = 0.3] (0.25,-0.933) circle (1pt);
        \filldraw[black] (0,-0.5) circle (1pt) ;
        \filldraw[black, draw opacity = 0.3, fill opacity = 0.3] (-0.433,-0.25) circle (1pt);
        
        \draw[black] (0,0) -- (0.5,0);
        \draw[black, draw opacity = 0.3] (0,0) -- (0.25,0.433);
        \draw[black, draw opacity = 0.3] (0.5,0) -- (0.25,0.433);
        
        \draw[black, draw opacity = 0.3] (0.5,0) -- (0.5,-0.5);
        \draw[black, draw opacity = 0.3] (0.5,0) -- (0.933,-0.25);
        \draw[black, draw opacity = 0.3] (0.5,-0.5) -- (0.933,-0.25);
        
        \draw[black] (0.5,-0.5) -- (0,-0.5);
        \draw[black, draw opacity = 0.3] (0.5,-0.5) -- (0.25,-0.933);
        \draw[black, draw opacity = 0.3] (0,-0.5) -- (0.25,-0.933);
        
        \draw[black, draw opacity = 0.3] (0,-0.5) -- (0,0);
        \draw[black, draw opacity = 0.3] (0,-0.5) -- (-0.433,-0.25);
        \draw[black, draw opacity = 0.3] (0,0) -- (-0.433,-0.25);
    \end{tikzpicture} & 
    \begin{tikzpicture}
        \filldraw[black] (0,0) circle (1pt) ;
        \filldraw[black, draw opacity = 0.3, fill opacity = 0.3] (0.25,0.433) circle (1pt);
        \filldraw[black] (0.5,0) circle (1pt) ;
        \filldraw[black, draw opacity = 0.3, fill opacity = 0.3] (0.933,-0.25) circle (1pt);
        \filldraw[black] (0.5,-0.5) circle (1pt) ;
        \filldraw[black, draw opacity = 0.3, fill opacity = 0.3] (0.25,-0.933) circle (1pt);
        \filldraw[black] (0,-0.5) circle (1pt) ;
        \filldraw[black, draw opacity = 0.3, fill opacity = 0.3] (-0.433,-0.25) circle (1pt);
        
        \draw[black, draw opacity = 0.3] (0,0) -- (0.5,0);
        \draw[black, draw opacity = 0.3] (0,0) -- (0.25,0.433);
        \draw[black, draw opacity = 0.3] (0.5,0) -- (0.25,0.433);
        
        \draw[black] (0.5,0) -- (0.5,-0.5);
        \draw[black, draw opacity = 0.3] (0.5,0) -- (0.933,-0.25);
        \draw[black, draw opacity = 0.3] (0.5,-0.5) -- (0.933,-0.25);
        
        \draw[black, draw opacity = 0.3] (0.5,-0.5) -- (0,-0.5);
        \draw[black, draw opacity = 0.3] (0.5,-0.5) -- (0.25,-0.933);
        \draw[black, draw opacity = 0.3] (0,-0.5) -- (0.25,-0.933);
        
        \draw[black] (0,-0.5) -- (0,0);
        \draw[black, draw opacity = 0.3] (0,-0.5) -- (-0.433,-0.25);
        \draw[black, draw opacity = 0.3] (0,0) -- (-0.433,-0.25);
    \end{tikzpicture} 
\\
Outer even & Outer odd & Inner even & Inner odd
    \end{tabular}
    \label{eq:bond_by_bond}
\end{equation}
This decomposition is slightly better than the analogous one shown in Fig.~\ref{fig:bond-by-bond_Trotter} for the 2D lattice, since this decomposition still preserves one important spatial symmetry, namely the $C_4$ rotation for the 8-spin plaquette and $C_6$ rotation for the 12-spin plaquette.

The triangle-by-triangle decomposition is also a possibility.  As shown below, as in 2D we have a decomposition $H=H_1+H_2$, but here $H_1$ corresponds to even-numbered triangles around the plaquette [left column in \eqref{eq:tri_by_tri}] and $H_2$ to odd-numbered triangles [right column].
\begin{equation}
    \begin{tabular}{ccccc}
    \raisebox{2.3em}{$N_\Delta=6$}\,\,\,\, & & 
    \begin{tikzpicture}
        \coordinate (A) at (-0.25,0.433);
        \coordinate (B) at (0,0.866);
        \coordinate (C) at (0.25,0.433);
        \coordinate (D) at (0.75,0.433);
        \coordinate (E) at (0.5,0);
        \coordinate (F) at (0.75,-0.433);
        \coordinate (G) at (0.25,-0.433);
        \coordinate (H) at (0,-0.866);
        \coordinate (I) at (-0.25,-0.433);
        \coordinate (J) at (-0.75,-0.433);
        \coordinate (K) at (-0.5,0);
        \coordinate (L) at (-0.75,0.433);
        \filldraw[black] (A) circle (1pt) ;
        \filldraw[black] (B) circle (1pt);
        \filldraw[black] (C) circle (1pt) ;
        \filldraw[black, draw opacity = 0.3, fill opacity = 0.3] (D) circle (1pt);
        \filldraw[black] (E) circle (1pt) ;
        \filldraw[black] (F) circle (1pt);
        \filldraw[black] (G) circle (1pt) ;
        \filldraw[black, draw opacity = 0.3, fill opacity = 0.3] (H) circle (1pt);
        \filldraw[black] (I) circle (1pt) ;
        \filldraw[black] (J) circle (1pt);
        \filldraw[black] (K) circle (1pt) ;
        \filldraw[black, draw opacity = 0.3, fill opacity = 0.3] (L) circle (1pt);
        
        \draw[black] (A) -- (C);
        \draw[black] (A) -- (B);
        \draw[black] (B) -- (C);
    
        \draw[black, draw opacity = 0.3] (C) -- (E);
        \draw[black, draw opacity = 0.3] (C) -- (D);
        \draw[black, draw opacity = 0.3] (D) -- (E);
    
        \draw[black] (E) -- (G);
        \draw[black] (E) -- (F);
        \draw[black] (F) -- (G);
    
        \draw[black, draw opacity = 0.3] (G) -- (I);
        \draw[black, draw opacity = 0.3] (G) -- (H);
        \draw[black, draw opacity = 0.3] (H) -- (I);
    
        \draw[black] (I) -- (K);
        \draw[black] (I) -- (J);
        \draw[black] (J) -- (K);
    
        \draw[black, draw opacity = 0.3] (K) -- (A);
        \draw[black, draw opacity = 0.3] (K) -- (L);
        \draw[black, draw opacity = 0.3] (L) -- (A);
        
    \end{tikzpicture}
     & \,\,\,\, & 
    \begin{tikzpicture}
        \coordinate (A) at (-0.25,0.433);
        \coordinate (B) at (0,0.866);
        \coordinate (C) at (0.25,0.433);
        \coordinate (D) at (0.75,0.433);
        \coordinate (E) at (0.5,0);
        \coordinate (F) at (0.75,-0.433);
        \coordinate (G) at (0.25,-0.433);
        \coordinate (H) at (0,-0.866);
        \coordinate (I) at (-0.25,-0.433);
        \coordinate (J) at (-0.75,-0.433);
        \coordinate (K) at (-0.5,0);
        \coordinate (L) at (-0.75,0.433);
        \filldraw[black] (A) circle (1pt) ;
        \filldraw[black, draw opacity = 0.3, fill opacity = 0.3] (B) circle (1pt);
        \filldraw[black] (C) circle (1pt) ;
        \filldraw[black] (D) circle (1pt);
        \filldraw[black] (E) circle (1pt) ;
        \filldraw[black, draw opacity = 0.3, fill opacity = 0.3] (F) circle (1pt);
        \filldraw[black] (G) circle (1pt) ;
        \filldraw[black] (H) circle (1pt);
        \filldraw[black] (I) circle (1pt) ;
        \filldraw[black, draw opacity = 0.3, fill opacity = 0.3] (J) circle (1pt);
        \filldraw[black] (K) circle (1pt) ;
        \filldraw[black] (L) circle (1pt);
        
        \draw[black, draw opacity = 0.3] (A) -- (C);
        \draw[black, draw opacity = 0.3] (A) -- (B);
        \draw[black, draw opacity = 0.3] (B) -- (C);
    
        \draw[black] (C) -- (E);
        \draw[black] (C) -- (D);
        \draw[black] (D) -- (E);
    
        \draw[black, draw opacity = 0.3] (E) -- (G);
        \draw[black, draw opacity = 0.3] (E) -- (F);
        \draw[black, draw opacity = 0.3] (F) -- (G);
    
        \draw[black] (G) -- (I);
        \draw[black] (G) -- (H);
        \draw[black] (H) -- (I);
    
        \draw[black, draw opacity = 0.3] (I) -- (K);
        \draw[black, draw opacity = 0.3] (I) -- (J);
        \draw[black, draw opacity = 0.3] (J) -- (K);
    
        \draw[black] (K) -- (A);
        \draw[black] (K) -- (L);
        \draw[black] (L) -- (A);
        
    \end{tikzpicture}
    \\
    \raisebox{1.8em}{$N_\Delta=4$}\,\,\,\, & & 
    \begin{tikzpicture}
        \filldraw[black] (0,0) circle (1pt) ;
        \filldraw[black] (0.25,0.433) circle (1pt);
        \filldraw[black] (0.5,0) circle (1pt) ;
        \filldraw[black, draw opacity = 0.3, fill opacity = 0.3] (0.933,-0.25) circle (1pt);
        \filldraw[black] (0.5,-0.5) circle (1pt) ;
        \filldraw[black] (0.25,-0.933) circle (1pt);
        \filldraw[black] (0,-0.5) circle (1pt) ;
        \filldraw[black, draw opacity = 0.3, fill opacity = 0.3] (-0.433,-0.25) circle (1pt);
        
        \draw[black] (0,0) -- (0.5,0);
        \draw[black] (0,0) -- (0.25,0.433);
        \draw[black] (0.5,0) -- (0.25,0.433);
        
        \draw[black, draw opacity = 0.3] (0.5,0) -- (0.5,-0.5);
        \draw[black, draw opacity = 0.3] (0.5,0) -- (0.933,-0.25);
        \draw[black, draw opacity = 0.3] (0.5,-0.5) -- (0.933,-0.25);
        
        \draw[black] (0.5,-0.5) -- (0,-0.5);
        \draw[black] (0.5,-0.5) -- (0.25,-0.933);
        \draw[black] (0,-0.5) -- (0.25,-0.933);
        
        \draw[black, draw opacity = 0.3] (0,-0.5) -- (0,0);
        \draw[black, draw opacity = 0.3] (0,-0.5) -- (-0.433,-0.25);
        \draw[black, draw opacity = 0.3] (0,0) -- (-0.433,-0.25);
    \end{tikzpicture} & \,\,\,\, & 
    \begin{tikzpicture}
        \filldraw[black] (0,0) circle (1pt) ;
        \filldraw[black, draw opacity = 0.3, fill opacity = 0.3] (0.25,0.433) circle (1pt);
        \filldraw[black] (0.5,0) circle (1pt) ;
        \filldraw[black] (0.933,-0.25) circle (1pt);
        \filldraw[black] (0.5,-0.5) circle (1pt) ;
        \filldraw[black, draw opacity = 0.3, fill opacity = 0.3] (0.25,-0.933) circle (1pt);
        \filldraw[black] (0,-0.5) circle (1pt) ;
        \filldraw[black] (-0.433,-0.25) circle (1pt);
        
        \draw[black, draw opacity = 0.3] (0,0) -- (0.5,0);
        \draw[black, draw opacity = 0.3] (0,0) -- (0.25,0.433);
        \draw[black, draw opacity = 0.3] (0.5,0) -- (0.25,0.433);
        
        \draw[black] (0.5,0) -- (0.5,-0.5);
        \draw[black] (0.5,0) -- (0.933,-0.25);
        \draw[black] (0.5,-0.5) -- (0.933,-0.25);
        
        \draw[black, draw opacity = 0.3] (0.5,-0.5) -- (0,-0.5);
        \draw[black, draw opacity = 0.3] (0.5,-0.5) -- (0.25,-0.933);
        \draw[black, draw opacity = 0.3] (0,-0.5) -- (0.25,-0.933);
        
        \draw[black] (0,-0.5) -- (0,0);
        \draw[black] (0,-0.5) -- (-0.433,-0.25);
        \draw[black] (0,0) -- (-0.433,-0.25);
    \end{tikzpicture}
    \\
    & & $H_1$ & & $H_2$
    \end{tabular}
    \label{eq:tri_by_tri}
\end{equation}
The triangle-by-triangle approach maintains its most important advantage over the bond-by-bond approach, namely the reduction in CNOT count from 9 to 8 per triangle.  Furthermore, the triangle-by-triangle approach has two additional advantages specific to the star plaquettes, one minor and one which is important to the success of our demonstration on real quantum hardware.  

The minor advantage comes from partial parallelization of Trotter steps to slightly reduce run time and thus noise.  Consider three successive triangles around the edge of the plaquette.  During one Trotter step, we first apply the circuit from Fig.~\ref{fig:tri-by-tri_circuit} on the first and third triangles, then on the second:
\begin{equation}
    \begin{tikzcd}[ampersand replacement=\&, row sep={0.4cm,between origins}, column sep=0.1cm]
\& \qw\& \ctrl{1}\gategroup[3,steps=8,style={rounded corners, inner sep=-0.5pt}]{} \& \ctrl{2} \& \qw \& \ctrl{1} \& \ctrl{1} \& \ctrl{2} \& \qw \& \ctrl{1} \& \qw  \& \qw
\& \qw \& \qw \& \qw \& \qw \& \qw \& \qw \& \qw \& \qw \& \qw \& \qw\\
\& \qw\& \targ{}  \& \qw \& \ctrl{1} \& \targ{}  \& \targ{}  \& \qw \& \ctrl{1} \& \targ{}  \& \qw \& \qw
\& \qw \& \qw \& \qw \& \qw \& \qw \& \qw \& \qw \& \qw \& \qw \& \qw\\
\& \qw\& \qw \& \targ{}  \& \targ{}  \& \qw      \& \qw \& \targ{}  \& \targ{}  \& \qw \& \qw \& \qw
\& \ctrl{1}\gategroup[3,steps=8,style={rounded corners, inner sep=-0.5pt}]{} \& \ctrl{2} \& \qw \& \ctrl{1} \& \ctrl{1} \& \ctrl{2} \& \qw \& \ctrl{1} \& \qw \& \qw\\
\& \qw\& \qw \& \qw \& \qw \& \qw \& \qw \& \qw \& \qw \& \qw \& \qw \& \qw
\& \targ{}  \& \qw \& \ctrl{1} \& \targ{}  \& \targ{}  \& \qw \& \ctrl{1} \& \targ{} \& \qw \& \qw\\
\& \qw\& \ctrl{1}\gategroup[3,steps=8,style={rounded corners, inner sep=-0.5pt}]{} \& \ctrl{2} \& \qw \& \ctrl{1} \& \ctrl{1} \& \ctrl{2} \& \qw \& \ctrl{1} \& \qw \& \qw
\& \qw \& \targ{}  \& \targ{}  \& \qw      \& \qw \& \targ{}  \& \targ{}  \& \qw \& \qw \& \qw\\
\& \qw\& \targ{}  \& \qw \& \ctrl{1} \& \targ{}  \& \targ{}  \& \qw \& \ctrl{1} \& \targ{}\& \qw \& \qw
\& \qw \& \qw \& \qw \& \qw \& \qw \& \qw \& \qw \& \qw \& \qw \& \qw\\
\& \qw\& \qw \& \targ{}  \& \targ{}  \& \qw      \& \qw \& \targ{}  \& \targ{}  \& \qw \& \qw \& \qw
\& \qw \& \qw \& \qw \& \qw \& \qw \& \qw \& \qw \& \qw \& \qw \& \qw
\end{tikzcd}
\label{eq:Trotter_parallelization}
\end{equation}
For visual clarity we have not drawn the single-qubit rotations and have grouped the gates belonging to each copy of the circuit from Fig.~\ref{fig:tri-by-tri_circuit}.  Evidently, the two pieces of the Trotter step can be made to overlap, resulting in a 1/8 reduction in the number of layers in the circuit.

More importantly, the triangle-by-triangle approach actually incurs exactly 0 error when time-evolving the pinwheel ground states, no matter the size of the Trotter time step.  This remarkable property follows from the frustration-free nature of the Hamiltonian.  To see this, consider one Trotter step of size $t$, given by the operator
\begin{equation}
    F_t = e^{-iH_1 t}e^{-iH_2t},
\end{equation}
where $H_1$ acts on the even triangles and $H_2$ on the odd triangles, as in \eqref{eq:tri_by_tri}.  Then, since the exact ground state of $H$ is also an eigenstate of the local Hamiltonians $H_\Delta$ acting on each triangle of the star plaquette, it is also an eigenstate of both $H_1$ and $H_2$, each with eigenvalue $-3(J/2)N_\Delta/2$.  The ground state is thus an eigenstate of $F$ with eigenvalue $e^{-it \times -3(J/2)N_\Delta }$, exactly the same as its eigenvalue under the full time evolution $U_t^{ } = e^{-iHt}$.  Intuitively, even though $e^{-iH_1 t}e^{-iH_2 t}\neq e^{-iHt}$, they are equal specifically when acting on the ground state because $H_1$ and $H_2$ each act on the state just as scalar multiplication.  (Note, however, that the ground state of $H$ is not an exact eigenstate of noisy approximations to $H_1$ and $H_2$, hence on a real quantum device some Trotter error will remain.)

The lack of Trotter error for the ground state suggests a means of testing out UVQPE and ODMD on the frustration-free star plaquettes with shorter circuit depth than is required when trying to minimize Trotter error by taking many small Trotter steps, hence with much lower noise when running on present-day and near-term quantum devices.  Considering UVQPE in particular, since we know that for any time step size $\Delta t$, $U_{\Delta t}$ and $F_{\Delta t}$ have the same eigenvalue for the ground state we want to find, we can solve a generalized eigenvalue problem for the eigenvalues of the operator $F_{\Delta t}$ rather than $U_{\Delta t}$.  

Then recall from Sec.~\ref{sec:algs_UVQPE} that we must choose $\Delta t$ such that the spectrum of $H\Delta t$ lies in an interval of size $2\pi$ in order to unambiguously find eigenvalues of $H$ from eigenvalues of $U_{\Delta t}$.  Let us further assume that the eigenvalues of $H\Delta t$ specifically lie in the interval $(-\pi,\pi)$, and let the minimal and maximal values be $E_0\Delta t$ and $E_{N-1}\Delta t$.  Then by the triangle inequality, the phases of the eigenvalues of $F_{\Delta t}$ will also lie in $[-E_{N-1}\Delta t,-E_0\Delta t]$.  As a result, the ground state of $H$ is guaranteed to still have the most positive phase for $F_{\Delta t}$ and thus will be found by the UVQPE variant where we look for eigenstates of $F_{\Delta t}$.

We can likewise generate the subspace used for the generalized eigenvalue problem by evolving the initial state $|\psi_0\rangle$ with $F$ rather than $U$:
\begin{align}
    |\psi_m\rangle &= U_{m\Delta t}|\psi_0\rangle \\ 
    &\mapsto F_{m\Delta t}|\psi_0\rangle = e^{-iH_1 m\Delta t}e^{-iH_2 m\Delta t}|\psi_0\rangle.\nonumber
\end{align}
Using this basis and solving for the ground state using $F$, the $(j,k)$ element of the overlap matrix on the right-hand side of Eq.~\eqref{eq:UVQPE} would then become
\begin{align}
    \langle e^{-iH(k-j)\Delta t}\rangle & = \langle e^{iHj\Delta t}e^{-iHk\Delta t}\rangle \\ & \mapsto \langle e^{iH_2j\Delta t}e^{iH_1(j-k)\Delta t}e^{-iH_2k\Delta t}\rangle.\nonumber
\end{align}
Likewise, the $(j,k)$ element of the time-evolution operator matrix on the left-hand side becomes
\begin{align}
    \langle e^{-iH(1+k-j)\Delta t}\rangle & = \langle e^{iHj\Delta t}e^{-iHt}e^{-iHk\Delta t}\rangle \label{eq:UVQPE_approx_mat_elt_v1}\\
    & \mapsto
    \langle F_{j\Delta t}^\dg F_{\Delta t} F_{k\Delta t}\rangle.\nonumber
\end{align}
Effectively, each matrix element requires at most three Trotter steps, substantially limiting the circuit depth while still producing a linear algebra problem that is, at least in theory, guaranteed to give the correct ground state energy.

We emphasize that this method, solving the generalized eigenvalue problem for $F_{\Delta t}$, is primarily useful in the special case that we already know the ground state.  Note, however, that for small enough $\Delta t$, $F_{\Delta t}$ and $U_{\Delta t}$ only differ by an error of order $\Delta t^2$, so even for real unsolved problems such as the full 2D kagome lattice Heisenberg model, it could potentially provide a good approximation to the eigenvalues.  

Even in the specialized setting of a frustration-free Hamiltonian, this approach still has some drawbacks.  First, we lose the Toeplitz structure of the matrices, since letting $j\rightarrow j+1$, $k\rightarrow k+1$ in Eq.~\eqref{eq:UVQPE_approx_mat_elt_v1} changes the value of the matrix element.  Then unlike in UVQPE as presented in Sec.~\ref{sec:algs_UVQPE}, we can no longer measure only the expectation values from the top row of the matrix, significantly increasing the number of different circuits that must be run.  Second, even for the 8-spin plaquette, there are still 48 CNOT gates per Trotter step, so even three steps may lead to a high degree of noise on present-day quantum devices.

To make hardware demonstrations feasible we make a further approximation: we simply replace all occurrences of $e^{-iHt}$ in Eq.~\eqref{eq:UVQPE} by a single Trotter step $F(t)$, giving the generalized eigenvalue problem
\begin{widetext}
    \begin{equation}
    \left(\begin{array}{cccc} 
    \langle F_{\Delta t}\rangle & \langle F_{2\Delta t}\rangle & \langle F_{3\Delta t}\rangle & \cdots \\
    1 & \langle F_{\Delta t}\rangle & \langle F_{2\Delta t}\rangle & \cdots \\
    \langle F_{-\Delta t}\rangle & 1 & \langle F_{\Delta t}\rangle & \cdots\\
    \vdots & \vdots & \vdots & \ddots
    \end{array}\right) \mathbf{v} = 
    \left(\begin{array}{cccc} 
    1 & \langle F_{\Delta t}\rangle & \langle F_{2\Delta t}\rangle & \cdots \\
    \langle F_{-\Delta t}\rangle & 1 & \langle F_{\Delta t}\rangle & \cdots \\
    \langle F_{-2\Delta t}\rangle & \langle F_{-\Delta t}\rangle & 1 & \cdots\\
    \vdots & \vdots & \vdots & \ddots
    \end{array}\right) \lambda \mathbf{v}\label{eq:UVQPE_m1}
    \end{equation}
\end{widetext}
with all expectation values again taken in an initial state $|\psi_0\rangle$.  This restores the Toeplitz structure and hence the efficiency of the method in terms of the number of circuits to be run, and it limits every expectation value to a short circuit with a constant depth corresponding to a single Trotter step, thus bringing hardware noise down to a manageable level.  On the other hand, this generalized eigenvalue problem no longer clearly represents energy minimization within a subspace, and we must reconsider why the method should work at all.

First, it is at least plausible that the approach \emph{could} give the correct ground state energy because, if the initial state $|\psi_0\rangle$ is chosen to be the exact ground state, then the matrix on the left-hand side is exact a scalar multiple $e^{-i E_0\Delta t}$ times the matrix on the right, giving the correct eigenvalue $\lambda$.  For a more general initial state with an overlap with the true ground state, we should consider how UVQPE works; as shown in Ref.~\cite{UVQPE_Yizhi}, with successively larger matrices the linear combinations of basis states allow phase cancellation of the coefficients of excited states, ultimately isolating the true ground state.  This process still occurs for the generalized eigenvalue problem in Eq.~\eqref{eq:UVQPE_m1}.  In fact, the effective increased phase randomness due to Trotter error actually makes the non-ground state components cancel out even faster, giving rapid convergence.

\subsection{Initial states\label{sec:star_plaquettes_psi0}}

For the full 2D kagome lattice, we argued in Sec.~\ref{sec:full_2D_state_prep} that dimer coverings that come as close as possible to having one dimer along an edge of each triangle are good initial states, likely having significant overlap with the ground state.  Applying the same logic on the star plaquettes gives the pinwheel states of \eqref{eq:pinwheel}, which as discussed above are in fact exact ground states.

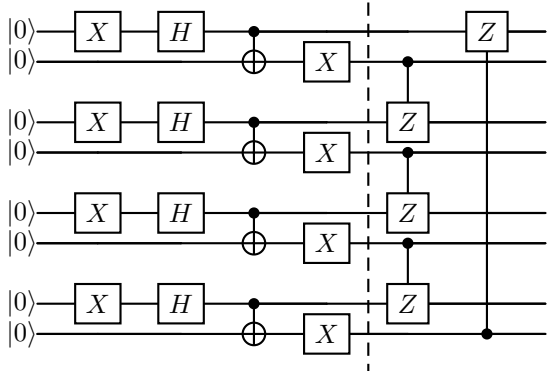
\begin{figure}
    \centering
    \begin{tikzcd}[ampersand replacement=\&, row sep={0.4cm,between origins}]
\ket{0} \& \gate{X} \& \gate{H} \& \ctrl{1} \& \qw      \slice[style=black]{} \& 
\qw \& \gate{Z} \& \qw \\
\ket{0} \& \qw      \& \qw      \& \targ{}  \& \gate{X} \& 
\ctrl{2} \& \qw  \& \qw \\ \\
\ket{0} \& \gate{X} \& \gate{H} \& \ctrl{1} \& \qw      \& 
\gate{Z} \& \qw  \& \qw \\
\ket{0} \& \qw      \& \qw      \& \targ{}  \& \gate{X} \& 
\ctrl{2} \& \qw  \& \qw \\ \\
\ket{0} \& \gate{X} \& \gate{H} \& \ctrl{1} \& \qw      \& 
\gate{Z} \& \qw  \& \qw \\
\ket{0} \& \qw      \& \qw      \& \targ{}  \& \gate{X} \& 
\ctrl{2} \& \qw  \& \qw \\ \\
\ket{0} \& \gate{X} \& \gate{H} \& \ctrl{1} \& \qw      \& 
\gate{Z} \& \qw \& \qw \\
\ket{0} \& \qw      \& \qw      \& \targ{}  \& \gate{X} \& 
\qw \& \ctrl{-10} \& \qw 
\end{tikzcd}
    \caption{Circuit to prepare initial states for generating the subspace bases for UVQPE and ODMD.  We show the circuit for the 8-spin plaquette; there is an analogous state for the 12-spin case.  The state at the dashed line is the pinwheel of \eqref{eq:pinwheel}, and the CZ gates to the right add entanglement between dimers.}
    \label{fig:plaquette_initial_state_Sz0}
\end{figure}

Since our goal is to use the star plaquettes as test cases for the hybrid algorithms, we need to pick different initial states that are close to, but not equal to, these known ground states.  One solution would be to use a different dimer covering, where dimers connect spins that are not adjacent.  However, we would like to use an initial state that is cheap to prepare on any quantum hardware with connectivity similar to that of the model itself.  Our solution is to take the known dimer pinwheel ground state, then apply controlled-$Z$ gates clockwise on the outer bonds that do not currently feature a dimer, as shown in Fig.~\ref{fig:plaquette_initial_state_Sz0}.  The total overlap of this state with the ground state subspace is 0.286 for the 8-spin star and just 0.001 for the 12-spin star.  

As noted in Sec.~\ref{sec:full_2D_scaling}, the algorithms require the noise level to be below the initial state overlap, so the latter initial state would require on the order of $10^6$ shots per expectation value even if the only noise comes from sampling.  We therefore also use a state where every other CZ operator is omitted.  The resulting state has an overlap of 0.016 with the ground state subspace.  Schematically, the two initial states look like
\begin{equation}
    \begin{tabular}{ccc}
\begin{tikzpicture}
\coordinate (A) at (-0.25,0.433);
\coordinate (B) at (0,0.866);
\coordinate (C) at (0.25,0.433);
\coordinate (D) at (0.75,0.433);
\coordinate (E) at (0.5,0);
\coordinate (F) at (0.75,-0.433);
\coordinate (G) at (0.25,-0.433);
\coordinate (H) at (0,-0.866);
\coordinate (I) at (-0.25,-0.433);
\coordinate (J) at (-0.75,-0.433);
\coordinate (K) at (-0.5,0);
\coordinate (L) at (-0.75,0.433);
    \filldraw[black] (A) circle (1pt) ;
    \filldraw[black] (B) circle (1pt);
    \filldraw[black] (C) circle (1pt) ;
    \filldraw[black] (D) circle (1pt);
    \filldraw[black] (E) circle (1pt) ;
    \filldraw[black] (F) circle (1pt);
    \filldraw[black] (G) circle (1pt) ;
    \filldraw[black] (H) circle (1pt);
    \filldraw[black] (I) circle (1pt) ;
    \filldraw[black] (J) circle (1pt);
    \filldraw[black] (K) circle (1pt) ;
    \filldraw[black] (L) circle (1pt);
    
    \draw[black] (A) -- (C);
    \draw[black] (A) -- (B);
    \draw[black, line width = 4pt] (B) -- (C);
    \draw[rotate around={60:($(A)!0.5!(B)$)},black] ($(A)!0.5!(B)$) ellipse (0.4 and 0.1);

    \draw[black] (C) -- (E);
    \draw[black] (C) -- (D);
    \draw[black, line width = 4pt] (D) -- (E);
    \draw[black] ($(C)!0.5!(D)$) ellipse (0.4 and 0.1);

    \draw[black] (E) -- (G);
    \draw[black] (E) -- (F);
    \draw[black, line width = 4pt] (F) -- (G);
    \draw[rotate around={-60:($(E)!0.5!(F)$)},black] ($(E)!0.5!(F)$) ellipse (0.4 and 0.1);

    \draw[black] (G) -- (I);
    \draw[black] (G) -- (H);
    \draw[black, line width = 4pt] (H) -- (I);
    \draw[rotate around={60:($(G)!0.5!(H)$)},black] ($(G)!0.5!(H)$) ellipse (0.4 and 0.1);

    \draw[black] (I) -- (K);
    \draw[black] (I) -- (J);
    \draw[black, line width = 4pt] (J) -- (K);
    \draw[black] ($(I)!0.5!(J)$) ellipse (0.4 and 0.1);

    \draw[black] (K) -- (A);
    \draw[black] (K) -- (L);
    \draw[black, line width = 4pt] (L) -- (A);
    \draw[rotate around={-60:($(K)!0.5!(L)$)},black] ($(K)!0.5!(L)$) ellipse (0.4 and 0.1);
    
\end{tikzpicture}
& &
\begin{tikzpicture}
\coordinate (A) at (-0.25,0.433);
\coordinate (B) at (0,0.866);
\coordinate (C) at (0.25,0.433);
\coordinate (D) at (0.75,0.433);
\coordinate (E) at (0.5,0);
\coordinate (F) at (0.75,-0.433);
\coordinate (G) at (0.25,-0.433);
\coordinate (H) at (0,-0.866);
\coordinate (I) at (-0.25,-0.433);
\coordinate (J) at (-0.75,-0.433);
\coordinate (K) at (-0.5,0);
\coordinate (L) at (-0.75,0.433);
    \filldraw[black] (A) circle (1pt) ;
    \filldraw[black] (B) circle (1pt);
    \filldraw[black] (C) circle (1pt) ;
    \filldraw[black] (D) circle (1pt);
    \filldraw[black] (E) circle (1pt) ;
    \filldraw[black] (F) circle (1pt);
    \filldraw[black] (G) circle (1pt) ;
    \filldraw[black] (H) circle (1pt);
    \filldraw[black] (I) circle (1pt) ;
    \filldraw[black] (J) circle (1pt);
    \filldraw[black] (K) circle (1pt) ;
    \filldraw[black] (L) circle (1pt);
    
    \draw[black] (A) -- (C);
    \draw[black] (A) -- (B);
    \draw[black, line width = 4pt] (B) -- (C);
    \draw[rotate around={60:($(A)!0.5!(B)$)},black] ($(A)!0.5!(B)$) ellipse (0.4 and 0.1);

    \draw[black] (C) -- (E);
    \draw[black] (C) -- (D);
    \draw[black] (D) -- (E);
    \draw[black] ($(C)!0.5!(D)$) ellipse (0.4 and 0.1);

    \draw[black] (E) -- (G);
    \draw[black] (E) -- (F);
    \draw[black, line width = 4pt] (F) -- (G);
    \draw[rotate around={-60:($(E)!0.5!(F)$)},black] ($(E)!0.5!(F)$) ellipse (0.4 and 0.1);

    \draw[black] (G) -- (I);
    \draw[black] (G) -- (H);
    \draw[black] (H) -- (I);
    \draw[rotate around={60:($(G)!0.5!(H)$)},black] ($(G)!0.5!(H)$) ellipse (0.4 and 0.1);

    \draw[black] (I) -- (K);
    \draw[black] (I) -- (J);
    \draw[black, line width = 4pt] (J) -- (K);
    \draw[black] ($(I)!0.5!(J)$) ellipse (0.4 and 0.1);

    \draw[black] (K) -- (A);
    \draw[black] (K) -- (L);
    \draw[black] (L) -- (A);
    \draw[rotate around={-60:($(K)!0.5!(L)$)},black] ($(K)!0.5!(L)$) ellipse (0.4 and 0.1);
\end{tikzpicture}
\\
Overlap 0.001 & \,\,\,\,\,\,\,\,\,\,\,\, & Overlap 0.016
\end{tabular}\label{eq:pinwheel_with_CZ_states}
\end{equation}
where the thick lines indicate the locations of CZ operators applied to the pinwheel state.  
In Sec.~\ref{sec:results_classical}, we use the two initial states with their dramatically different overlaps with the ground state to explore in more depth how the algorithms depend on the quality of the initial state.

Note that applying the layer of CZ gates takes us out of the spin-0 sector, but we do remain in the $S^z_{ }=0$ sector.  If we were aiming to find the ground state energy of an unsolved model, an initial state that is less than fully symmetric would be a bad choice.  In this case it simply makes the computational task more challenging and hence actually makes for a better test of the algorithms.  

\begin{figure*}
    \centering
    \input{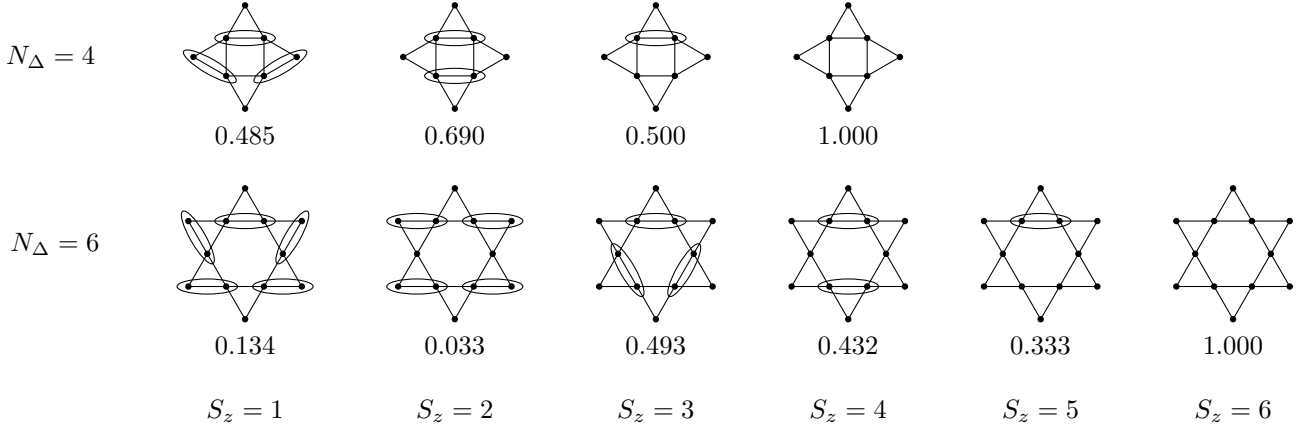}
    \caption{Initial states for the 8- and 12-spin plaquettes in different symmetry sectors with $S^z>0$.  For larger $S^z$, the states are chosen to be similar to the pinwheel states of Eq.~\eqref{eq:pinwheel} while being in the correct $S^z$ sector and reasonably spatially symmetric.  The circled dimers are in a spin singlet state, while the remaining spins are in the state $|\!\su\rangle$.  The number under each state gives the overlap with the true ground state in that symmetry sector, or if there are multiple degenerate ground states, the total of the overlap with the ground subspace.}
    \label{fig:initial_states}
\end{figure*}

We can also consider good initial states for other $S^z_{ }$ sectors, which will be relevant for finding the overall ground state when we turn on the magnetic field, $h$, and hence for finding the magnetization curve.  To find initial states that have reasonable overlap with the ground state in each $S^z$ symmetry sector, we use a partial dimer covering, leaving enough free spins to reach the $S^z$ value for the sector.  We illustrate our initial states for the 8- and 12-spin stars and give their overlaps with the true ground states of their respective $S^z_{ }$ sectors in Fig.~\ref{fig:initial_states}.

\subsection{Error mitigation details\label{sec:star_plaquettes_error_miti}}

Implementations of quantum algorithms on near-term devices benefit from error mitigation strategies to reduce noise.  Standard methods include zero noise extrapolation~\cite{Li2017_ZNE,Temme2017,Kandala2019,He2020_ZNE,Krebsbach2022,Kim2023}, especially via Richardson extrapolation~\cite{Richardson}, 
dynamical decoupling~\cite{Viola1999,Ezzell2023}, and probabilistic error cancellation~\cite{Temme2017,Kim2023}. In addition to these problem-agnostic methods, for Hamiltonian simulation we can use the symmetries of the Hamiltonian to reduce and detect errors~\cite{BonetMonroig2018,Tran2021,Cai2021quantumerror}.  As noted in Sec.~\ref{sec:full_2D_model}, we use two error mitigation strategies based on the conservation of total $S^z$: detection of bit strings inconsistent with conservation of $S^z$ during time evolution and twirling with single-qubit gates to average away leakage into incorrect symmetry sectors.

\paragraph{$S^z_{ }$ error detection:} First, we use symmetry-based error detection.  When $h=0$, our initial state $|\psi_0\rangle$ consisting of CZ operators applied to the pinwheel state \eqref{eq:pinwheel} has total $S^z=0$.  The triangle-by-triangle Trotterized time evolution conserves $S^z$, so if measured bit strings after time evolution have a different total $S^z$, they can be safely discarded.  

However, we have overlooked a significant complication.  Recalling the mirror circuit method from Sec.~\ref{sec:algs_mirror}, we sample bit strings not from the state $e^{-iHt}|\psi_0\rangle$ but rather from $|0(t)\rangle = U_0^{\dg}e^{-iHt}U_0^{ }|\mathbf{0}\rangle$ when measuring $F_1$ and the analogous states [Eqs.~\eqref{eq:0Rt} and \eqref{eq:0Rit}] when measuring $F_2$ and $F_3$.  Focusing first on the state for $F_1$, we note that the state-preparation circuit implementing $U_0$ explicitly does \emph{not} conserve $S^z_{ }$, since it must convert from the all-0 state, which is in the $S^z=L$ sector, to a state in the $S^z=0$ sector.  Thus although the time evolved state $e^{-iHt}U_0^{ }|\mathbf{0}\rangle$ should be in the $S^z=0$ sector, allowing for error detection, the state $|0(t)\rangle$ that we actually measure will not be.

Fortunately, we can overcome this by careful scrutiny of the implementation of $U_0^{ }$ given in Eq.~\eqref{eq:U0_circuit}.  On each two-qubit dimer, $U_0^{ }$ implements the transformation 
\begin{equation}
   \frac{1}{\sqrt{2}}\left(\begin{array}{cccc}
    0 & 1 & 0 & 1\\
    1 & 0 & 1 & 0\\
    -1 & 0 & 1 & 0\\
    0 & -1 & 0 & 1\end{array}\right).\label{eq:dimer_mapping}
\end{equation}
Then when we apply $U_0^\dg$ to a state that lies entirely in the $S^z=0$ sector, we will map each pair of spins with labels $(j,j+1)$ by the Hermitian conjugate of \eqref{eq:dimer_mapping}.  Looking at the rows of the matrix, we see that a pair of spins in the state $|00\rangle$ will map to a linear combination of $|01\rangle$ and $|11\rangle$, as will spins in the state $|11\rangle$.  On the other hand, a state of the form $|01\rangle$ or $|10\rangle$ will map to a linear combination of $|00\rangle$ and $|10\rangle$.  

But if we are correctly in the $S^z_{ }=0$ sector, each basis state with weight in $e^{-iHt}U_0^{ }|\mathbf{0}\rangle$ will have an equal number of 0s and 1s, and thus when the state is divided into pairs of spins, the number of pairs in the $|00\rangle$ state will be equal to the number of pairs in the $|11\rangle$ state.  Thus the total number of pairs in either the state $|00\rangle$ or the state $|11\rangle$ will be \emph{even}.  Thus, after applying $U_0^\dg$, any measured bit string in the state $|0(t)\rangle$ that has an \emph{odd} total number of spin pairs in the states $|01\rangle$ and $|11\rangle$ corresponds to a symmetry sector error, and that measurement can be discarded.  On any star plaquette with $4n$ spins, this condition eliminates exactly half of the possible bit strings as invalid measurement outcomes.

Other related approaches could further limit the number of valid bitstrings, thus allowing more errors to be detected and the corresponding shots discarded.  For example~\footnote{As suggested by one of the anonymous referees.}, $U_0$ could be divided into a layer of $X$ gates acting on some qubits to transform $|\mathbf{0}\rangle$ into a product state in the target $S^z$ sector, followed by local entangling gates that transform $|01\rangle$, rather than $|00\rangle$, into a singlet dimer.  The layer of $X$ gates could be skipped when applying $U_0^\dg$, with bit-flips done in classical post-processing of measured bistrings.  Then $F_1$ would be the probability to measure the product state from after the initial layer of $X$ gates, rather than the all-0 state.  In this case, only the $n$ choose $n/2$ bitstrings with $S^z_{ }=0$ would be valid, and any other measured bitstring would be discarded.

This approach to error detection scales well with system size and with error rate.  For an $n$-qubit system, checking each measured bit string has cost linear in $n$, so the total cost is $nM$ where $M$ is the number of shots; there is negligible overhead compared with the existing cost of classically recording measurement outcomes.  Unlike other error detection schemes, this method also does not rely on adding ancillas or extra CNOT gates.  The method will also scale well with decreasing error rates as quantum hardware improves: the number of shots discarded will go to 0.  Thus although this method is based on ``post-selection,'' in principle it incurs no overhead whatsoever in quantum resources.  

We can perform a similar analysis for the circuits for $F_2$ and $F_3$ to find a condition to filter out invalid measurements on the states $|0_R(t)\rangle$ and $|0_{Ri}(t)\rangle$.  However, the analysis of how basis states propagate through the GHZ preparation circuit is sufficiently complicated that we do not make use of filtering for $F_2$ and $F_3$.  Empirically, it remains true that around half of the possible bit strings are not valid measurement outcomes if the symmetry sector is conserved during time evolution. 

\paragraph{$S^z_{ }$-preserving twirling:} Our second symmetry-based mitigation method is a form of twirling, or randomized compiling, introduced in Ref.~\cite{Tran2021}.  We use the fact that a well-chosen layer of single-qubit $R_z^{ }$ rotation gates can leave desired symmetry sectors invariant while adding a phase to others, so that in averaging over various runs, the erroneous symmetry sector components will cancel out.

To be concrete, suppose we have a product state of $2n$ spins where half are in the state $|0\rangle$ and half are in the state $|1\rangle$.  Then if we apply $R_z^{ }(\theta)^{\otimes 2n}$ to the state, where
\begin{equation}
    R_z^{ }(\theta) = \left(\begin{array}{cc}e^{i\theta} & 0 \\ 0 & e^{-i\theta}\end{array}\right)
\end{equation}
and importantly $\theta$ is the same for each qubit, the net phase acquired will be exactly zero.  The same will be true for any superposition of such product states, i.e. any state, not just a product state, with $S^z_{ }=0$.  
On the other hand, a state with total spin of $S^z_{ }=2$, with $n+1$ qubits in the state $|0\rangle$ and $n-1$ in the state $|1\rangle$, would acquire an overall phase of $2\theta$.  

Now let's suppose we have a superposition of states with $S^z_{ }=0$ and $S^z_{ }=2$, $a|0\rangle + b|2\rangle$.  If we run half our shots with an $R_z^{ }$ layer inserted with $\theta=\pi/2$, our final measurement outcomes will be effectively sampled from the density matrix
\begin{align}
    \rho = & \frac{1}{2}(a|0\rangle + b|2\rangle)(a^\ast\langle 0| + b^\ast\langle 2|) \nonumber\\
    &+ \frac{1}{2}(a|0\rangle - b|2\rangle)(a^\ast\langle 0| - b^\ast\langle 2|) \nonumber\\
    = & |a|^2|0\rangle\langle 0| + |b|^2|2\rangle\langle 2|.\label{eq:twirled_rho}
\end{align}
Thus if the state was supposed to be in the $S^z_{ }=0$ sector, but there was a small leakage of order $\epsilon$ into the $S^z_{ }=2$ sector due to hardware noise, this twirling procedure, which costs only an inconsequential layer of 1-qubit gates, reduces the contribution from the erroneous symmetry sector to be of order $\epsilon^2$.  Note that, although a direct measurement of $\rho$ either with or without twirling would give symmetry sector $S^z=0$ with probability $|a|^2$, when the measurements come after the further unitary operations such as inverted state preparation circuits, the linear-in-$\epsilon$ off-diagonal elements of the density matrix that are present without twirling would have an effect on the final measurements.

This twirling method works even when we take the superposition with a reference state for measuring $F_2$ and $F_3$.  In that case, the state during time evolution should be a superposition of $S^z_{ }=0$ and $S^z_{ }=L$, so as long as we choose $\theta$ to be an integer multiple of $2\pi/L$, both symmetry sectors will be unaffected.  On the other hand, the most likely leakage will be to $S^z = \pm 2$, $L-2$, and we can choose $\theta=\pi/2$ so that state components in all of these symmetry sectors will acquire a relative negative sign.  We can thus cancel out some errors by running each of our circuits half the time with a layer of $R_z^{ }(\pi/2)$ gates after the Trotter evolution and half without.


\begin{figure*}
    \centering
    \includegraphics[width=0.98\textwidth]{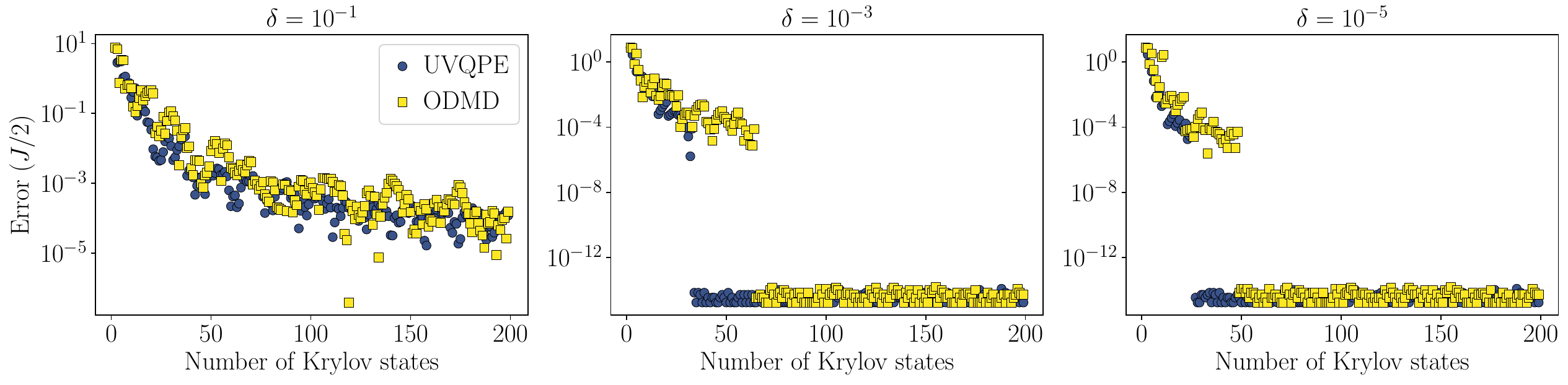}
    \caption{Convergence of UVQPE and ODMD estimate of the ground state energy with the number of time steps/Krylov basis states used for the 8-spin star plaquette.  Time evolution and computation of expectation values are numerically exact, with no error or shot noise.  The initial state is the pinwheel plus a layer of CZ operators, as shown in Fig.~\ref{fig:plaquette_initial_state_Sz0}.  Even though the simulations are exact, we use noise filtering with three levels of singular value threshold, because we need to invert nearly singular matrices so even noise at the level of machine precision can be problematic, and because we want to understand the effect of the filtering on convergence.  We see that in the noiseless case, more aggressive noise filtering (larger $\delta$) leads to slower convergence.}
    \label{fig:GS_results_exact_8}
\end{figure*}

\begin{figure*}
    \centering
    \includegraphics[width=0.98\textwidth]{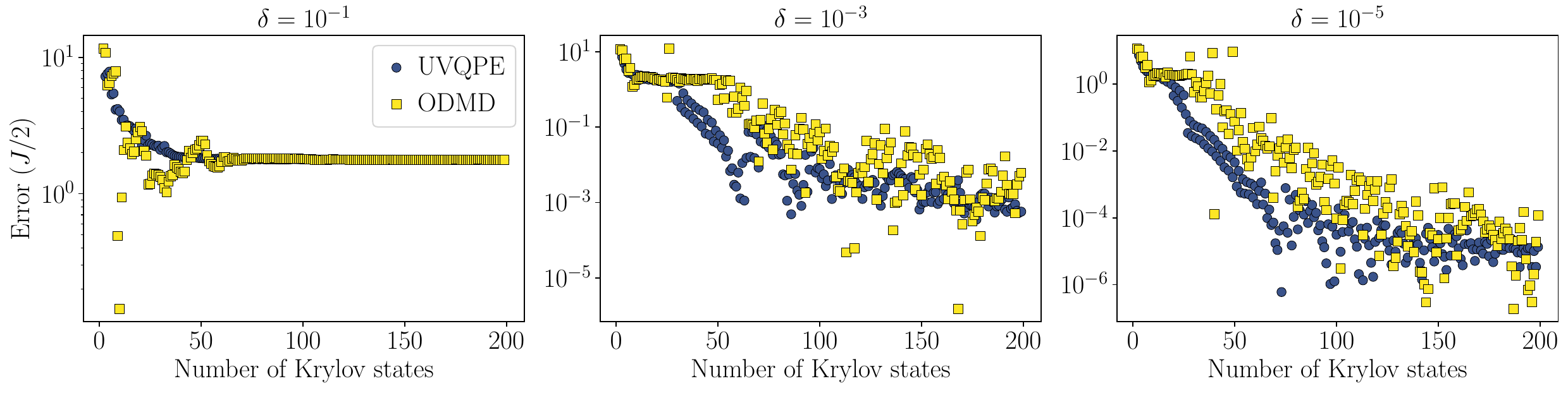}
    \caption{We perform the same numerical experiment as in Fig.~\ref{fig:GS_results_exact_8}, but for the 12-spin plaquette using six-CZ initial state from the left of Eq.~\eqref{eq:pinwheel_with_CZ_states}, which has just a 0.1\% overlap with the ground subspace.  As a result of this low overlap, both algorithms initially converge towards a low-lying excited state instead of to the ground state.  With small $\delta$, both algorithms eventually escape the local minimum and find the ground state energy; with agressive noise filtering using $\delta=0.1$, the algorithms never converge to the ground state.}
    \label{fig:GS_results_exact_12}
\end{figure*}

\section{Ground state energy results\label{sec:results}}

Finally, we present our results from running both UVQPE and ODMD on the star plaquettes to find the ground state energy, focusing only on the case $h=0$.  Following Lieb's theorem~\cite{Lieb1989}, we therefore consider only the $S^z=0$ symmetry sector.  We show results using exact classical state vector simulation both with and without shot noise due to taking a finite number of measurements to determine $F_1$, $F_2$, and $F_3$ for each expectation value.  These simulations use the exact time evolution operator $e^{-iHt}$.  For the 8-spin plaquette, we also show results using both a noisy classical emulator and real quantum hardware, namely the Quantinuum H1-1 processor; these simulations use the single Trotter step approximation discussed in Sec.~\ref{sec:star_plaquettes_Trotter} where $U_t$ is replaced by $F_t$.

\subsection{Exact classical simulation results\label{sec:results_classical}}

For both the 8-spin and 12-spin star plaquettes, we use an initial state $|\psi_0\rangle$ given by the pinwheel state \eqref{eq:pinwheel} augmented by a layer of CZ gates as described in Sec.~\ref{sec:star_plaquettes_psi0}.  For the 8-spin plaquette we use four CZ gates as illustrated in Fig.~\ref{fig:plaquette_initial_state_Sz0}; the overlap of this state with the ground state subspace is 0.286.  For the 12-spin plaquette we consider two different initial states of this type, shown in Eq.~\eqref{eq:pinwheel_with_CZ_states}; their respective overlaps with the ground state subspace are 0.001 and 0.016.  
We then compute the expectation values of Eq.~\eqref{eq:U_expecs} in two ways: (1) direct computation of the expectation value via matrix multiplication and (2) exactly generating the states $|0(t)\rangle$, $|0_R(t)\rangle$, and $|0_{Ri}(t)\rangle$ from Eqs.~\eqref{eq:0t}, \eqref{eq:0Rt}, and \eqref{eq:0Rit}, respectively, then sampling from the states and using Eq.~\eqref{eq:mirror_circuit_computation} to find the expectation values.  

We process the expectation values using both UVQPE and ODMD, truncating small singular values from the matrices on the right-hand side of Eq.~\eqref{eq:UVQPE} and Eq.~\eqref{eq:ODMD}, respectively.  Specifically, we truncate all singular values below a threshold that is a factor $\delta$ times the largest singular value of the matrix; empirically, we need to truncate singular values up to around 10x the noise level on the individual matrix elements in order for the algorithms to converge correctly.

For the time step size in the simulations, $\Delta t$, we recall from Sec.~\ref{sec:full_2D_H_bounds} that for $h=0$ the Hamiltonian spectrum is bounded by $\norm{H}_2 = (3J/2)N_\Delta$, thus requiring $\Delta t < 2\pi/(3JN_\Delta)$.  The 8- and 12-spin plaquettes have $N_\Delta$ = 4 and 6, respectively.  We choose a relatively conservative value of $\Delta t/(J/2) = 0.1$, satisfying the bound for both plaquettes.  The resulting (unitless) time-step operator is $U_{\Delta t} = \exp\left(-i\times 0.1\sum \boldsymbol{\sigma}\cdot\boldsymbol{\sigma}\right)$.

\begin{figure}
    \centering
    \includegraphics[width=0.98\columnwidth]{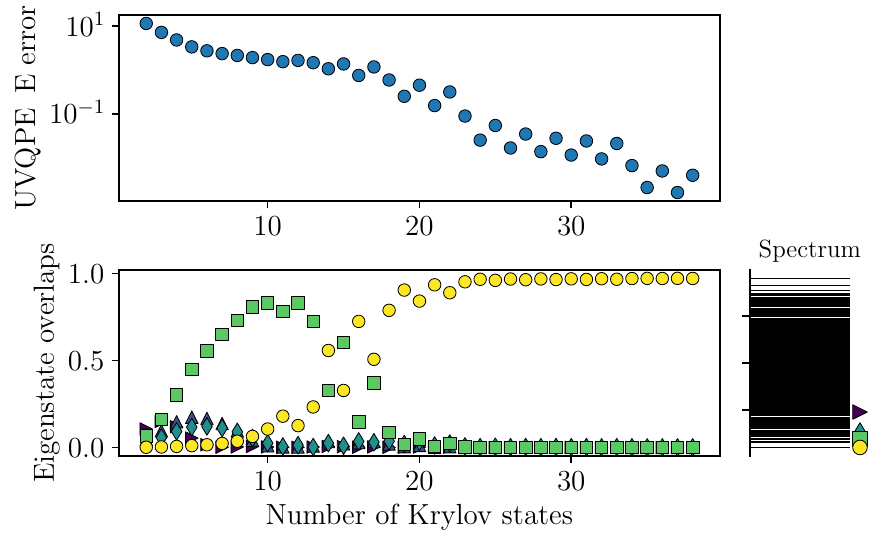}
    \caption{A closer examination of the energy convergence for the 12-spin plaquette with the six-CZ initial state, using UVQPE with $\delta=10^{-6}$.  The upper panel shows the energy estimate as a function of the number of time steps/basis states, similar to the UVQPE data from Fig.~\ref{fig:GS_results_exact_12}, rightmost panel; energy error is measured in units of $J/2$.  In the lower panel, we show the convergence of $|\tilde{v}_0\rangle$, the UVQPE approximation to the ground state, by finding its overlap with each eigenstate of $H$.  We plot the overlaps for the five eigenstates that contribute most to the approximation to the ground state across the simulation.  On the right we show where the energies of those eigenstates fall on the spectrum of $H$; the five states include the ground state and several low-lying excited states.  Evidently, in the first 10 or so time steps, UVQPE converges towards the excited state marked with a green square.  With this small $\delta$, the algorithm escapes the local minimum and converges to the true ground state; with $\delta=0.1$ as on the left of Fig.~\ref{fig:GS_results_exact_12}, the algorithm gets stuck at the excited state.}
    \label{fig:UVQPE_GS_12_spin_low_overlap}
\end{figure}

The results of the simulations are summarized in a series of figures, as follows:
\begin{itemize}
    \item In Fig.~\ref{fig:GS_results_exact_8}, we show the convergence of the ground state energy for both UVQPE and ODMD for the 8-spin plaquette, with expectation values computed exactly.  We specifically plot, on a log scale, the error in the energy estimate as a function of time step, or in other words of the number of Krylov states in the basis \eqref{eq:basis_states}.  Although the expectation values are accurate to machine precision, the linear algebra problems in Eqs.~\eqref{eq:UVQPE} and \eqref{eq:ODMD} are ill-conditioned, so some noise filtering is still needed; we use three singular value thresholds, $\delta = 10^{-1}$, $10^{-3}$, and $10^{-5}$.  In this case, because the simulation is noiseless so not much singular value filtering is needed, both algorithms converge faster when $\delta$ is smaller.  
    
    \item In Fig.~\ref{fig:GS_results_exact_12}, we show the same calculation but for the 12-spin plaquette, using the initial state with six CZ operators as shown on the left of Eq.~\eqref{eq:pinwheel_with_CZ_states}, which has an overlap of only $10^{-3}$ with the true ground states.  When the level of noise filtering $\delta$ is greater in magnitude than the initial state overlap, the algorithm converges to a low-lying excited state instead of to the ground state---essentially, the contribution of the ground state to the initial state gets filtered out.  When $\delta$ is comparable to or smaller than the overlap, as in the middle panel, both algorithms plateau at the energy of the low-lying excited state before eventually converging.
    
\begin{figure*}
    \centering
    \includegraphics[width=0.98\textwidth]{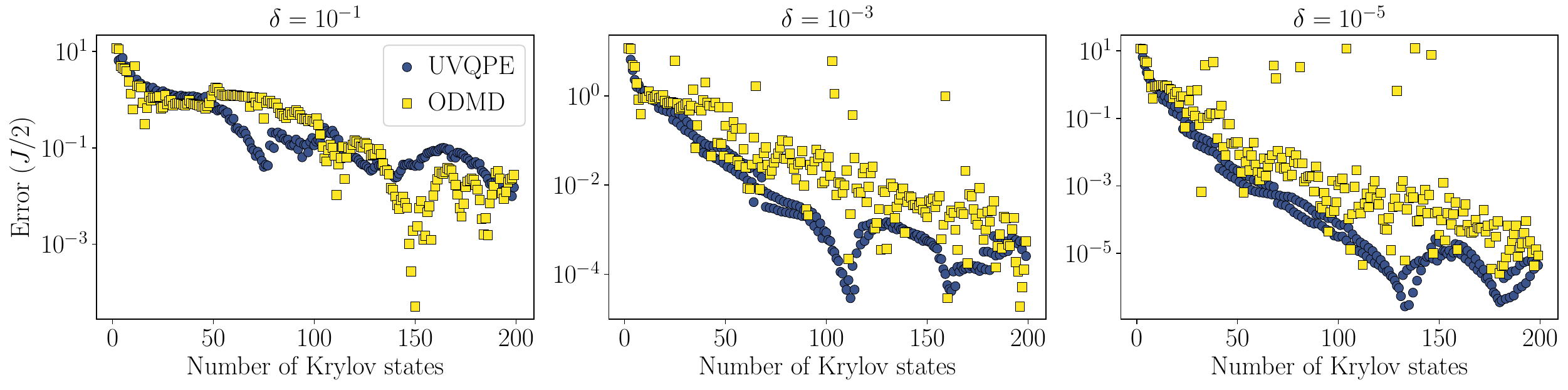}
    \caption{We perform the same numerical experiment on the 12-spin plaquette as in Fig.~\ref{fig:GS_results_exact_12}, but using the three-CZ initial state from the right of Eq.~\eqref{eq:pinwheel_with_CZ_states}.  This state has a much larger overlap with the ground subspace, 0.016 rather than 0.001.  As a result, both algorithms converge to the ground state even when $\delta$ is large.}
    \label{fig:GS_results_exact_12_v2}
\end{figure*}  

\begin{figure*}
    \centering
    \includegraphics[width=0.98\textwidth]{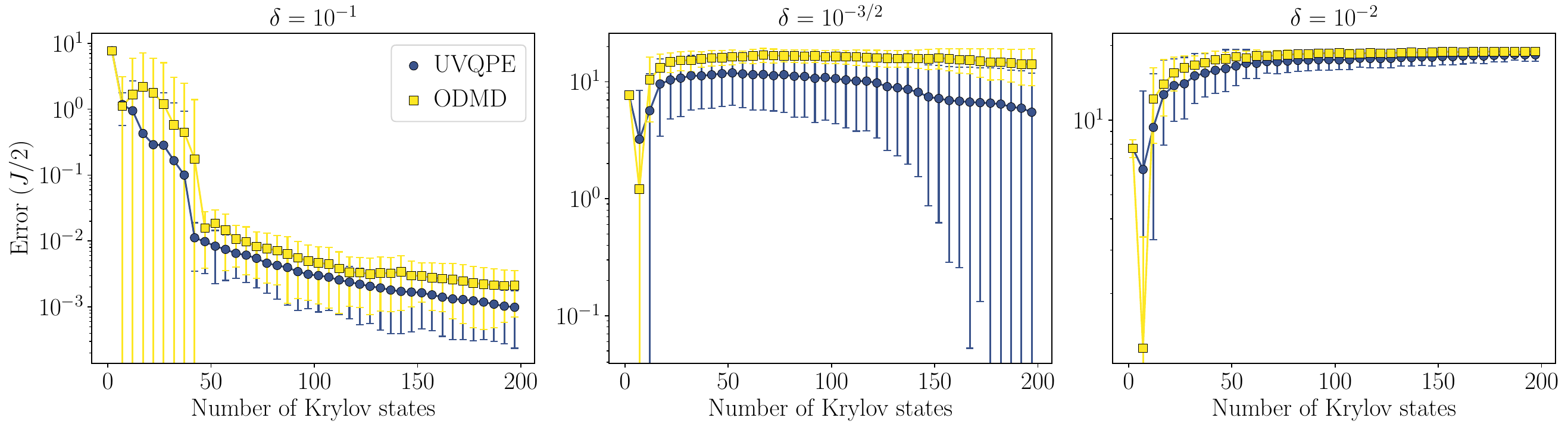}
    \caption{We perform the noisy version of the numerical experiment on the 8-spin plaquette from Fig.~\ref{fig:GS_results_exact_8}.  Each expectation value from Eq.~\eqref{eq:U_expecs} is computed by sampling the states \eqref{eq:0t}, \eqref{eq:0Rt}, and \eqref{eq:0Rit} to find $F_1$, $F_2$, and $F_3$; We use $10^3$ shots in total per expectation value, distributed among $F_1$, $F_2$, and $F_3$.  We perform the sampling 100 times at each time step, and we show here the mean energy estimate from UVQPE and ODMD across the 100 shot noise realizations.  Error bars show the standard deviation across noise realizations.  A larger $\delta$ is needed to filter out the noise so that the algorithms can converge.}
    \label{fig:GS_results_exact_sampled_8}
\end{figure*}

\begin{figure*}
    \centering
    \includegraphics[width=0.98\textwidth]{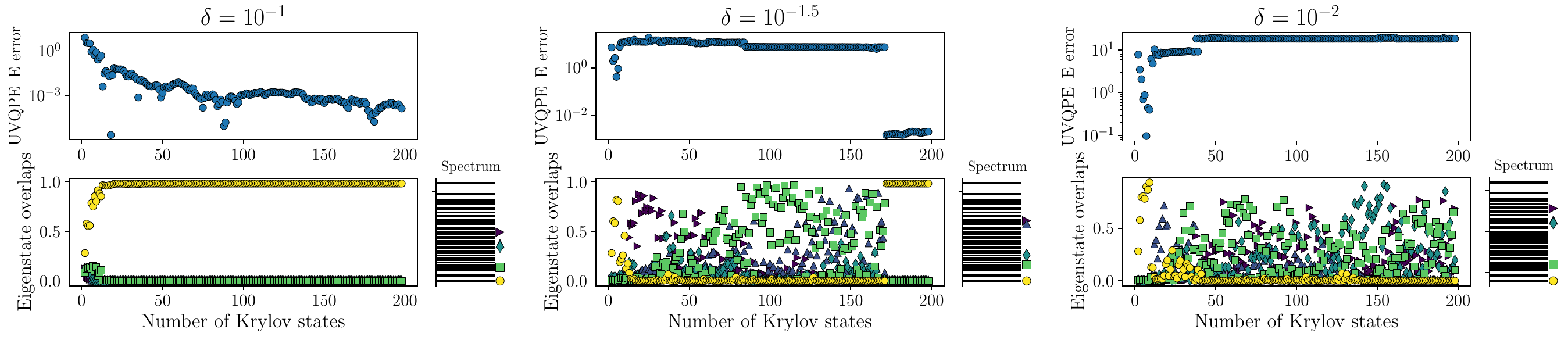}
    \caption{For each singular value threshold for the numerical experiment on the 8-spin plaquette (Fig.~\ref{fig:GS_results_exact_sampled_8}), we consider a single noise realization and show the convergence of energy and eigenvector in UVQPE in the style of Fig.~\ref{fig:UVQPE_GS_12_spin_low_overlap}.  By looking at a single noise realization, we clearly see that when the singular value threshold is too low relative to the noise level, in this case $\delta=10^{-2}$, the algorithm fails to converge at all.  In contrast, convergence occurs quickly for a sufficiently large noise threshold, and occurs after a long energy plateau in the case of intermediate noise filtering.}
    \label{fig:GS_results_exact_sampled_8_one_shot}
\end{figure*}
    
    \item In Fig.~\ref{fig:UVQPE_GS_12_spin_low_overlap}, we focus on UVQPE (with $\delta = 10^{-6}$) to demonstrate that the plateau in energy does indeed correspond to finding a low-lying excited state.  Recall from Sec.~\ref{sec:algs_UVQPE} that UVQPE gives an approximation to the ground state eigenvector, $|\tilde{v}_0\rangle$, in the form of a linear combination of the basis states \eqref{eq:basis_states}.  For the small star plaquettes, we can explicitly generate the basis states and thus the specified linear combination, and we can find the overlap of $|\tilde{v}_0\rangle$ with the true ground state and other eigenvectors of the Hamiltonian.  In the figure, we show the convergence of energy in the upper panel and the decomposition of the corresponding ground state approximation into Hamiltonian eigenstates in the lower panel.  After around 10 time steps, the UVQPE approximation has a very high overlap with the excited state marked with a green square.  With a small value of $\delta$, the algorithm eventually finds the true ground state, marked with a yellow circle, but with a larger $\delta$ it converges to the excited state instead.
    
    \item In Fig.~\ref{fig:GS_results_exact_12_v2}, we show the same calculations as in Fig.~\ref{fig:GS_results_exact_12}, but using the initial state with three CZ operators as shown on the right of Eq.~\eqref{eq:pinwheel_with_CZ_states}, which has a higher overlap with the ground state, around 0.016.  In this case, the energy converges for all three levels of noise filtering, just as for the 8-spin plaquette.  (We also show the convergence of the eigenvector for this initial state, analogous to Fig.~\ref{fig:UVQPE_GS_12_spin_low_overlap}, in the upper left panel of Fig.~\ref{fig:UVQPE_GS_12_spin}.)
    
    \item In Fig.~\ref{fig:GS_results_exact_sampled_8}, we show the convergence of the energy for both UVQPE and ODMD for the 8-spin plaquette with shot noise taken into account.  We use 1000 shots in total for each expectation value, distributed with 40\% to the circuit for $F_1$ and 30\% each to the circuits for $F_2$ and $F_3$.  We run the sampling 100 times at each time step, and we view the $n$th experiment, across all time steps, as one noise realization.  For each noise realization we compute all the expectation values and then the UVQPE and ODMD energy estimates as a function of time step.  In the figure, the plotted value of the energy at each time step is the mean of the energies at that time step across all noise realizations, and the corresponding error bar is the standard deviation, indicating the range of energy estimates consistent with this level of shot noise. We use three singular value thresholds, $\delta=10^{-1}$, $10^{-3/2}$, and $10^{-2}$.  The largest $\delta$ is above the scale of the shot noise, and both algorithms converge.  The smallest $\delta$ is below the noise level, so the energy estimates do not converge.  With an intermediate level of noise filtering, the energy starts to converge after a long plateau.
    
	One question looking at the right panel of Fig.~\ref{fig:GS_results_exact_sampled_8} is whether the algorithms fail to converge at all, or instead converge to some excited state rather than the ground state.  The apparent convergence of the energy to a fixed value above that of the ground state is in fact a result of averaging over high-energy states and over 100 noise realizations.  To show this, in Fig.~\ref{fig:GS_results_exact_sampled_8_one_shot} we plot the convergence in UVQPE of both energy and eigenvector, in the style of Fig.~\ref{fig:UVQPE_GS_12_spin_low_overlap}, for one representative shot noise realization for each of the three singular value thresholds.  The convergence is clear for $\delta=10^{-1}$.  For the too-small value $\delta=10^{-2}$, the estimated eigenvector continues to have large components of multiple high-energy eigenstates even at long times.  In the case of intermediate $\delta$, the estimated eigenvector fluctuates for a significant time before suddenly converging.  (In some cases the convergence occurs through a series of such jumps after long plateaus in the energy estimate.)  Thus the appearance in Fig.~\ref{fig:GS_results_exact_sampled_8} of a long plateau followed by slow convergence apparently represents an average over many trajectories in which the sudden convergence appears at different numbers of Krylov steps used.
    
    \item In Fig.~\ref{fig:GS_results_exact_sampled_12}, we show the same calculations but for the 12-spin plaquette.  We use the higher-overlap initial state with three CZ operators; however, because the overlap is still smaller than for the the 8-spin plaquette, we use $10^4$ total shots per expectation value to reduce the noise level.  The energy converges with $\delta=0.1$.  However, it converges to a low-lying excited state for $\delta=0.3$, which is too large relative to the initial overlap, and it fails to converge at all for $\delta=0.01$, which is too small relative to the shot noise.
    
\end{itemize}

\begin{figure*}
    \centering
    \includegraphics[width=0.98\textwidth]{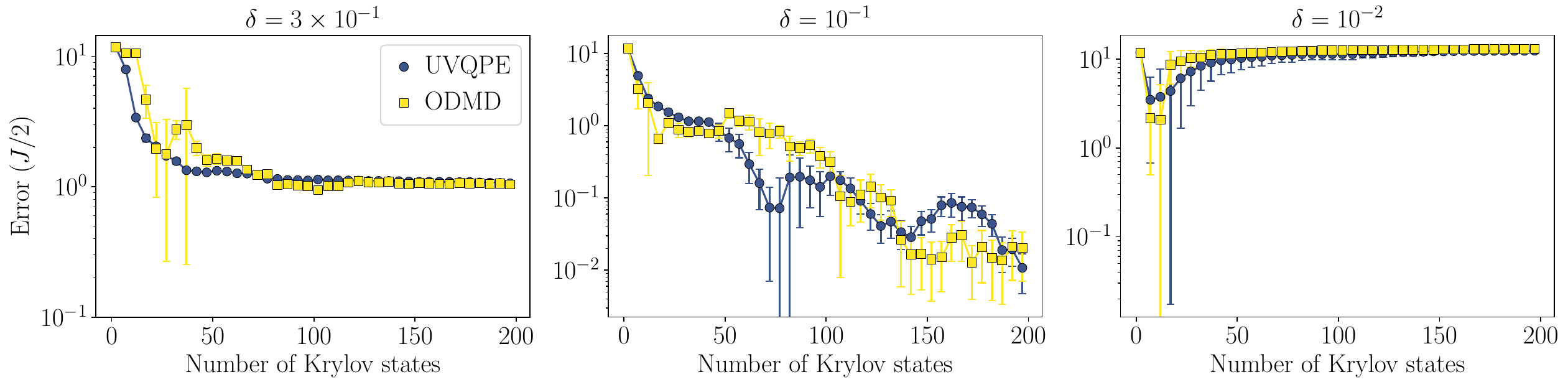}
    \caption{We perform the noisy version of the numerical experiment on the 12-spin plaquette from Fig.~\ref{fig:GS_results_exact_12_v2}, using the same approach as in Fig.~\ref{fig:GS_results_exact_sampled_8} with $10^4$ shots per expectation value.  We observe convergence for a range of $\delta$ around 0.1---large enough to filter out the shot noise, but small enough not to filter out the signal from the overlap between the initial state and the ground state.}
    \label{fig:GS_results_exact_sampled_12}
\end{figure*}

To summarize, we find that both algorithms rapidly converge to the ground state energy even in the presence of (normally distributed) statistical noise, so long as we use a noise-filtering technique to regularize the linear algebra problems of Eqs.~\eqref{eq:UVQPE} and \eqref{eq:ODMD}.  

The appropriate level of noise filtering is a balancing act: if the threshold is below the true noise level (e.g. Fig.~\ref{fig:GS_results_exact_sampled_8}, right panel), the noise has too strong an effect and the algorithm doesn't converge; on the other hand, if the threshold is larger than the overlap between the initial state and the ground state (e.g. Fig.~\ref{fig:GS_results_exact_12}, left panel),  the desired signal is also filtered out and the algorithm may converge to a low-lying excited state instead of to the ground state.  However, as long as the initial state overlap is larger than the noise level, there will be a filtering threshold that allows both UVQPE and ODMD to converge.  

\subsection{Noisy emulator and hardware results\label{sec:results_Quantinuum}}

Finally, we turn to a demonstration of UVQPE and ODMD on a real quantum device, namely the Quantinuum H1-1 processor.  We also use the corresponding noisy classical emulator provided by Quantinuum. By running on real hardware and on a noisy emulator, we show that the algorithm is robust not only to shot noise, as demonstrated above, but also to a variety of other coherent and incoherent noise sources such as state preparation and measurement errors, amplitude damping, and dephasing~\footnote{The quantity extracted from each circuit in our algorithms is just the probability of $|\mathbf{0}\rangle$ in each of the states $|0(t)\rangle$, $|0_R^{ }(t)\rangle$, and $|0_{R_i}^{ }(t)\rangle$.  These probabilities are affected by dephasing, which can be converted into bit flip errors both during time evolution and when applying the inverse state preparation unitaries $U_0^\dg$, $U_R^\dg$, and $U_{R_i}^\dg$.  Note that dephasing that occurs near the end of the circuit will not impact the final measurement, so in some cases dephasing will be less impactful than other noise sources.  On the other hand, dephasing just before a layer of $R_Z^{ }$ gates used for twirling will not be reduced, so such errors could be more problematic than expected despite error mitigation.}.  

As we saw in Fig.~\ref{fig:GS_results_exact_sampled_12} above, for the 12-spin plaquette there is only a narrow range of noise-filtering thresholds $\delta$ that give robust convergence, even when we use $10^4$ shots with no other source of noise.  Given the high per-shot cost on the Quantinuum device, we therefore perform the hardware demonstration only on the 8-spin plaquette, where we can use $10^3$ shots per expectation value.  We use the initial state from Sec.~\ref{sec:star_plaquettes_psi0} and a time step size of $0.1(J/2)$, the same as in the classical simulations reported in Fig.~\ref{fig:GS_results_exact_sampled_8}.  

Unlike the noiseless simulations, where we computed the states \eqref{eq:0t}, \eqref{eq:0Rt}, and \eqref{eq:0Rit} exactly, here we are limited by (real or emulated) hardware noise to shorter circuits and thus make the $U_t\rightarrow F_t$ approximation discussed in Sec.~\ref{sec:star_plaquettes_Trotter}, so that our longest circuits have at most 78 CNOT gates.  For running on the Quantinuum hardware and emulator, we transpile our circuit to the correct gate set using BQSKit~\cite{BQSKit, younis2022transpile}.

\begin{figure}
    \centering
    \includegraphics[width=0.98\columnwidth]{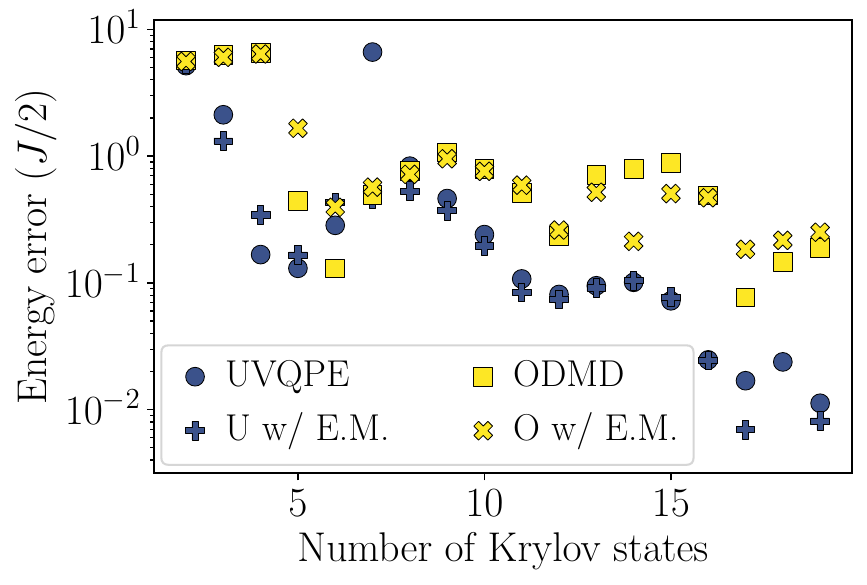}
    \caption{Convergence of energy estimate from UVQPE and ODMD for the 8-spin plaquette using expectation values measured with a noisy emulator that approximates the behavior of the Quantinuum quantum processor.  Color corresponds to the algorithm used.  Convex points (circle, square) show the results when no error mitigation is applied, while non-convex points ($\times$, +) show the results with two forms of symmetry-based error mitigation: symmetry sector post-selection and symmetry-based twirling.  The effects of error mitigation are small, likely because shot noise is the dominant noise source (see Fig.~\ref{fig:overlap_errors}).}
    \label{fig:emulator_convergence}
\end{figure}

\begin{figure}
    \centering
    \includegraphics[width=0.98\columnwidth]{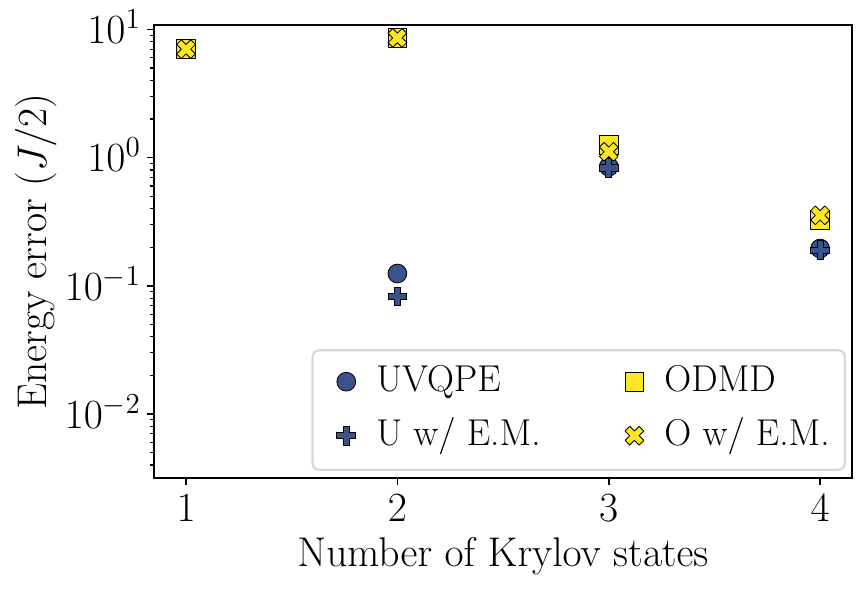}
    \caption{Convergence of energy estimate from UVQPE and ODMD for the 8-spin plaquette using expectation values measured on the Quantinuum H1-1 processor, i.e. on quantum hardware.  Even when measuring just five expectation values of the form $\langle F_{k\Delta t}\rangle$, we can obtain the correct ground state energy up to an error on the order of $10^{-1} (J/2)$.  As shown in Fig.~\ref{fig:overlap_errors}, with the number of shots used in these experiments, expectation values computed on hardware are statistically indistinguishable from those computed using the noisy emulator, so with more time steps the energy would further converge as in Fig.~\ref{fig:emulator_convergence}.}
    \label{fig:hardware_convergence}
\end{figure}

Using this approximation, we compute the expectation values $\langle F_{k\Delta t}\rangle$ up to 20 time steps using the noisy classical emulator, which as we show in Fig.~\ref{fig:emulator_convergence} is enough to converge the ground state energy to an accuracy of $10^{-2}(J/2)$.  We run on hardware for five of the time steps, with spacing $2\Delta t=0.2(J/2)$ through $10\Delta t=1.0(J/2)$.  Because the spacing $2\Delta t$ is still small enough to satisfy the bound $\Delta t < 2\pi/(3JN_\Delta)$, we can also estimate the ground state energy purely from the hardware results, with an accuracy of $10^{-1}(J/2)$ as shown in Fig.~\ref{fig:hardware_convergence}.  

\begin{figure*}
    \centering
    \includegraphics[width=0.98\textwidth]{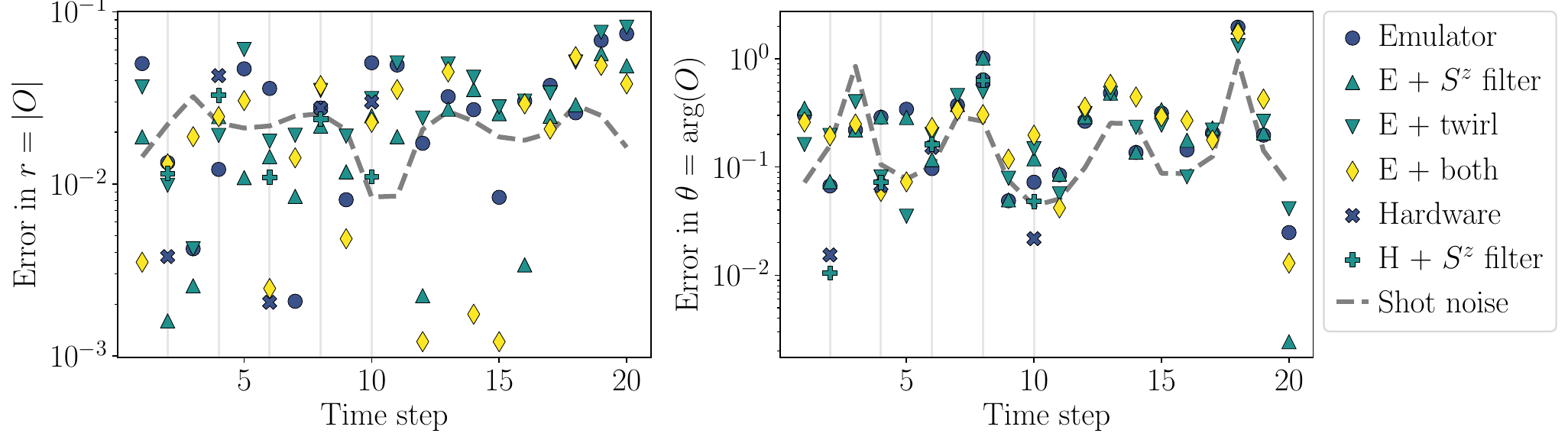}
    \caption{Error in computing the expectation values $O=\langle F_{k\Delta t}\rangle$ at time steps $k=1$ through 20, using the Quantinuum H1-1 processor and the corresponding noisy classical emulator for the 8-spin plaquette, with 1000 shots per expectation value.  Letting $O\equiv re^{i\theta}$, the left panel shows the error in the magnitude and the right panel the error in the angle.  Emulator results are shown by convex symbols (circle, triangle, diamond), while hardware results at the five time steps indicated by vertical lines are denoted by non-convex markers ($\times$ and +).  Color shows the number of symmetry-based error mitigation techniques applied: none for blue, one for green, and two for yellow.  For comparison, a classical estimate of the shot noise is shown as a dashed gray line.  Shot noise appears in this case to be the dominant noise source. }
    \label{fig:overlap_errors}
\end{figure*}

For both the noisy emulator and the hardware implementation, we use a total of $10^3$ shots per expectation value: 400 for $F_1$ and 300 each for $F_1$ and $F_2$.  With the classical emulator, we also run half as many shots (200 for $F_1$, 150 each for $F_2$ and $F_3$) with a layer of $R_Z$ gates inserted as described in Sec.~\ref{sec:star_plaquettes_error_miti}; replacing (a randomly selected) half of the original measurements with the $R_Z$-inserted, or ``twirled'' measurements slightly reduces the error in the expectation values as shown in Fig.~\ref{fig:overlap_errors}.  We also show, for both noisy emulator and hardware, the effect of symmetry sector-based post selection.  

Figs.~\ref{fig:emulator_convergence} and \ref{fig:hardware_convergence} show the energy estimates both with and without the symmetry-based error mitigation.  The effect on the convergence appears to be minimal, which we attribute to the fact that shot noise is actually the dominant noise source.  This can be seen in Fig.~\ref{fig:overlap_errors} where, alongside the actual error in expectation values from noisy emulator and hardware, we also show an estimate of shot noise given 1000 total shots per time step.  In particular, for each time step we run an exact classical simulation (including the $U\rightarrow F$ approximation) 100 times, and we take the standard deviation among the 100 experiments as the approximate shot noise.  The large shot noise explains why, for some expectation values, the error mitigation strategies actually increase rather than decrease the error.

For the longer circuits and greater number of shots that would be needed to study a much larger two-dimensional system, we expect that the relative importance of noise channels like amplitude damping and dephasing would increase.  In the noise filtering step of classical post-processing, in order to allow for a singular value threshold $\delta$ that is simultaneously larger than the error in the matrix elements and smaller than the overlap of the initial state with the ground state, an early fault-tolerant quantum computer might be needed to reduce such noise channels to manageable levels.  A thorough investigation of the relative importance of each type of error would require much larger systems and many more shots, and thus is beyond the scope of the current paper.


\section{Magnetization curve results\label{sec:mag_curve}}

As previously noted, for an antiferromagnetic model such as the Heisenberg model considered in this paper, when there is no magnetic field ($h=0$) Lieb's theorem~\cite{Lieb1989} guarantees that the ground state will be in the $S^z=0$ sector.  In  other words, there is no net magnetic moment.  On the other hand, if the field $h$ is much larger than the interaction strength $J$, the ground state will be the fully polarized state with all spins pointing along the direction of the field, giving the maximum possible magnetization.  At intermediate field strengths, there will be some tendency for spins to align with field, giving rise to the magnetization curve: the mean of $\langle S^z\rangle$ over all spins, measured in the ground state, as a function of the applied field $h$.

For the kagome lattice Heisenberg model, the magnetization curve is believed to have a series of plateaus, where the magnetization remains constant as the applied field increases~\cite{Hida2001,Schulenburg2002,Honecker_2004,Shimokawa2013,Nishimoto2013,Nakamura2020,Schluter2022,Liu2022}.  The plateaus correspond to various phases of matter that are stable to small changes in the applied field.  Most plateaus likely correspond to partial magnetic ordering, but some could be new exotic spin liquids different from the one expected at $h=0$.  Quantum computers may help to conclusively determine the nature of these plateau states, and finding the magnetization curve itself is the first step.

Fortunately, for a Hamiltonian that conserves $S^z$, such as the Heisenberg model, we can find the magnetization curve simply by computing the $h=0$ ground state energy within each $S^z$ symmetry sector.  The reason is that the field term in $H$ is exactly the conserved quantity, total $S^z$.  Therefore every eigenstate at $h=0$ is also an eigenstate at finite $h$, with its change in energy just given by $-h\sum S^z$.  In particular, the energies of all states in a given symmetry sector change in the same way as $h$ is increased, so for each sector the $h=0$ ground state remains the lowest in energy for any $h$.  Furthermore, the energy eigenvalues themselves are easily computed from the $h=0$ eigenvalues.

\begin{figure}
    \centering
    \includegraphics[width=0.98\columnwidth]{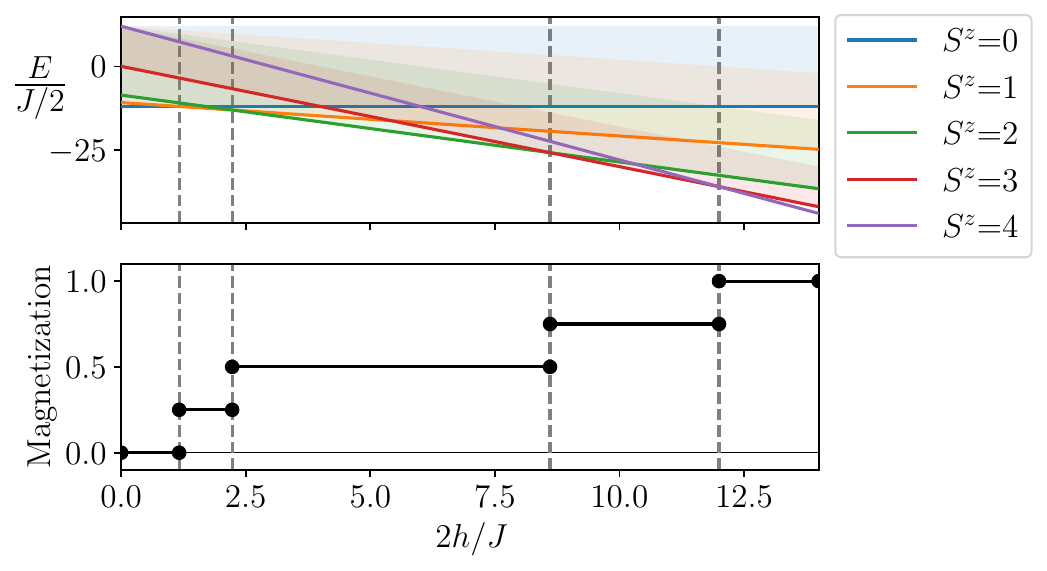}
    \caption{Demonstration of how to derive the magnetization curve from $h=0$ ground state energies, illustrated using the 8-spin plaquette.  Top: Each color corresponds to one $S^z$ symmetry sector: the solid line indicates the ground state within that sector, and the shaded region shows the full range of eigenvalues for the sector.  The energies simply vary linearly in $h$ starting with $h=0$ on the left.  The ground state switches from one magnetization sector to the next when the ground state lines cross at the locations indicated by dashed vertical lines.  Bottom: We find the magnetization curve by plotting the magnetization of the symmetry sector with the lowest-energy ground state for each $h$.  The value jumps at the crossing points because there are only a few possible values of magnetization, but the curve could be continuous for an infinite 2D system.}
    \label{fig:mag_curve_8}
\end{figure}

\begin{figure}
    \centering
    \includegraphics[width=0.98\columnwidth]{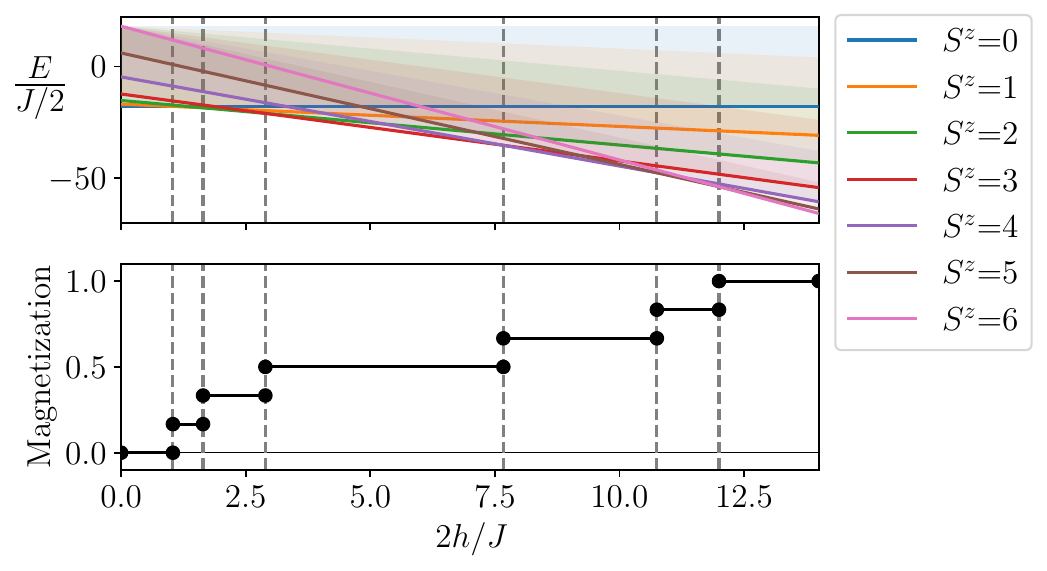}
    \caption{Same as Fig.~\ref{fig:mag_curve_8} but for the 12-spin plaquette.}
    \label{fig:mag_curve_12}
\end{figure}

We show this computation of the magnetization curve using the exact eigenvalue spectra of the 8- and 12-spin plaquettes in Figs.~\ref{fig:mag_curve_8} and \ref{fig:mag_curve_12}, respectively.  In each figure, the lines in the upper panel show how the energy of the ground state within each spin sector evolves with increasing $h$.  At each crossover point, where a new spin sector becomes the ground state, the magnetization jumps, giving the magnetization curve shown in the lower panel.

Here we used the exact eigenvalues to illustrate the approach to finding the magnetization curve.  However, our goal is to show that this calculation is also feasible using UVQPE and/or ODMD.  Once we have the ground state energy in each $S^z$ symmetry sector at $h=0$, the computation of the magnetization curve is identical to the exact calculation above.  We therefore simply demonstrate finding the ground state energy in all symmetry sectors using UVQPE.

\begin{figure*}
    \centering
    \includegraphics[width=0.98\textwidth]{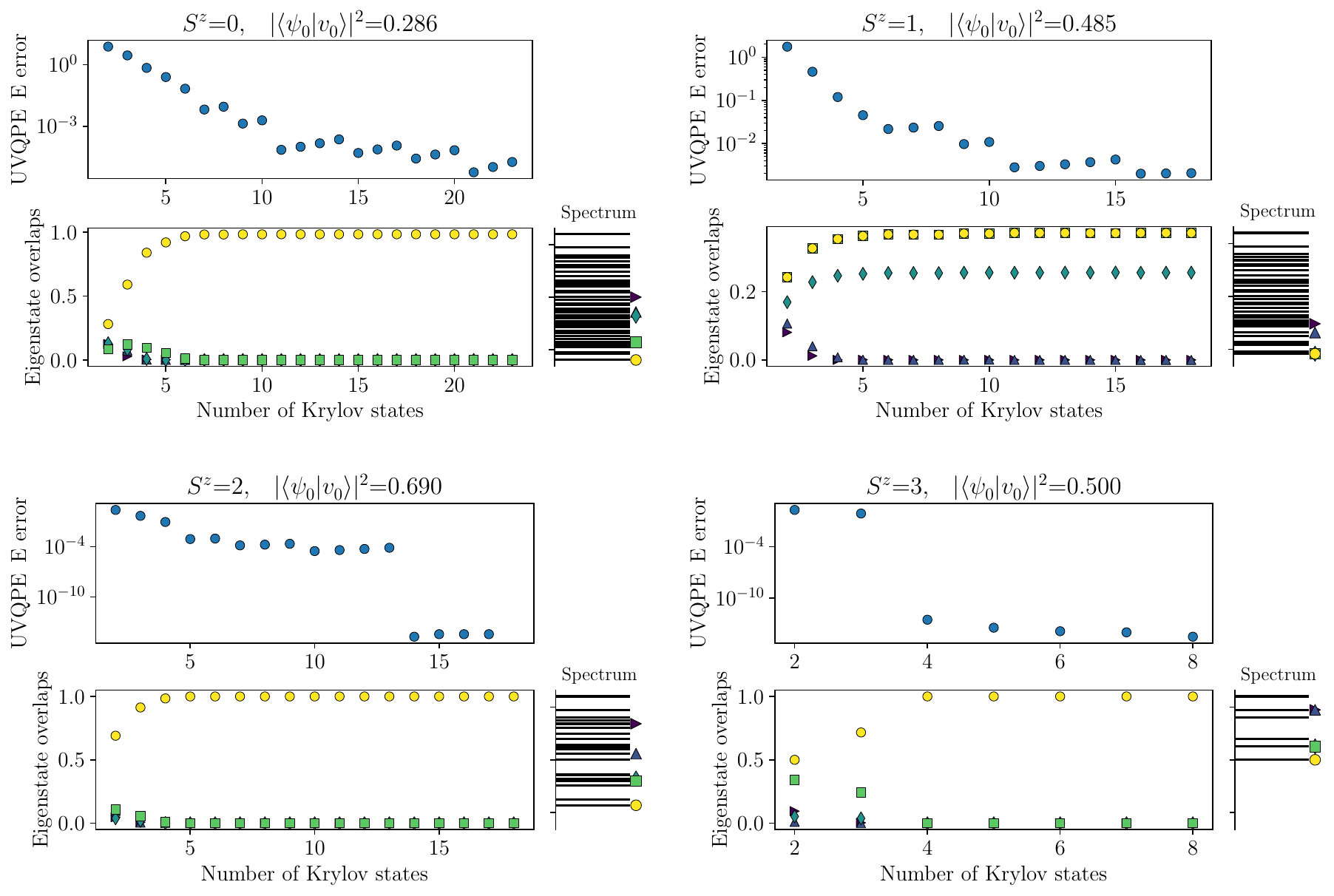}
    \caption{Convergence of ground state energy and corresponding approximation to the ground state eigenvector in the different spin sectors of the 8-spin star plaquette.  Each panel corresponds to one $S^z$ symmetry sector, and is also labeled with the overlap between the initial state $|\psi_0\rangle$ and the true ground state in that sector, $|v_0\rangle$.  We measure the convergence of the UVQPE approximation to the ground state by finding its overlap with each eigenstate of $H$, and we plot the overlaps for the five eigenstates that contribute most to the approximation to the ground state across the simulation.  The energies of those eigenstates are shown at the right of each panel along with the full spectrum in the corresponding $S^z$ symmetry sector.  Energy error is measured in units of $J/2$.}
    \label{fig:UVQPE_GS_8_spin}
\end{figure*}

In Fig.~\ref{fig:UVQPE_GS_8_spin}, in the upper half of each panel, we show the convergence of ground state energy for the four distinct non-trivial $S^z$ symmetry sector of the 8-spin plaquette.  We use the initial states discussed in Sec.~\ref{sec:star_plaquettes_psi0}; each panel is labeled by the overlap between the initial state $|\psi_0\rangle$ and the ground state(s) $|v_0\rangle$.  We run UVQPE using exact classical simulation, with no Trotter error or shot noise, and we use a singular value threshold of $\delta=10^{-6}$.  All symmetry sectors converge to the true ground state energy with high precision within 20 steps.

\begin{figure*}
    \centering
    \includegraphics[width=0.98\textwidth]{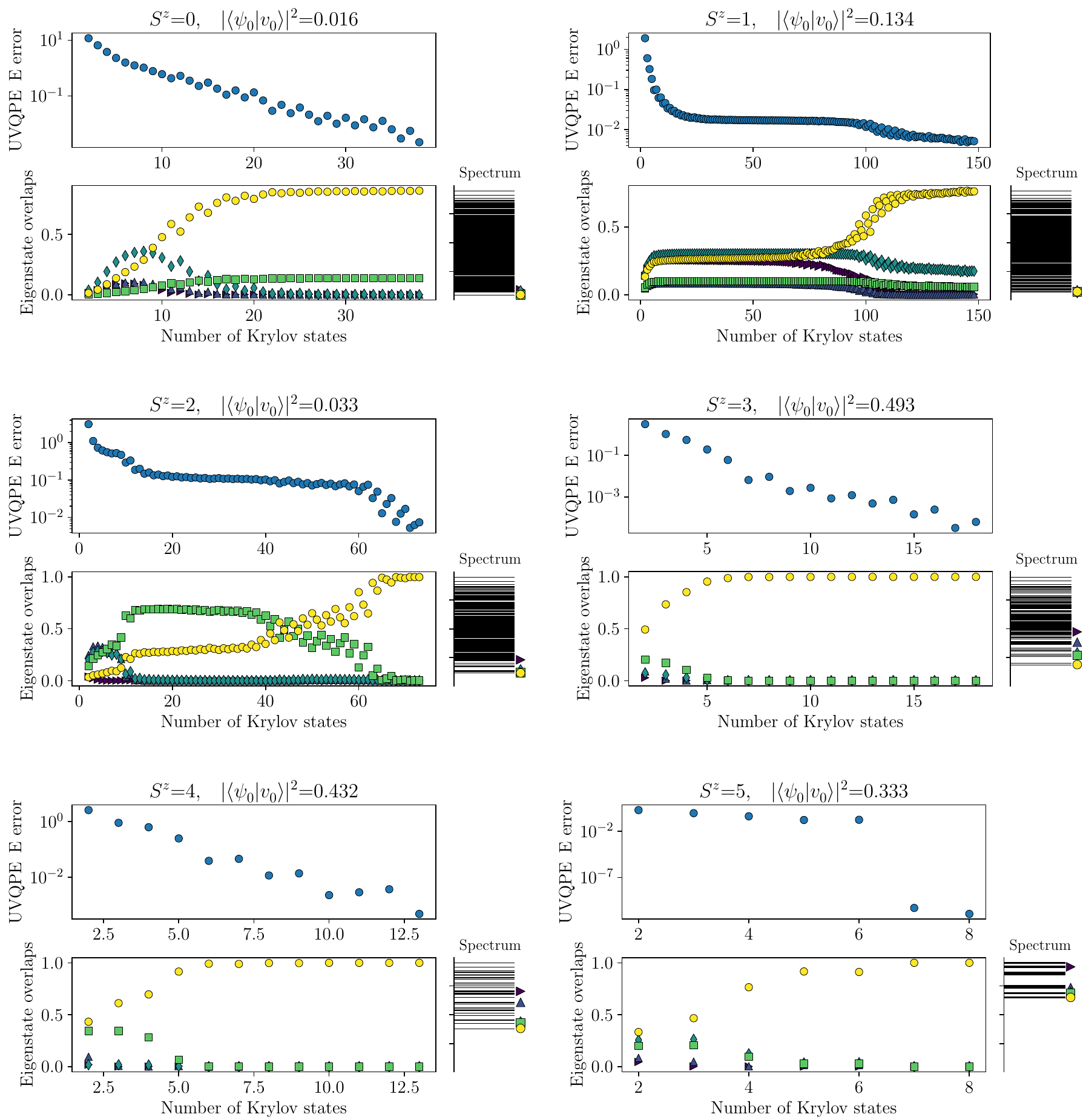}
    \caption{Here we show the same analysis as in Fig.~\ref{fig:UVQPE_GS_8_spin}, but for the 12-spin plaquette.  Note that convergence is slower both when the initial overlap is smaller and when the initial state has significant overlap with several eigenstates that all have relatively low energy.}
    \label{fig:UVQPE_GS_12_spin}
\end{figure*}  

Also recall from Sec.~\ref{sec:algs_UVQPE} that UVQPE gives an approximation to the ground state eigenvector, $|\tilde{v}_0\rangle$, in the form of a linear combination of the basis states \eqref{eq:basis_states}.  For the small 8- and 12-spin plaquettes, we can explicitly generate the basis states and thus the specified linear combination, and we can find the overlap of $|\tilde{v}_0\rangle$ with the true ground state and other eigenvectors of the Hamiltonian.  To help understand the convergence of the ground state energy from UVQPE, we also show in Fig.~\ref{fig:UVQPE_GS_8_spin}, in the lower half of each panel, the convergence of the approximate ground state eigenvector.  Specifically, we plot the overlap of five eigenvectors of $H$ with $|\tilde{v}_0\rangle$, observing how the overlap with the true ground state $|v_0\rangle$ grows as the number of time steps/Krylov basis vectors increases, while the overlaps with the other eigenvectors decrease. 

To gain additional insight into the relationship between the convergence of the approximations to the energy and to the eigenvector, it is helpful to consider worse initial states that have a much lower overlap with the true ground state.  As we show for the $S^z=1$ and $S^z=2$ symmetry sectors in App.~\ref{appendix:UVQPE_eigvect_convergence}, the algorithm will often converge towards a low-lying excited state first, leading to a plateau in the energy convergence, before ultimately reaching a good approximation to the ground state.

We also test out UVQPE for finding the ground state in each $S^z$ sector, and hence the magnetization curve, for the 12-spin plaquette, with results shown in Fig.~\ref{fig:UVQPE_GS_12_spin}.  As in the 8-spin case, the UVQPE simulations are carried out using exact time evolution and with no shot noise, using initial states from Sec.~\ref{sec:star_plaquettes_psi0}.  Note that for $S^z=0$ we use the initial state with three CZ operators, from the left of Eq.~\eqref{eq:pinwheel_with_CZ_states}.  Here the convergence behavior, which we can understand by looking at the eigenvector convergence in the lower half of each panel, shows some interesting features.  In the $S^z=0$ sector, the eigenvector initially has a significant component of an excited state, though less prominently than in the case of the six-CZ initial state as shown in Fig.~\ref{fig:UVQPE_GS_12_spin_low_overlap} above.  Although it looks like weight remains in two different eigenstates in the long-time limit, those are actually two orthogonal states both within the ground state subspace.  Interestingly, convergence is slower for $S^z=1$ even though the initial state overlap is much larger, which occurs because the initial state $|\psi_0\rangle$ has similarly high overlap with several low-lying excited states.

The most important takeaway message, however, is that for both plaquettes and in all symmetry sectors, UVQPE eventually converges to the correct ground state energy.  As a result, the algorithm (likewise, ODMD) can be used to compute magnetization curves effectively following the prescription from Figs.~\ref{fig:mag_curve_8} and \ref{fig:mag_curve_12}.


\section{Discussion\label{sec:discussion}}

In this paper, we have proposed that hybrid quantum-classical algorithms based on real-time evolution provide a promising route to resolving open problems in frustrated magnetism and quantum spin liquid physics, including on the challenging kagome lattice Heisenberg model.  We focused particularly on two recent algorithms, UVQPE~\cite{UVQPE_Klymko} and ODMD~\cite{ODMD}, that use a series of expectation values of the time evolution operator, $\langle\psi_0|e^{-iHt}|\psi_0\rangle$ for $t=\Delta t$, $2\Delta t$, $\cdots$, to find an approximation to the ground state energy and give access to the ground state itself as a superposition of the states $e^{-ik\Delta t}|\psi_0\rangle$.  The expectation values are measured on a quantum computer and are post-processed classically via a small generalized eigenvalue or linear least squares problem to find the ground state.

We have made three main contributions.  First, we provided a compact and practical summary of UVQPE and ODMD in Sec.~\ref{sec:algs}.  We also give a detailed description of the mirror circuit method~\cite{Cortes2022} for computing expectation values and wave function overlaps on a quantum computer.  We show how to resolve a phase ambiguity in previous versions of the mirror circuit method, and we perform a careful numerical analysis of how to distribute a measurement budget among the various required quantum circuits.

Second, we outlined a detailed approach to running UVQPE and ODMD on the 2D kagome lattice Heisenberg model.  We specifically highlighted the uses of $S^z$ spin symmetry, namely to enable the mirror circuit method, to reduce the effective Hilbert space size, and to perform error mitigation.  We then described an efficient ``triangle-by-triangle'' implementation of Trotterized time evolution, and we showed how to choose and prepare a good initial state $|\psi_0\rangle$ in which to calculate the expectation values $\langle\psi_0|e^{-iHt}|\psi_0\rangle$.  Finally, we argued that the cost of the algorithms does not scale too quickly with system size: the scaling is exponential due to translation invariance of the ground state, but the exponential has a very small base.  

Third, we provided an empirical demonstration of UVQPE and ODMD applied to a single 12-spin star plaquette of the kagome lattice and an analogous 8-spin plaquette.  Remarkably, these apparently frustrated small systems are actually ``frustration-free'' with an exact classically solvable ground state.  We make use of this exact solvability to design good initial states and reduce the required circuit depth for Trotter evolution.  For the 8-spin plaquette this allows a compact and efficient simulation on both quantum hardware, namely the Quantinuum H1-1 processor, and a corresponding noisy classical emulator.  We find that even in the presence of both shot noise and (real or emulated) hardware noise, both algorithms converge rapidly to the true ground state energy.

Although the latter two points are specific to one model, UVQPE and ODMD are effective in a wide variety of systems.  Even in models with fewer symmetries to exploit, the basic algorithms of UVQPE and ODMD are still effective, for example by using the Hadamard test to find matrix elements, rather than using the mirror circuit approach.  

Conclusively determining the true ground state of the kagome lattice Heisenberg model and other models of frustrated magnetism will require more qubits than are available in present-day quantum processors, as well as lower noise levels.  However, we have shown, with theoretical analysis of 2D systems and practical demonstrations on small systems, that hybrid algorithms based on real-time evolution such as UVQPE and ODMD are strong candidates for solving these important problems in condensed matter and materials physics.  Even before the advent of fully fault-tolerant error-corrected quantum computers, these algorithms may allow for new physical insights that have not been possible via classical computation.


\begin{acknowledgments}
We thank Katie Klymko, Yizhi Shen, Roel Van Beeumen, Daan Camps, and Siva Darbha for their insights into real-time evolution-based hybrid algorithms.  The use of the dimer ground states was inspired partially by a conversation with Ehud Altman.  We thank Yan Wang for discussions of the exact ground state of the star plaquette model and Mike Kolodrubetz for a discussion about the resulting Floquet evolution.  Circuit diagrams were prepared using Quantikz~\cite{quantikz}.  This work was supported by the “Embedding QC into Many-body Frame-works for Strongly Correlated Molecular and Materials Systems” project, which is funded by the U.S. Department of Energy, Office of Science, Office of Basic Energy Sciences (BES), the Division of Chemical Sciences, Geosciences, and Biosciences, and by the Office of Science, Office of Advanced Scientific Computing Research Accelerated Research for Quantum Computing Program of the U.S. Department of Energy. This research used resources of the Oak Ridge Leadership Computing Facility, which is a DOE Office of Science User Facility supported under Contract DE-AC05-00OR22725. Our hardware demonstration used the Quantinuum H1-1 processor.  
\end{acknowledgments}

\appendix

\section{Shot noise distribution for overlap from mirror circuits\label{appendix:mirror_shot_noise}}

As noted in the main text, Sec.~\ref{sec:algs_mirror}, when computing expectation values using the mirror circuit method, there are two important choices we need to make.  First, with a fixed measurement budget, we must decide how to allocate shots between the three circuits for measuring $F_1$, $F_2$, and $F_3$.  Second, we can choose whether to compute the magnitude of the expectation value using Eq.~\ref{eq:mirror_circuit_computation} or instead to use $\sqrt{F_1}$.  

\begin{figure*}
    \centering
    \includegraphics[width=\textwidth]{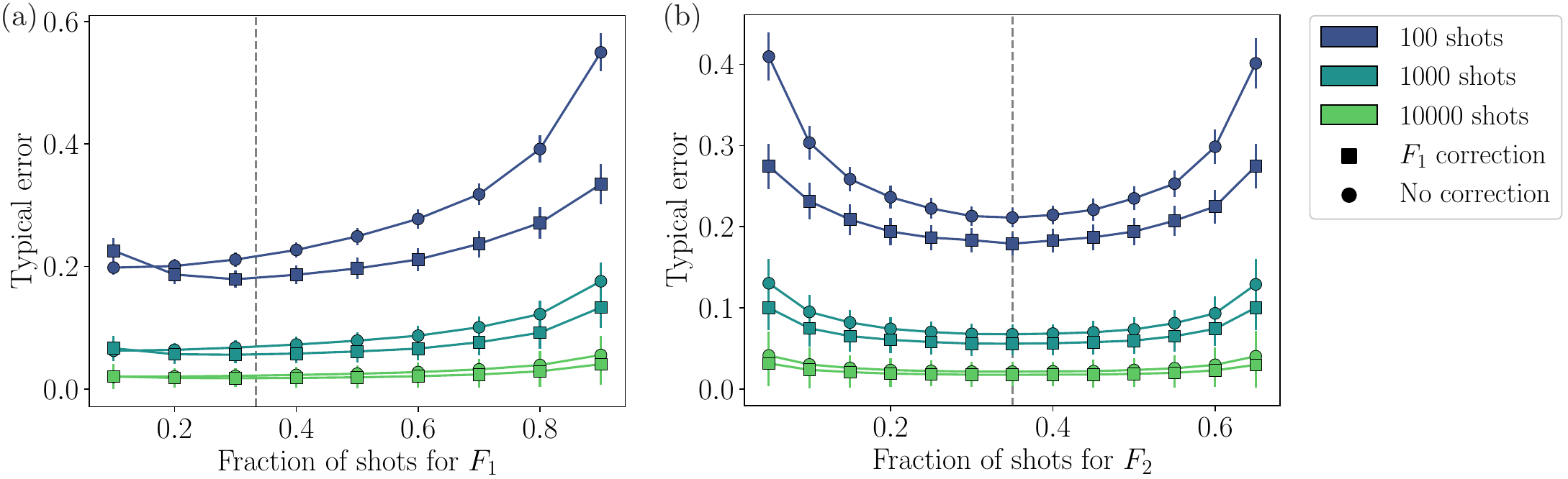}
    \caption{Investigation of the optimal allocation of shots between the circuits for $F_1$, $F_2$, and $F_3$ used in the mirror circuit with reference state approach to computing matrix elements.  For each data point, we fix a total number of shots to be divided among the three circuits, indicated by color in the figure.  Error is measured as $|O - O_m|$, where $O$ is the exact overlap and $O_m$ the estimated one from the sampled circuit output.  For each data point, we run the experiment for 100 different time steps and for each one perform the sampling with the specified number of shots 100 times; the ``typical error'' is the standard deviation of the error over all $10^4$ shots.  Error bars indicate the standard deviation of the typical error over batches of such experiments.  In (a), we vary the fraction of shots assigned to the circuit for $F_1$, with the $F_2$ and $F_3$ circuits evenly dividing the remaining measurement budget.  Squares indicate that we have used only $F_1$ to compute the magnitude, while circles compute $O_m$ using Eq.~\eqref{eq:mirror_circuit_computation}.  Evidently the optimal choice is to recompute the magnitude using just $F_1$, and furthermore to use approximately the same number of shots for each of the three circuits, indicated by the dashed vertical line.  In (b), we fix the fraction of shots assigned to $F_1$ at 0.3, then vary the fraction assigned to $F_2$ with $F_3$ taking the remainder.  The optimal choice is clearly to take an equal number of shots for $F_2$ and for $F_3$.}
    \label{fig:mirror_meas_budget}
\end{figure*}

\begin{figure*}
    \centering
    \includegraphics[width=0.98\textwidth]{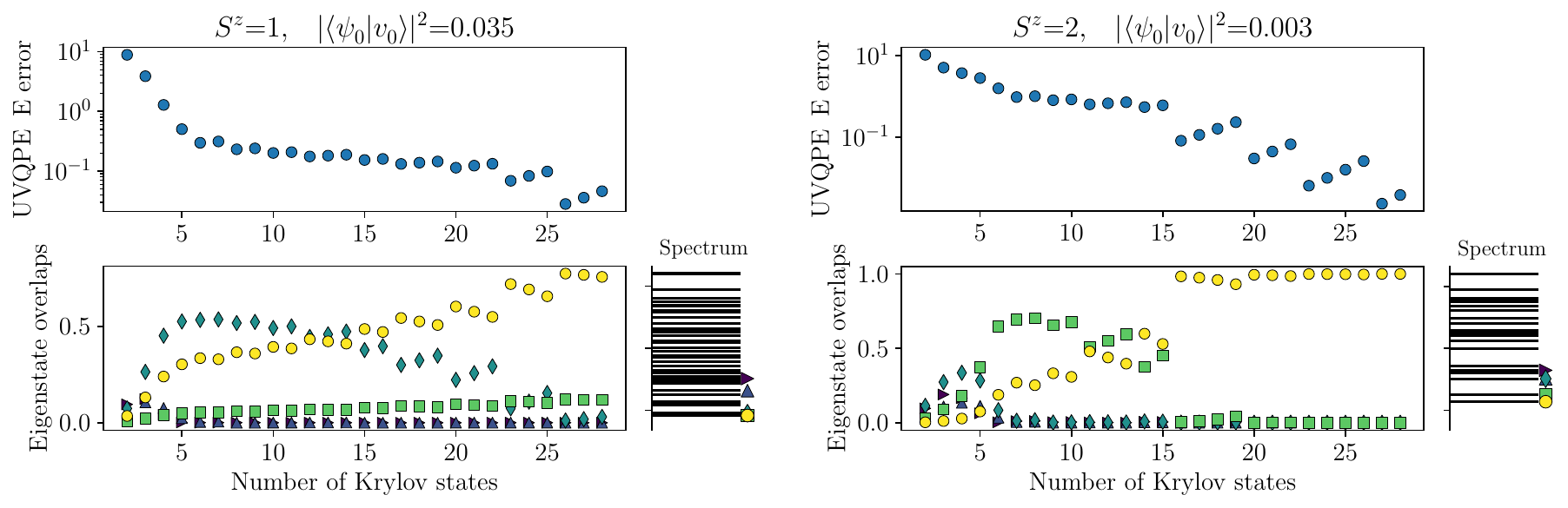}
    \caption{Here we show the same analysis as in Fig.~\ref{fig:UVQPE_GS_8_spin}, but with initial states for the 8-spin plaquette that have much lower overlap with the ground states of their respective symmetry sectors.  Convergence takes longer because UVQPE first populates multiple low-lying states, after which the energy estimate plateaus while weight shifts from low excited states to the ground state.}
    \label{fig:UVQPE_GS_8_spin_low_overlap}
\end{figure*}

We empirically investigate both of these questions using the expecatation values $\langle \psi_0|e^{-iHt}|\psi_0\rangle$ for the 8-spin plaquette with the $S^z=0$ initial state from Sec.~\ref{sec:star_plaquettes_psi0}, with 10 different times $t$.  For each time, we compute the states from which $F_1$, $F_2$, and $F_3$ are sampled.  We then run 100 experiments for each $t$, where in each one we sample from these states $m$ times to measure the three quantities and hence compute the overlap, $O_m$.  We estimate the typical error due to shot noise as the standard deviation of the error in the overlap, $|O-O_m|$, over all experiments and all $t$ for a given number of shots $m$.  (Note that $O$ and $O_m$ are complex, and this error takes into account the errors in both magnitude and phase.)  We use $m=10^2$, $10^3$, and $10^4$.

As we show in Fig.~\ref{fig:mirror_meas_budget}, the most accurate results for the expectation values are found when the number of shots devoted to $F_1$, $F_2$, and $F_3$ are roughly equal.  This is true regardless of which method is used to compute $|O|$.  Furthermore, we find that the typical error is substantially lower when computing $|O|$ as $\sqrt{F_1}$ rather than with Eq.~\eqref{eq:mirror_circuit_computation}, which is reasonable since a convolution of probability distributions has a larger variance than the constituent distributions.


\section{UVQPE ground state eigenvector convergence with bad initial states\label{appendix:UVQPE_eigvect_convergence}}

In Sec.~\ref{sec:mag_curve}, we saw that the convergence of ground state energy from UVQPE can be understood via the convergence of the approximation to the corresponding eigenvector.  For the 8-spin plaquette, where the initial state in each $S^z$ symmetry sector had a high overlap with the true ground state, both the energy and corresponding approximate eigenvector converged quickly.  In contrast, on the 12-spin plaquette we observed more complex behavior, especially when the initial state had larger overlap with low-lying excited states than with the ground state.  Here we show similar behavior on the smaller 8-spin plaquette, in the $S^z=1$ and $S^z=2$ sectors, by using initial states whose overlaps with the true ground states of those symmetry sectors are just 3\% and 0.3\%, respectively.  We observe that the energy first decreases as UVQPE populates multiple low-lying states, plateaus for some time, then ultimately decreases again as the weight shifts towards only the overall ground state.  If a larger threshold $\delta$ is used, the convergence takes much longer; for $S^z=2$ with $\delta=0.1$, UVQPE converges to the first excited state rather than to the ground state.  Since a cutoff $\delta$ of this magnitude is needed in a noisy simulation, a very small overlap of less than a percent may be insufficient for UVQPE to be used in practice.


\bibliographystyle{modified-apsrev4-2}
%


\end{document}